\documentclass[12pt,a4paper]{article}
\pdfoutput=1
\usepackage[latin1]{inputenc}
\usepackage{ragged2e}
\usepackage{amsfonts,amsbsy,bm,euscript,mathrsfs}
\usepackage{amssymb,stmaryrd,faktor,slashed}
\usepackage{color}
\usepackage[tbtags]{amsmath}
\usepackage[bookmarks=true,colorlinks=true,linkcolor=black,citecolor=black,urlcolor=black,bookmarksnumbered]{hyperref}

\usepackage[a4paper,text={170mm,245mm},centering]{geometry}

\usepackage{authblk, physics, bbm, multirow, hhline}

\usepackage{graphicx}
\usepackage[verbose]{wrapfig}
\usepackage{caption}

\numberwithin{equation}{section}

\renewcommand{\arraystretch}{1.3}

\makeatletter
\renewcommand\section{\@startsection {section}{1}{\z@}%
	{-3.5ex \@plus -1ex \@minus -.2ex}%
	{2.3ex \@plus.2ex}%
	{\normalfont\large\bfseries}}
\renewcommand\subsection{\@startsection{subsection}{2}{\z@}%
	{-3.25ex\@plus -1ex \@minus -.2ex}%
	{1.5ex \@plus .2ex}%
	{\normalfont\normalsize\bfseries}}
\makeatother

\expandafter\def\expandafter\bfseries\expandafter{\bfseries\ifmmode\else\boldmath\fi}
\expandafter\def\expandafter\mdseries\expandafter{\mdseries\ifmmode\else\unboldmath\fi}
\expandafter\def\expandafter\normalfont\expandafter{\normalfont\ifmmode\else\unboldmath\fi}

\providecommand{\href}[2]{#2}

\DeclareMathOperator\arsinh{arsinh}
\DeclareMathOperator\Str{Str}

\DeclareMathOperator\Pf{Pf}
\DeclareMathOperator\diag{diag}

\date{15.06.2023}

\begin{document}
\phantom{-}
\vspace{2.0cm}
\begin{center}
	
	{\Large\bf Integrability treatment of AdS/CFT orbifolds
		\vspace{0.3cm}
	}
	
	\vspace{1.0cm}
	
	{Torben Skrzypek\footnote{t.skrzypek20@imperial.ac.uk}
	}	
	
	\vspace{0.5cm}	
	
	{\em
		Blackett Laboratory, Imperial College, London SW7 2AZ, U.K.
	}

\end{center}

\vspace{0.5cm}

\begin{abstract}
	We elaborate on the treatment of orbifolds of type IIB string theory on $AdS_5\cross S^5$ and their dual gauge theories with integrability techniques. The implementation of orbifolds via twisted spin-chains, thermodynamic Bethe Ansatz equations with chemical potentials and $Y$- and $T$-systems with modified asymptotics is confronted with twisted boundary conditions of the string sigma-model. This allows us to consistently twist the quantum spectral curve, which is believed to bridge the two sides of the AdS/CFT duality. We discuss Abelian orbifolds of $PSU(2,2|4)$ and treat the special cases of $\mathcal{N}=2$ supersymmetric \linebreak {$\mathbb{Z}_2$-orbifolds} and type 0B string theory on $AdS_5\cross S^5$ as primary examples. This opens a pathway to probe the validity of the duality and to study the long-standing question of tachyon stabilisation in non-supersymmetric AdS/CFT. We comment on the current understanding of this issue and point out the next steps in this challenge.
\end{abstract}

\newpage
\tableofcontents
\setcounter{footnote}{0}
\setcounter{section}{0}
\

\section{Introduction}\label{intro}

One of the most intriguing discoveries in string theory has been the AdS/CFT correspondence \cite{Maldacena:1997re} that relates type IIB string theory on an $AdS_5\cross S^5$ background to $\mathcal{N}=4$ super Yang-Mills theory (SYM) with gauge group $SU(N)$. The duality is controlled by two parameters. On the gauge theory side these can be identified as the rank of the gauge group $N$ and the gauge coupling $g_{YM}$, while on the string theory side they consist of the $AdS$-radius $R$ in string units $\sqrt{\alpha'}$ and the string coupling $g_s$. They are related via the t' Hooft coupling
\begin{equation}\label{thooft}
\lambda:=g_{YM}^2N=\frac{R^4}{\alpha'^2}\,,\qquad g_{YM}^2=4\pi g_s\,.
\end{equation}

The situation simplifies significantly if we take the large-$N$ limit while keeping $\lambda$ fixed \cite{tHooft:1973alw}, as only planar diagrams contribute in the gauge theory and the string becomes weakly coupled. The parameter $\lambda$ can be chosen arbitrarily, but for our following discussions it will be useful to instead define the quantity 
\begin{equation}\label{coupling}
g:=\frac{R^2}{4\pi\alpha'}=\frac{\sqrt{\lambda}}{4\pi}\,,
\end{equation}
which we will often just refer to as the ``coupling constant". If we take $g$ to be small, we are in a perturbative regime of the gauge theory. On the other hand, we can take $g$ to be large and find the large-tension limit of string theory. Thus, we have a ``weak-strong duality" between these two well-controlled domains.
Though, despite a plethora of impressive matchings across this duality, e.g. for BPS operators, the full AdS/CFT conjecture remains unproven to this day. 

Type IIB string theory on $AdS_5\cross S^5$ can be described as a non-linear sigma-model with target space
\begin{equation}
\frac{PSU(2,2|4)}{SO(4,1)\cross SO(5)}\supset AdS_5\cross S^5\,.
\end{equation}
Due to it's high amount of symmetry, this sigma-model is classically integrable \cite{Bena:2003wd}. In the last 20 years, this property has been exploited to compute the spectrum of non-BPS operators in the CFT dual, leading to the development of a number of techniques now commonly summarised under the name of ``Integrability". These advances were in part inspired by the insight of Berenstein, Maldacena and Nastase (BMN) \cite{Berenstein:2002jq} that a string with large angular momentum $J$ along an angle of $S^5$ is dual to a single-trace operator 
\begin{equation}\label{BMNlike}
\mathcal{O}_J= \Tr(Z^J)=\Tr\,(\overbrace{ZZ\dots Z}^J)\,,
\end{equation}
where $Z$ is a complex scalar field in the adjoint representation of $SU(N)$, which has charge $J=1$ under one of the three Cartan elements of the $SU(4)$ $\mathcal{R}$-symmetry. The full spectrum of single-trace operators could be described as excitations on top of this ``ground state". It was observed that their dynamics are very constrained and that we can describe single-trace operators as integrable spin-chains \cite{Minahan:2002ve, Beisert:2003yb, Beisert:2006ez}. 

Using Bethe Ansatz techniques, the spectrum of single-trace operators could be determined in the weak-coupling limit \cite{Beisert:2005fw}. However, at finite coupling and finite size, the original spin-chain computations failed to account for wrapping corrections, which arise because the spin-chain has to be closed to a circle and interaction effects can extend around the entire chain, an effect which has been described already by L\"uscher \cite{Luscher:1985dn} in the 80s. 

A framework to account for these finite-size effects has been developed with the thermodynamic Bethe Ansatz (TBA) \cite{Zamolodchikov:1989cf,Arutyunov:2007tc,Bombardelli:2009ns, Arutyunov:2009ur}. The idea is to Wick-rotate the finite-size string/spin-chain moving along the infinite time-dimension, such that it becomes an infinitely extended string moving in compact time, i.e. at finite temperature.
The resulting TBA equations describe the ground state of the original theory and with some effort also some excited states. 

Taking advantage of the underlying $PSU(2,2|4)$ symmetry, it was possible to simplify the TBA equations to their core, introducing the $Y$- and $T$-systems \cite{Gromov:2009tv}. This led to the proposal of a new method to arrange the $T$-system in a number of $Q$-functions, whose algebraic relationships are determined by the symmetries and make manifest the ambiguities that arise with the arbitrary choice of ground state and ordering of excitations that were necessary in the spin-chain and TBA description. Together with the appropriate asymptotics and analyticity constraints, these $Q$-functions build up the quantum spectral curve (QSC) \cite{Gromov:2014caa}, which currently seems to be the most powerful technique for solving the spectral problem. In principle, it is applicable to the entire single-trace spectrum \cite{Marboe:2017dmb}. Moreover, it makes contact to the algebraic curve description of the semi-classical string dual \cite{Gromov:2007aq}. That being said, the strong-coupling limit of the QSC is still an open problem. Some guidance to this problem is provided by certain semi-classical string states that could already be matched across the duality \cite{Schafer-Nameki:2005yyn,Roiban:2007jf}. 

\

Parallel to the development of these techniques, the question of whether the $AdS_5\cross S^5$ background could be deformed while preserving integrability was raised. This would open up the possibility to study less supersymmetric theories using the same toolbox. Indeed a number of deformations have been successfully studied with TBA and QSC, including the $\beta$- and $\gamma$-deformations, which are $\mathcal{N}=1$ and $\mathcal{N}=0$ supersymmetric models built from a TsT-transformation of the $S^5$ on the string theory side and a non-commutative deformation on the gauge theory side \cite{Lunin:2005jy, Frolov:2005dj, Frolov:2005iq,Alday:2005ww}. 

Even simpler deformations can be constructed by orbifolding the original theory by a discrete subgroup of the symmetry group $PSU(2,2|4)$. This paper is devoted to their description. Although, in principle, one can think of arbitrary discrete subgroups $G$ of $PSU(2,2|4)$, we will restrict our attention to Abelian orbifolds, as their treatment presents all the relevant details without the need for too involved book-keeping \cite{Hanany:1999sp,Solovyov:2007pw}. The generating groups $G$ of these orbifolds are therefore part of the Cartan subgroup $C$
\begin{equation}
G\subset C \subset PSU(2,2|4)\,,
\end{equation} 
which has 6 generators. The Cartan subgroup also classifies the possible states as they derive from representations of $PSU(2,2|4)$ whose quantum numbers correspond to the Cartan generators.
We introduce four twist angles $\alpha,\dot\alpha,\vartheta,\dot{\vartheta}$ for the four subgroups
\begin{equation}\label{cartan}
SU(2)\cross SU(2)\cross SU(2)\cross SU(2)~\subset PSU(2,2|4)\,,
\end{equation}
keeping two Cartan generators untwisted so we can use light-cone gauge on the string theory side\footnote{From a purely gauge theoretic point of view this restriction to four Cartan generators is not always necessary, see \cite{Beisert:2005he,Beccaria:2011qd} for a discussion.}. 

The orbifolding procedure is straight-forward: we project onto states invariant with respect to $G$ which gives us the untwisted sector, the subset of states of $\mathcal{N}=4$ SYM that survive the orbifolding. Then, we realise that the strings only have to close up to a $G$-transformation, so we have to introduce twisted sectors for each element of $G$. In the spin-chain, we can realise these twisted boundary conditions by inserting a twist operator into the trace \cite{Berenstein:2004ys,Beisert:2005he}. This twist operator then becomes a chemical potential when we move to the TBA \cite{vanTongeren:2013gva}. Finally, in the context of QSC, the twist only shows up in the asymptotics of the $Q$-functions. 

A lot of these steps have been described in the literature before with varying conventions. Below we streamline the notation using the twist angles $\alpha,\dot\alpha,\vartheta,\dot{\vartheta}$. The twisted QSC was first described in \cite{Kazakov:2015efa} but its application to orbifolds is a new result of this paper. 

The simplest examples of this construction are orbifolds by $G=\mathbb{Z}_2$, which conserve either half ($\mathcal{N}=2$) or none ($\mathcal{N}=0$) of the supersymmetry. We will discuss both cases in turn.

\

Type IIB string theory can be studied on an $\mathbb{C}^2/\mathbb{Z}_n$ orbifold background, where half of the supersymmetry is preserved \cite{Douglas:1996sw}. If we place D3-branes at the orbifold locus and take the same scaling limit as in \cite{Maldacena:1997re}, we can construct a duality between type IIB string theory on an orbifold of $AdS_5\cross S^5$ and the corresponding orbifold of $\mathcal{N}=4$ SYM \cite{Kachru:1998ys,Lawrence:1998ja,Gadde:2010zi}. We restrict our attention to $G=\mathbb{Z}_2$, with the orbifold action on the string theory side rotating two orthogonal planes in the embedding space of $S^5$. This results in the twist angle $\vartheta=\pi$. We find two sectors of string states, the untwisted sector, which descends from $\mathbb{Z}_2$-invariant states of type IIB string theory on regular $AdS_5\cross S^5$, and the twisted sector which arises from strings that close only up to a $\mathbb{Z}_2$-transformation.

In the dual gauge theory we have to take the $\mathcal{N}=4$ SYM with gauge group $SU(2N)$ and orbifold it. We find an $\mathcal{N}=2$ quiver gauge theory with gauge group $SU(N)\cross SU(N)$ of equal coupling and 2 hypermultiplets in the bifundamental representation \cite{Kachru:1998ys}. In the large-$N$ limit, closed string states map to single-trace operators in this gauge theory and we can match the $\mathbb{Z}_2$-symmetry we used in the string construction to the $\mathbb{Z}_2$-symmetry that exchanges the two $SU(N)$ factors. Instead of the BMN ground state \eqref{BMNlike}, we have to analyse states that are either even or odd under this exchange
\begin{equation}\label{groundstates}
\mathcal{O}_{\pm J}=\Tr(Z^J)\pm\Tr(\tilde Z^J)\,,
\end{equation}
where the $Z$ and $\tilde{Z}$ are in the adjoint representations of the two $SU(N)$ factors. These are still BPS-states, so their conformal dimension $\Delta$ should satisfy $\Delta=J$. We will indeed find that the integrability techniques yield no correction to their conformal dimension. To actually test the AdS/CFT correspondence in this model, we would have to look at a more complicated operator as for example the twisted-sector Konishi operator (see \cite{deLeeuw:2011rw} for a parallel discussion in the case of other orbifolds). The Konishi operator is the primary operator
\begin{equation}\label{twistedKonishi}
\mathcal{O}_{\text{tw. Konishi}}=\Tr\big(X\bar X+ Y\bar Y+ Z\bar Z\big)-\Tr\big(\tilde X\bar{\tilde X}+\tilde Y\bar{\tilde Y}+\tilde Z\bar{\tilde Z}\big)
\end{equation}
with classical conformal dimension $\Delta_0=2$. Here $X$ and $Y$ are the two other complex scalar fields in the adjoint of the first $SU(N)$ with $\mathcal{R}$-symmetry charges $J_1=1$ and $J_2=1$, respectively. This specific combination forms a singlet under the full $\mathcal{R}$-symmetry group $SU(4)$. As the Konishi operator belongs to a long multiplet of the supersymmetry algebra, its conformal dimension is not protected and becomes anomalous
\begin{equation}
\Delta_{\text{tw. Konishi}}=2+\gamma(g)\,.
\end{equation} 
Further supersymmetry descendants of \eqref{twistedKonishi} such as $\Tr([Z,Y]^2)-\Tr \small([\tilde Z,\tilde Y]^2\small)$ could be connected to small-charge limits of semi-classical string states as in \cite{Beccaria:2012xm}, making a precise QSC calculation a worthwhile challenge.

The exploration of $\mathcal{N}=2$ orbifold theories with integrability techniques probes the larger landscape of 4D $\mathcal{N}=2$ SCFTs, which are themselves a vibrant area of research (see \cite{Pomoni:2019oib} for a review in the context of integrability and \cite{Pomoni:2021pbj} for some recent progress). These theories can also be studied using supersymmetric localisation (see e.g. \cite{Pestun:2007rz, Mitev:2014yba,Mitev:2015oty,Pestun:2016zxk, Niarchos:2019onf,Galvagno:2020cgq, Niarchos:2020nxk, Billo:2021rdb,Beccaria:2022ypy,Beccaria:2022kxy}). It would be interesting to see how integrability and localisation can work together to give a more complete description of the observables in 4D $\mathcal{N}=2$ SCFTs.

\

The second example we discuss below is type 0B string theory on $AdS_5\cross S^5$. Type 0B string theory in flat space arises from an alternative GSO projection in the RNS formalism. It can also be constructed via a $\mathbb{Z}_2$-orbifolding of the Green-Schwarz (GS) string in light-cone gauge by the space-time fermion number \cite{Dixon:1986iz,Seiberg:1986by}.

In flat space the $\mathbb{Z}_2$-orbifold in question arises when we take an arbitrary plane in the target space and rotate it by $2\pi$. This leaves all bosonic states invariant, but spacetime fermionic states pick up a minus-sign and therefore get projected out of the spectrum. As in the previous case we have to add a twisted sector, which can be constructed using GS string but with antiperiodic boundary conditions for the worldsheet fermionic fields. The resulting spectrum is purely bosonic and therefore not supersymmetric. 

In more general curved backgrounds with flux a similar orbifolding can be attempted. A geometric construction via Melvin-twist was discussed in \cite{Takayanagi:2001jj, Skrzypek:2021eue} and can be extended to fluxed backgrounds like the pp-wave spacetime. In $AdS_5\cross S^5$ we can perform the orbifolding procedure by rotating one of the planes in the embedding space of the $S^5$. In our previous parametrisation this corresponds to a twisting by $\vartheta=\dot{\vartheta}=\pi$.

The dual gauge theory is again an $SU(N)\cross SU(N)$ gauge theory with $2\times 6$ scalars, 6 in the adjoint representation of each of the two $SU(N)$ factors, and $2\times4$ Weyl-fermions, 4 in each bi-fundamental representation $(\mathbf{N},\mathbf{\bar{N}})$ and $(\mathbf{\bar{N}},\mathbf{N})$ \cite{Nekrasov:1999mn}. We will rewrite the 6 real scalars in the adjoint of the first $SU(N)$ as 3 complex scalars $\{X,Y,Z\}$ and similarly for the other $SU(N)$ as $\{\tilde X,\tilde Y,\tilde Z\}$. The interactions of these fields have been studied in \cite{Klebanov:1999ch,Tseytlin:1999ii,Klebanov:1999um} and as their form is not dictated by supersymmetry, they are not protected against renormalisation. The single-trace operators are again graded by the $\mathbb{Z}_2$-symmetry exchanging the two $SU(N)$ factors. However, the operators $\mathcal{O}_{\pm J}$ \eqref{groundstates} are not BPS anymore. The untwisted operator $\mathcal{O}_{+ J}$ inherits a trivial conformal dimension from the original theory, but $\mathcal{O}_{- J}$ receives an anomalous dimension, which can be computed perturbatively in the TBA formalism. Following \cite{Arutyunov:2010gu} (eq. (4.67) in \cite{vanTongeren:2013gva}) we find the conformal dimension
\begin{equation}\label{teaser}
\Delta=J-\frac{8\Gamma\left(J-\frac{1}{2}\right)\zeta(2J-3)}{\sqrt{\pi}\Gamma(J)}(2g)^{2J}+\order{g^{2J+2}}\,.
\end{equation}

\

In the context of type 0B string theory, we are primarily interested in the fate of the closed-string tachyon. In flat space the lowest energy mode of the twisted sector is tachyonic. On $AdS_5 \cross S^5$ background, however, it was suggested that this tachyon might get stabilised \cite{Klebanov:1999ch}. Some evidence supporting this suggestion has been provided in \cite{Skrzypek:2021eue}. In the strong coupling limit the $AdS_5 \cross S^5$ space becomes approximately flat and the tachyon is definitely present, so the stabilisation can only occur at finite coupling. 

At weak coupling an a priori unrelated ``tachyon" arises: As the gauge theory is not protected by supersymmetry, we have to ensure renormalisability. Therefore, we have to add all possible operators to the action that contribute at leading order in large-$N$. This includes a double-trace operator which has non-vanishing $\beta$-function and thus spoils conformality \cite{Tseytlin:1999ii,Klebanov:1999um,Adams:2001jb,Dymarsky:2005nc,Dymarsky:2005uh, Pomoni:2009joh,Fokken:2013aea}. We could try to tune the double-trace coupling to a conformal fixed point, but it turns out that the only fixed point is at complex values.\footnote{A similar behaviour has been found in the fishnet model \cite{Gurdogan:2015csr, Gromov:2017cja}, which is a non-supersymmetric truncation of $\mathcal{N}=4$ SYM.} If we compute the conformal dimension of $\mathcal{O}_{\pm 2}$ at this complex fixed point, we find that it has imaginary anomalous dimension. Therefore it could be described as a tachyon when applying a naive supergravity logic. 

When discussing the integrability treatment of this model, one finds a divergence of the conformal dimension \eqref{teaser} for $J=2$. This breakdown of the TBA involves precisely the operator $\mathcal{O}_{-2}$ which becomes tachyonic at the complex fixed point. The QSC on the other hand is supposed to include all quantum effects and describe the spectrum of a theory at its conformal fixed point, so it should find precisely the imaginary anomalous dimension we expect for $\mathcal{O}_{-2}$. When the QSC was studied for the $\gamma$-deformed theory \cite{Levkovich-Maslyuk:2020rlp}, the anomalous dimension indeed turned out imaginary. Extrapolation of these results and comparison to some preliminary perturbative calculations (see appendix \ref{appendix3}) lead us to the conjecture that the conformal dimension of $\mathcal{O}_{-2}$ has the weak coupling expansion
 \begin{equation}\label{conj.1}
\Delta_\pm=2\pm i\left[8g^2-64(3 \zeta_3+1)g^6 +\order{g^{8}}\right]\,,
\end{equation}
This suggests that the QSC captures the weak coupling tachyon and might be capable of interpolating it to finite coupling.

Building on these observations, we hope that a future study of the type 0B QSC can resolve the question whether the weak and strong coupling tachyons we described are stabilised in an intermediate regime of coupling or not. If yes we would have found a possible contender for non-supersymmetric holography (with potential lessons for a string description of QCD), if not we can ask the question whether these a priori independent tachyonic instabilities are related and can be matched.  This question has been discussed in perturbation theory, but the full range of coupling is not yet accessible \cite{Dymarsky:2005nc,Dymarsky:2005uh, Fokken:2013aea}. We hope that integrability will be able to extend our knowledge to finite coupling and give a definite answer to the question of stability of non-supersymmetric orbifolds.

\

The paper is organised as follows. In section \ref{sigma} we set up our discussion by summarising the non-linear sigma-model description of the $AdS_5\cross S^5$ side of the AdS/CFT duality. Then we introduce the geometric implementation of the orbifolding by specifying the appropriate twist factors. This will guide our further analysis of the dual gauge theory operators, which is carried out in section \ref{spin-chain}. There we discuss the spin chain realisation, asymptotic Bethe Ansatz (ABA) and thermodynamic Bethe Ansatz (TBA) of the orbifolded gauge theory. 

In section \ref{QSC} we match the dual perspectives of orbifold theories and describe the quantum spectral curve (QSC), which links both descriptions. More specifically, \ref{new} introduces the twisting prescriptions for the orbifold QSC, which are one of the main results of this paper. The simplest $\mathbb{Z}_2$-orbifolds are addressed in section \ref{examples}, where we first discuss the $\mathcal{N}=2$ supersymmetric orbifold and then type 0B string theory om $AdS_5\cross S^5$.  Finally, we comment on our results and possible future directions in section \ref{Discussion}. A few technicalities are left to the appendices. Appendix \ref{appendix} discusses the relation between $T$-system and $Q$-functions appearing in the QSC, while appendix \ref{appendix2} specifies the prefactors of the $Q$-function asymptotics. Appendix \ref{appendix3} summarises the perturbative approach to the QSC, which we used to back up our conjecture \eqref{conj.1}.

\section{Orbifolding the sigma-model}\label{sigma}

The type IIB string on $AdS_5\cross S^5$ has been described in \cite{Metsaev:1998it} as a special sigma-model of $PSU(2,2|4)$ gauged by its $SO(4,1)\cross SO(5)$ subgroup. It has been shown to be classically integrable in \cite{Bena:2003wd}, where a Lax connection was identified (for a pedagogic introduction to these coset models, see \cite{Zarembo:2017muf}). 

We can represent an element $\mathfrak{g}$ in the superalgebra $\mathfrak{psu}(2,2|4)$ by complex $8\cross 8$ matrices of the form
\begin{equation}
\mathfrak{g}=\begin{pmatrix}
A&B\\C&D
\end{pmatrix},
\end{equation}
where $A$ and $D$ represent the bosonic subgroups $u(2,2)$ and $u(4)$ and $B$ and $C$ are related via 
\begin{equation}
C=B^\dagger \begin{pmatrix}
\mathbbm{1}_{2\cross 2}&0\\0&-\mathbbm{1}_{2\cross 2}
\end{pmatrix}.
\end{equation}
We furthermore require that the supertrace $\Str \mathfrak{g} =\Tr A - \Tr D$ vanishes and mod out an overall scale factor. 

This construction reveals an additional $\mathbb{Z}_4$ symmetry given by the transformation
\begin{equation}\label{Z4}
\begin{pmatrix}
A&B\\C&D
\end{pmatrix}\to\begin{pmatrix}
EA^TE&-EC^TE\\EB^TE&ED^TE
\end{pmatrix},\qquad E=\begin{pmatrix}
0&-1&0&0\\
1&0&0&0\\
0&0&0&-1\\
0&0&1&0
\end{pmatrix}.
\end{equation}
One can show that the subgroup generated by the $\mathbb{Z}_4$-invariant matrices is precisely  $SO(4,1)\cross SO(5)$, which we have to gauge.  

The sigma-model describes an embedding $G:\Sigma\to\mathcal{M}$ of the string worldsheet $\Sigma$ into the just defined coset space $\mathcal{M}=\frac{PSU(2,2|4)}{SO(4,1)\cross SO(5)}$. The action is formulated in terms of the Lie algebra-valued current 
\begin{equation}
j_\mu=-G^{-1}\partial_\mu G=j_\mu^{(0)}+j_\mu^{(1)}+j_\mu^{(2)}+j_\mu^{(3)}
\end{equation}
as 
\begin{equation}
S=g\int_\Sigma\dd^2\sigma~ \Str\left(\sqrt{-h}h^{\mu\nu}j_\mu^{(2)} j_\nu^{(2)}+\epsilon^{\mu\nu}j_\mu^{(1)}j_\nu^{(3)}\right),
\end{equation}
where we have split the current $j$ into eigenvectors of the $\mathbb{Z}_4$-symmetry such that $j^{(n)}$ has eigenvalue $i^n$. The constant prefactor $g$ is the coupling constant defined in \eqref{coupling}.
This action descends from the WZW model and is independent of $j^{(0)}$, which only describes gauge degrees of freedom. 
The equations of motion are 
\begin{equation}
\dd\star k=0\,, \qquad k=G\left(j^{(2)}+\frac{1}{2}\star j^{(1)}-\frac{1}{2}\star j^{(3)}\right)G^{-1}
\end{equation} 
and $j$ satisfies the flatness condition
\begin{equation}
\dd j-j\wedge j=0
\end{equation}
by construction. These two conditions can be combined with an arbitrary ``spectral factor" $z$ in one flatness condition
\begin{equation}\label{lax}
\dd L-L\wedge L =0
\end{equation}
for the Lax connection
\begin{equation}
L(z)=j^{(0)}+\frac{z^2+1}{z^2-1}j^{(2)}-\frac{2z}{z^2-1}\star j^{(2)}+\sqrt{\frac{z+1}{z-1}}j^{(1)}+\sqrt{\frac{z-1}{z+1}}j^{(3)}\,.
\end{equation}

Note that this Lax connection has a branch cut between $z=\pm 1$, which will become important when we look at the analytic structure of the QSC. Since $L$ is flat, we can define a ``Wilson-loop" around the closed string called the monodromy matrix
\begin{equation}\label{monodromy}
T(z)=\mathcal{P}\exp\int\dd\sigma~L_\sigma(z)\,,
\end{equation}
which is independent of the coordinate choice on $\Sigma$. Classical integrability follows from the fact that $\Str T(z)$ is conserved due to \eqref{lax}
\begin{equation}
\partial_\tau\Str T(z)=0\,,
\end{equation}
so a Taylor expansion in $z$ yields infinitely many conserved charges. We can also diagonalise 
\begin{equation}
T(z)\to \diag\left(e^{ip_{\hat 1}},e^{ip_{\hat 2}},e^{ip_{\hat 3}},e^{ip_{\hat 4}},e^{ip_1},e^{ip_2},e^{ip_3},e^{ip_4}\right)\,,
\end{equation}
where we defined pseudomomenta $p_{\hat{\imath}}$ for the $AdS_5$ subspace and $p_i$ for the $S^5$. In an appropriate parametrisation (see \cite{Gromov:2007aq} for details) we can relate the large-$z$ asymptotics of these pseudomomenta to the Noether charges of the classical solution.

 To characterise a classical string in $AdS_5\cross S^5$ we need to specify six charges, which denote ``angular momenta" in the various independent $SU(2)$ subgroups of $PSU(2,2|4)$. We can picture this by looking at the embedding space of $AdS_5\cross S^5$
\begin{equation}
AdS_5\cross S^5 \subset \mathbb{R}^{1,5}\cross \mathbb{R}^6\,.
\end{equation}
We can choose 6 orthogonal planes in this space 
\begin{equation}\label{embedding}
\mathbb{R}^{1,5}\cross \mathbb{R}^6= \left(\mathbb{R}^{1,1}\cross\mathbb{R}^2\cross\mathbb{R}^2\right)\cross\left(\mathbb{R}^2\cross\mathbb{R}^2\cross\mathbb{R}^2\right)\,.
\end{equation}
The angular momenta in these planes are denoted by 
\begin{equation}\label{charge}
(\Delta,S_1,S_2|J_1,J_2,J_3)\,.
\end{equation}
The physical interpretation of $\Delta$ is slightly different due to the Lorentzian nature of its subspace, $\Delta$ denoting the energy of the solution. It is precisely these charges \eqref{charge} that are encoded in the large-$z$ asymptotics of the pseudomomenta as
\begin{equation} \label{charges}
\begin{pmatrix}
p_{\hat 1}\\p_{\hat 2}\\p_{\hat 3}\\p_{\hat 4} \\\hline p_1\\p_2\\p_3\\p_4 
\end{pmatrix}
\simeq\frac{1}{2gz}\begin{pmatrix}
+\Delta-S_1+S_2\\+\Delta+S_1-S_2\\-\Delta-S_1-S_2\\-\Delta+S_1+S_2\\\hline+J_3+J_1-J_2\\+J_3-J_1+J_2\\-J_3+J_1+J_2\\-J_3-J_1-J_2
\end{pmatrix}.
\end{equation}

\

Now we want to move on to Abelian orbifolds of $AdS_5\cross S^5$. Going back to the embedding space \eqref{embedding} and the orthogonal planes therein, we want to identify which subspace we can orbifold while keeping the string theory accessible to known techniques. By solving the dynamics of the string theory, we find a dispersion relation of the form $\Delta(S_i,J_i)$. This is so far only achievable by choosing light-cone gauge. A standard choice for the light-cone directions are the time $t$ in $\mathbb{R}^{1,1}$ and the angle $\phi_{J_3}$ in the $\mathbb{R}^2$ plane conjugate to the momentum $J:=J_3$ \cite{Berenstein:2002jq}. We therefore reserve the subspace $\mathbb{R}^{1,1}\cross\mathbb{R}^2$ containing these coordinates for fixing light-cone gauge and restricting from Embedding space to $AdS_5\cross S^5$. This leaves us with an 8-dimensional subspace 
\begin{equation}
\left(\mathbb{R}^2\cross\mathbb{R}^2\right)\cross\left(\mathbb{R}^2\cross\mathbb{R}^2\right)\,.
\end{equation}
In each of these planes we can orbifold by a discrete group $\mathbb{Z}_n$ by requiring physical states to be left unchanged by rotation of the plane by an angle $\phi={2\pi\over n}$. We can therefore introduce 4 a priori independent orbifolding angles $(\phi_{S_1},\phi_{S_2},\phi_{J_1},\phi_{J_2})$. This corresponds precisely to twisting the Cartan subgroup of $SU(2)^4\subset PSU(2,2|4)$ we identified in \eqref{cartan}.

In the sigma-model the orbifolding procedure consists not only of the restriction to $\mathbb{Z}_n$-invariant states but we also find additional ``twisted sectors", which are strings that close only up to a $\mathbb{Z}_n$-transformation. These states need to be added by hand, as they do not appear in the untwisted model. As there are $n$ elements in the $\mathbb{Z}_n$ group that we could twist the string by, any $\mathbb{Z}_n$ orbifolding yields $n$ distinct sectors, enumerated by integers $k\in\{0,\dots, n-1\}$. In terms of the sigma model, the twist can be achieved quite easily by introducing a twist operator $A\in C$ in every trace we take over the string. Accordingly, we can also define a ``twisted monodromy matrix" 
\begin{equation}
T'(z)=A T(z)\,, \qquad A\in C\,.
\end{equation}
This results in shifts of the pseudomomenta by integer multiples of the twist angles. Let us denote
\begin{equation}\label{angles}
\begin{split}
\alpha&=-\frac{k_{S_1}\phi_{S_1}+k_{S_2}\phi_{S_2}}{2}\,,\qquad \dot{\alpha}=\frac{k_{S_2}\phi_{S_2}-k_{S_1}\phi_{S_1}}{2}\,,\\
\vartheta&=-\frac{k_{J_1}\phi_{J_1}+k_{J_2}\phi_{J_2}}{2}\,\,,\qquad\, \dot{\vartheta}=\frac{k_{J_2}\phi_{J_2}-k_{J_1}\phi_{J_1}}{2}\,\,,
\end{split}
\end{equation}
where the integers $k$ parametrise the specific twisted sector. Note that we chose symmetric and antisymmetric combinations of the twist angles for later convenience. In the following we will only look at a general twisted sector and therefore keep the angles $\alpha$, $\dot{\alpha}$, $\vartheta$, $\dot\vartheta$ as arbitrary rational multiples of $\pi$. Then the pseudomomenta become
\begin{equation}\label{tcharges}
\begin{pmatrix}
p_{\hat 1}\\p_{\hat 2}\\p_{\hat 3}\\p_{\hat 4} \\\hline p_1\\p_2\\p_3\\p_4 
\end{pmatrix}
\simeq\frac{1}{2gz}\begin{pmatrix}
+\Delta-S_1+S_2\\+\Delta+S_1-S_2\\-\Delta-S_1-S_2\\-\Delta+S_1+S_2\\\hline+J_3+J_1-J_2\\+J_3-J_1+J_2\\-J_3+J_1+J_2\\-J_3-J_1-J_2
\end{pmatrix}+\begin{pmatrix}
+\dot{\alpha}\\-\dot{\alpha}\\+\alpha\\-\alpha\\\hline -\dot{\vartheta}\\+\dot{\vartheta}\\-\vartheta\\+\vartheta
\end{pmatrix}.
\end{equation}
We can see that the coordinate choice for diagonalisation of the pseudomomenta motivated the definitions \eqref{angles}. Going back to the matrix representation of the supergroup, one can convince oneself that supersymmetry is generally broken. Only in the special case that at least one $AdS_5$ twist and one $S^5$ twists coincide in all twisted sectors ($\alpha=\pm\vartheta$ or any dotted version thereof) we retain half of the supersymmetry.  \\

In summary, the orbifolding procedure results in restriction of solutions and additional twisted sectors. Both effects do not disturb the classical integrability. The local Lax connection remains unchanged, only the appropriate boundary conditions are required. Taking the supertrace of $T'(z)$ still results in an infinite tower of conserved charges. The only difficulty is the increased amount of book-keeping required to consider all twisted sectors and restrict to physical states.

\section{Orbifolding the spin-chain}\label{spin-chain}

In the large-$N$ limit, the AdS/CFT duality links closed-string states of the bulk theory to single-trace operators in the dual gauge theory. As discussed in the introduction, the AdS/CFT duality is a ``weak-strong" duality, which makes it difficult to interpolate quantities explicitly. However, if we restrict our attention to states with large quantum numbers, we can use these for approximations on both sides of the duality.

The most intuitive case is the BMN limit \cite{Berenstein:2002jq} where the angular momentum $J:=J_3$ is taken to infinity together with the coupling $g$ \eqref{coupling}, keeping the ratio $\mathcal{J}=\frac{J}{g}$ fixed. On the string side this corresponds to a point-like string moving in a pp-wave background while on the gauge theory side this describes a ``string" of $J$ scalar operators
\begin{equation}\label{BMN}
\mathcal{O}_J=\Tr(Z^J)\,,
\end{equation}
where $Z$ is the complex scalar field charged under the appropriate $R$-symmetry. Excitations of this BMN string can either be described directly on the string worldsheet or by adding ``inhomogeneities" to the single-trace operator. We could for example excite a string mode in the plane associated to the angular momentum $J_2$, which would correspond to adding the complex scalar $Y$ with charge $J_2=1$ to the single-trace operator and averaging over position. 

From this intuitive picture, we can already imagine how twisted sectors are to be translated to the gauge theory - we need to insert the twist operator into the trace. The description of single-trace operators of the gauge theory has been explored in the integrability program and we will run through the main stages of this development, highlighting at each step the influence of the twisting. We follow closely the notation of \cite{vanTongeren:2013gva} and direct interested readers there for a much more detailed review.  
\subsection{Twisted asymptotic Bethe Ansatz}
When we let $J$ become very large at finite coupling $g$, we can think of the string as having an almost infinitely extended worldsheet. Let $t$ be the $AdS_5$ time direction and $\phi$ the angle associated to $J$. Then we can choose light-cone coordinates 
\begin{equation}\label{l.c.}
x_-=\phi-t\,, \quad x_+=t\,,\quad P_-=\int_{0}^{L}\dd\sigma p_-=J-\Delta\,,\quad P_+=\int_{0}^{L}\dd\sigma p_+=J
\end{equation} 
and fix the gauge
\begin{equation}
x_+=\tau\,, \quad p_+=1\,,
\end{equation}
such that the string length $L$ is given by the angular momentum $J$ along $\phi$. The operator $\mathcal{H}=-P_-$ now becomes the Hamiltonian of the light-cone gauged theory. In the following we always aim at computing its spectrum, which provides us with ``energy" eigenvalues $E$, most importantly the groundstate energy $E_0$. We can then extract the conformal dimension by relating 
\begin{equation}\label{conformal}
\Delta=J+E\,.
\end{equation}
If we take $J$ to be large, the string worldsheet effectively decompactifies to a plane where we can perform a Bethe Ansatz analysis of the resulting two-dimensional QFT. The symmetry algebra $\mathfrak{psu}(2,2|4)$
is broken to $\mathfrak{psu}(2|2)\oplus\mathfrak{psu}(2|2)\oplus\mathcal{H}$. If we drop the level-matching condition, we can analyse excitations of the string with arbitrary momenta and consistently define asymptotic states and S-matrices (see \cite{Arutyunov:2009ga} for a review on this construction). This should be thought of as an off-shell calculation, imposing the level matching at the very end. As a result, the symmetry group gets centrally extended to two copies of $\mathfrak{psu}(2|2)_{c.e.}=\mathfrak{psu}(2|2)\oplus\{\mathcal{H},\mathcal{C},\mathcal{C}^\dagger\}$, where we can think of $\mathcal{C}$ as a generalisation of the string momentum. The central elements of both copies of $\mathfrak{psu}(2|2)_{c.e.}$ must match. The physical fields live in a short representation of these algebras, constrained by a shortening condition 
\begin{equation}
E^2+4CC^\dagger=1\,,
\end{equation}
where $\{E,C,C^\dagger\}$ are the eigenvalues of the operators $\{\mathcal{H},\mathcal{C},\mathcal{C}^\dagger\}$. 
This results in the dispersion relation between energy $\epsilon$ and momentum $p$
\begin{equation}\label{dis}
\epsilon^2=1+16g^2 \sin^2\frac{p}{2}\,.
\end{equation}

The 16 worldsheet fields factorise into two multiplets (or ``wings")
\begin{equation}\label{wings}
(w_1,w_2,\theta_3,\theta_4)\cross(\dot w_1,\dot w_2,\dot\theta_3,\dot\theta_4)
\end{equation} 
each consisting of 2 bosonic ($w_1$, $w_2$) and 2 fermionic fields ($\theta_3$,
$\theta_4$) (see \cite{Arutyunov:2009ga} for an explicit matrix representation). Since the S-matrices factorise as well, it suffices to
analyse this smaller system. 

To characterise different string states, we imagine a chain of well-separated particles with different momenta $p_i$. Since the decompactified field theory is integrable all physical processes factorise into 2-to-2 scattering events. Thus, specifying the particle content and initial kinematics, we immediately know all physically interesting quantities of the prepared state and the entire time evolution thereof.

\begin{figure}
	\centering
	\includegraphics[width=0.8\textwidth]{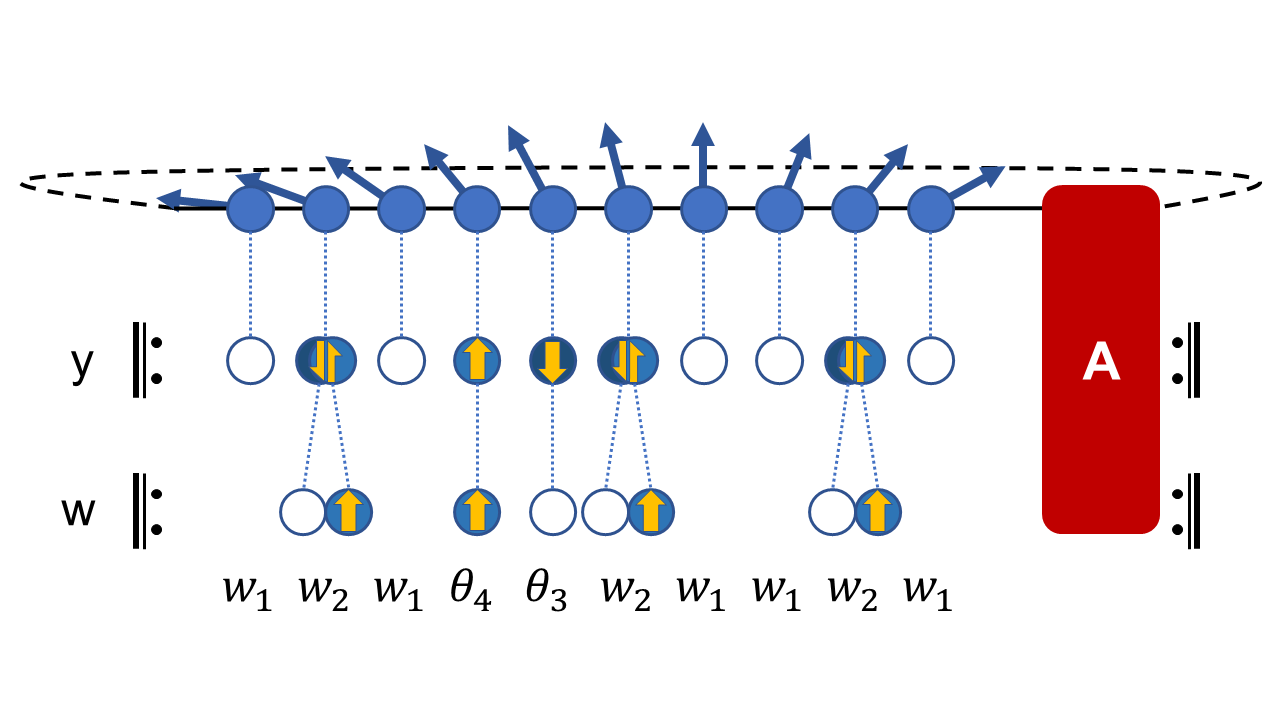}
	\caption{\small{Schematic representation of the nested $\mathfrak{psu}(2|2)_{c.e.}$ spin-chain. The fundamental particles (top) come with physical momenta and are distributed along the string, while the $y$-particles live on the spin-chain made up of the physical particle sites. The $w$-particles in turn live on the spin-chain spanned by the $y$-particle sites. Together these three types of excitations describe the full particle content. Solving the nested set of Bethe\,-Yang equations \eqref{BY1}-\eqref{BY3} related to all three chains then provides us with the appropriate wave function and energy. In case of the twisted spin-chain, a twist operator $A$ is introduced on closing the spin-chain.}}\label{figure1}
\end{figure}

In order to specify the particle content, we could count excitations of the fundamental fields like e.g. $n(w_1)$, but these are not conserved
under scattering. The process $w_1+w_2 \leftrightarrow\theta_3+\theta_4$
introduces an ambiguity, since we cannot discern these states from each other.
Thus, instead of counting 4 particle types, we can describe states by only 3
numbers. Instead of treating all particle types simultaneously, we construct a nested spin chain by introducing an artificial hierarchy of states (see fig. \ref{figure1}). 
For a state with total excitation number 
\begin{equation}
K^I=n(w_1)+n(w_2)+n(\theta_3)+n(\theta_4)
\end{equation}
the ``ground state" would be given by $K^I$ excitations of $w_1$ (without loss of generality we can assume
$n(w_1)\geq n(w_2)$). Then we treat fermionic excitations as further excitations of
this ground state, counting
\begin{equation}
K^{II}=2n(w_2)+n(\theta_3)+n(\theta_4)\,,
\end{equation}
where $w_2$ is treated as the combination $\theta_3\theta_4$. Finally, we take this
new ground state of fermionic excitations and count one specific type of
fermionic excitations on top of it
\begin{equation}
K^{III}=n(w_2)+n(\theta_4)\,.
\end{equation}
These numbers are conserved under scattering and are ordered $K^I\geq K^{II}
\geq K^{III}$. We have chosen a bosonic vacuum consisting of $w_1$ excitations,
but we could have chosen a fermionic vacuum by exchanging
$w\leftrightarrow\theta$.\footnote{Indeed the  need for an arbitrary choice of ground state will be fixed by the quantum spectral curve, where the dualities between different hierarchies build the algebraic framework of $QQ$-relations.}

These different levels of excitations are called fundamental particles,
$y$-particles and $w$-particles. While fundamental particles have momenta $p_i$
the auxiliary particles are parametrised by pseudomomenta $y_i$ and $w_i$
respectively. 

Knowing the particle content of our theory, we need to determine the 2-to-2 S-matrix that governs their interactions. For the $\mathfrak{psu}(2|2)_{c.e.}$ theory the S-matrix has been worked out (up to a phase), see e.g. \cite{Beisert:2005fw,Beisert:2006ez,Arutyunov:2009ga,vanTongeren:2013gva}.  In this paper we keep the S-matrices $S^{A,B}$ symbolic, referring to the previous references for explicit expressions. 

Now that we have completely characterised the decompactified theory, we need to remember that despite being large, the string is necessarily closed. Therefore we need to impose periodicity conditions on the string state. Take a fundamental particle with momentum $p_k$ in some arbitrary state. If we move it around the string of length $L$ once, we expect the resulting state to be the same as before this translation. Therefore, if we multiply all scattering matrices $S^{I,I}$ that the particle picks up along the way, they should only add up to a trivial phase factor $e^{-ip_kL}$
\begin{equation}
e^{-ip_kL}\sim\prod_{l\neq k}^{K^I} S^{I,I}(p_k, p_l)\,.
\end{equation}

This is the prototypical Bethe\,-Yang equation, which restricts the (sets of) momenta $p_i$ a state can sample of.  Now to capture the different particle types, we use the hierarchy we introduced earlier. Instead of introducing S-matrices for every type of particle, we treat all fundamental particles equally and think of higher excitations as living on a lattice or spin-chain with each node given by one of the fundamental particles. Again, this spin-chain has to be periodic, therefore scattering of different particle types can be described as a system of ``nested" Bethe\,-Yang equations
\begin{align}
e^{-ip_kL}&=\prod_{l\neq
	k}^{K^I}S^{I,I}(p_k,p_l)\prod_{l=1}^{K^{II}}S^{I,II}(p_k,y_l)\,,\label{BY1}\\
(-1)^m&=\prod_{l=1}^{K^I}S^{II,I}(y_k,p_l)\prod_{l=1}^{K^{III}}S^{II,III}(y_k,w_l)\,,\label{BY2}\\
1&=\prod_{l=1}^{K^{II}}S^{III,II}(w_k,y_l)\prod_{l\neq
	k}^{K^{III}}S^{III,III}(w_k,w_l)\,,\label{BY3}
\end{align}
where $m$ denotes the winding number of the string around $\phi$. 
The different S-matrices can be determined in either the algebraic or the coordinate
Bethe Ansatz framework \cite{deLeeuw:2007akd}, imposing the usual constraints from integrability,
unitarity, analyticity and crossing relations. It turns out that the auxiliary (2\textsuperscript{nd} and 3\textsuperscript{rd}) Bethe\,-Yang equations for $\mathfrak{psu}(2|2)_{c.e.}$ are related to the Lieb-Wu equations of the Hubbard model \cite{Mitev:2012vt}. 

A careful analysis of the Bethe\,-Yang equations reveals two caveats we have to be aware of. Firstly, we have to take into account that the individual momenta may turn out complex, which is due to the appearance of bound states. Secondly, the $y$-particles come in sets of two solutions $y_\pm$, related by inversion of the corresponding pseudorapidity. The Bethe\,-Yang equations can be modified to fix these issues by summing over all types of bound states and $y$-particles, resulting in strictly real momenta and unambiguous solutions. 

Solving the Bethe\,-Yang equations gives us the possible momenta and thus the energy spectrum of the string. This application of spin-chain techniques to the AdS/CFT problem has been dubbed the asymptotic Bethe Ansatz (ABA) and was formulated in \cite{Beisert:2005fw,Beisert:2006ez}. \\

For the orbifolded string we will have to add twisted sectors and, as in the sigma-model discussion, this simply results in twisted boundary conditions or, equivalently, the introduction of the twist operator $A$ in the trace of operators \cite{Berenstein:2004ys,Beisert:2005he} (also see \cite{Cavaglia:2020hdb} for a discussion of related twist operators).\footnote{It is useful to compare our conventions to the notation in \cite{Beisert:2005he}. There, $\mathbb{Z}_n$ orbifolds of the $S^5$ are discussed. They are parametrised by a twist angle $\omega=\exp(\frac{2\pi ik}{n})$, which acts diagonally on the $SU(4)$  representation of $S^5$ via $A=\text{diag}(\omega^{-t_1},\omega^{t_1-t_2},\omega^{t_2-t_3},\omega^{t_3})$. The $t_i$ parametrise the specific angles which get twisted. $t_2$ corresponds to twisting around the light-cone direction $J$, so we do not consider this case (see however \cite{Beccaria:2011qd} for a treatment of this case). For orbifolds along the other directions we find the mapping $\omega^{-t_1}=e^{i\dot{\vartheta}}$ and $\omega^{-t_3}=e^{i\vartheta}$.} We therefore expect the various Bethe\,-Yang equations to be modified by a constant factor. The coordinate choice we made for the sigma-model already anticipated this. The two pairs of twist angles $(\alpha,\vartheta)$ and $(\dot{\alpha},\dot\vartheta)$ correspond exactly to the two $\mathfrak{psu}(2|2)_{c.e.}$ subsectors. Therefore we get the twisted Bethe\,-Yang equations
\begin{align}
e^{-ip_kL}&=e^{i\dot\alpha}\prod_{l\neq
	k}^{K^I}S^{I,I}(p_k,p_l)\prod_{l=1}^{K^{II}}S^{I,II}(p_k,y_l)\,,\label{tBY1}\\
(-1)^m&=e^{i(\dot\alpha-\dot\vartheta)}\prod_{l=1}^{K^I}S^{II,I}(y_k,p_l)\prod_{l=1}^{K^{III}}S^{II,III}(y_k,w_l)\,,\label{tBY2}\\
1&=e^{-2i\dot\vartheta}\prod_{l=1}^{K^{II}}S^{III,II}(w_k,y_l)\prod_{l\neq
	k}^{K^{III}}S^{III,III}(w_k,w_l)\label{tBY3}
\end{align}
and similarly the second copy with undotted angles. In the end we need to combine the two subgroups, which we shall denote by left $(l)$ and right $(r)$ wing or by undotted and dotted indices. The auxiliary Bethe\,-Yang equations have to be considered in both sectors, while the fundamental Bethe\,-Yang equations get glued together 
\begin{equation}
e^{-ip_kL}=e^{i(\alpha+\dot\alpha)}\prod_{l\neq
	k}^{K^I}S^{I,I}(p_k,p_l)\prod_{l=1}^{K^{II}_{(l)}}S^{I,II}(p_k,y_l)\prod_{m=1}^{K^{II}_{(r)}}S^{I,II}(p_k,\dot y_m)\,.
\end{equation}
Furthermore, the level-matching condition has to be reintroduced in the form of a vanishing total momentum
\begin{equation}
1=\prod_{k=1}^{K^I}e^{ip_kL}\,.
\end{equation}
This set of equations describes the spectrum of states in the large-$J$ limit at weak coupling, so we expect them to describe the spectrum of single-trace operators sufficiently close to the BMN vacuum \eqref{BMN}. Indeed explicit gauge theory calculations confirmed this \cite{Fiamberti:2008sh}.

For finite-size strings however, the ABA is not sufficiently accurate, as the decompactification picture breaks down and the particles are not well separated anymore. Although the computation of certain ``wrapping contributions" as corrections to the ABA has been attempted \cite{Bajnok:2008bm, Ahn:2011xq}, we can use the mirror trick to get rid of the finite-size problem. This will be discussed in the next subsection.

\subsection{Twisted thermodynamic Bethe Ansatz}
 The main idea is to exchange time and space dimensions on the infinite cylinder. Thus instead of describing the QFT of excitations living on a spacelike circle, we describe a thermodynamic ensemble on the infinite line with temperature anti-proportional to the circumference of a timelike circle.

To perform this mapping one compactifies the original $\tau$ direction on a circle of radius $\beta$, performs a Wick rotation to Euclidean metric, exchanges the coordinates $(\tau,\sigma)$, performs another Wick rotation and then decompactifies $\tilde\sigma$. 

The resulting ``mirror" theory is again integrable and can be solved via the same methods, the new S-matrices being analytic continuations of the original ones. The dispersion relation (for bound states of $Q$ fundamental particles) is now given by
\begin{equation}\label{disp}
\tilde{\epsilon}_Q=2\arsinh\frac{\sqrt{Q^2+\tilde p^2}}{4g}\,.
\end{equation}
A more useful way to encode these kinematics is given by a ``rapidity" variable $u$ and the Zhukovski-map
\begin{equation}\label{Zhukovski}
x_{ph}(u)=\frac{u}{2g}+\sqrt{{u\over2g}-1}\,\sqrt{{u\over2g}+1}\,,\quad x_{mir}(u)=\frac{u}{2g}+i\sqrt{1-{u^2\over4g^2}}\,,
\end{equation}
which conveniently encapsulates the magnon kinematics and also appears in the S-matrices of the $\mathfrak{psu}(2,2|4)$ spin-chain. The main benefit of these at first glance quite arbitrary functions is that we can encode the kinematics of the physical model and the mirror model on one Riemann-surface by simply shifting their arguments $x^{[\pm n]}(u):=x(u\pm\frac{in}{2})$ (for $n=1$ we simply write $x^\pm:=x^{[\pm1]}$). When we identify
\begin{equation}
\epsilon=2 ig (x^-_{ph}-x^+_{ph})-1\,,\quad \frac{x^+_{ph}}{x^-_{ph}}=e^{i p}\,,
\end{equation} 
we see that the dispersion relation \eqref{dis} becomes
\begin{equation}
x^+_{ph}+\frac{1}{x^+_{ph}}-x^-_{ph}-\frac{1}{x^-_{ph}}=\frac{i}{g}\,.
\end{equation}
For higher order $Q$-strings we simply need to exchange $x^{\pm}_{ph}\to x^{[\pm Q]}_{ph}$ and get a similar relation.
In the mirror model we instead have to identify
\begin{equation}
\tilde p= 2 g (x^-_{mir}-x^+_{mir})+i\,,\quad e^{\tilde{\epsilon}}=\frac{x^+_{mir}}{x^-_{mir}}\,.
\end{equation}
The two maps $x_{ph}$ and $x_{mir}$ coincide above the real axis, but have different branch cuts. $x_{ph}$ has a ``short" branch cut on the interval $(-2g,2g)$ while $x_{mir}$ has a ``long" branch cut on $(-\infty,-2g)\cup(2g,\infty)$. Furthermore, under complex conjugation of $u$ we find that
\begin{equation}
\bar x_{ph}=x_{ph}\,,\quad \bar x_{mir}=\frac{1}{x_{mir}}\,.
\end{equation}
The two maps thus patch different parts of a double-layered Riemann surface and are each other's analytic continuation, tailored towards describing the physical or the mirror model.\\

The Euclidean partition function has to match in both models, so we can relate the ground state energy $E_0$ of the original model to the Helmholtz free energy $F$ of the mirror model in the decompactification limit $\beta\to\infty$ 

\begin{equation}
Z=\sum_{n}e^{-\beta E_n}\sim e^{-\beta E_0}\quad \Leftrightarrow\quad Z=e^{-LF}\,.
\end{equation} 
Thus, we can remain oblivious to finite-size effects at the cost of having to sum all states of the mirror theory in order to describe even one state in the string theory. 

When we analyse a twisted sector, we have to introduce the twist operator in the time-like circle now, so instead of boundary conditions along the space-like dimension it introduces a chemical potential for the various particles involved.

In the ABA for the mirror theory we start with a fermionic vacuum that yields similar Bethe\,-Yang equations. The original string is wound one time around the time direction so $m=1$ in the mirror theory. It turns out that bound states can consist of various combinations of fundamental or auxiliary excitations:
\begin{itemize}
	\item $Q$-strings consisting of $Q$ fundamental excitations,
	\item $vw$-strings consisting of twice as many $y$ particles as $w$ particles,
	\item $w$-strings consisting of arbitrary numbers of $w$ particles,
\end{itemize} 
whose poles form strings in the rapidity plane (this is called the string-hypothesis \cite{Arutyunov:2009zu}). Single excitations of fundamental particles and $w$-particles can be included as $1$-strings, while $y_\pm$-particles have to be added separately.
Now instead of looking at individual solutions, we need to characterise the ``solution density" of the system, together with the corresponding energies. To this end we introduce a smooth counting functions $c$ depending on the rapidity $u$. $c(u)$ grows monotonically and takes integer values at the rapidities $u_k$, which appear in a solution of the Bethe\,-Yang equations. 
In logarithmic form the fundamental Bethe\,-Yang equation becomes

\begin{equation}\label{counting}
c^I(u)=\frac{\beta}{2\pi} p(u)+\frac{1}{2\pi i}\sum_{l\neq
	k}^{K^I}\log S^{I,I}(p(u),p_l)+\frac{1}{2\pi i}\sum_{l=1}^{K^{II}}\log S^{I,II}(p(u),y_l)
\end{equation} 
and similarly for the auxiliary Bethe\,-Yang equations. Conversely, when $c(u)$ becomes an integer at rapidity $u_k$, this specific rapidity may correspond to a particle in a given solution or not, so we can introduce a particle density $\rho(c(u))$ and a conjugate ``hole density" $\bar\rho(c(u))$ such that for integer $c(u)$ we have $\rho+\bar{\rho}=1$. This solution density can be pulled back to rapidity space and becomes effectively continuous in the large-$\beta$ limit. Taking the $u$-derivative of \eqref{counting} and rescaling by $\beta$ we end up with the density Bethe\,-Yang equation
\begin{equation}\label{density}
\rho^I(u)+\bar{\rho}^I(u)=\frac{1}{2\pi}\frac{\dd p(u)}{\dd u}+K^{I,I}\star\rho^I+K^{I,II}\star\rho^{II}
\end{equation}
with 
\begin{equation}\label{kernel}
K^{A,B}(u,v)=\pm\frac{1}{2\pi i}\frac{\dd}{\dd u}\log S^{A,B}(u,v)
\end{equation}
and the star denoting convolution. In this form, we have dropped the information about specific particles to get an overall density, which was precisely what we wanted. We can now introduce some thermodynamic quantities that describe our system.

From the solution density and the dispersion relation \eqref{disp} we can build the total energy density $e$ (per unit length)
\begin{equation}
e=\int\dd u~ \rho (u)\epsilon(\tilde{p}(u))\,,
\end{equation}
while the entropy density is given by 
\begin{equation}
s =\beta^{-1} \int\dd u~\log\frac{\Gamma(\beta\Delta u (\rho+\bar\rho)+1)}{\Gamma(\beta\Delta u \rho+1)\Gamma(\beta\Delta u \bar{\rho}+1)}\,,
\end{equation}
Finally, each particle type has a particle density $n^i$ and a chemical potential $\mu_i$ that corresponds to the logarithm of the respective eigenvalue of the twist operator $A$
\begin{equation}
\mu_Q=iQ(\alpha+\dot\alpha)\,,\quad \mu_{M|vw}^{(r)}=2iM\dot\alpha\,,\quad\mu_{M|w}^{(r)}=2iM\dot\vartheta\,,\quad\mu_\pm^{(r)}=i(\dot\alpha-\dot\vartheta)\,,
\end{equation}
where we introduced an index $(a)$ to discern between the two copies (wings) of $\mathfrak{psu}(2|2)_{c.e.}$, the left $(l)$ or right $(r)$ wing.\footnote{This naming anticipates the structure of the ``T-hook" we will encounter later (see fig.\ref{figure2}).} For the left wing the chemical potentials take the same form but with undotted twist angles.

The $y$-particles pick up a chemical potential from their fermionic character and the winding number $m$
\begin{equation}
\mu_\pm=i\pi(m+1)\,.
\end{equation}
However, we will set $m=0$ in the following.

The quantity we want to compute in the end is the Helmholtz free energy, which is given by 
\begin{equation}\label{free}
F=\beta(e+Ts+\mu_in^i)\,,
\end{equation}
where $T=L^{-1}$. The equilibrium condition $\delta F=0$ results in the TBA equations, where \eqref{density} and the auxiliary versions thereof are used to derive expressions for the hole density in terms of the particle density. We introduce the functions $Y_i=\rho_i/\bar{\rho}_i$ for states of type $i$ ($Y_\pm^{-1}$ for y-particles, absorbing the $i\pi$ chemical potential), which makes the (canonical) TBA equations take the form \cite{vanTongeren:2013gva}
\begin{align}
\begin{split}
\log Y_Q=&-\mu_{Q}-L\tilde{\epsilon}_Q+\log(1+Y_M)\star K^{MQ}_{\mathfrak{sl(2)}}\label{can1}\\
&+\sum_{(a)}\log\left(1+\frac{1}{Y^{(a)}_{M|vw}}\right)\star K^{MQ}_{vwx}+\sum_{(a),b=\pm}\log\left(1-\frac{1}{Y^{(a)}_{b}}\right)\,\hat\star\, K^{yQ}_{b}\,,
\end{split}\\
\log Y^{(a)}_{M|vw}=&-\mu_{M|vw}^{(a)}+\log\left(1+\frac{1}{Y^{(a)}_{N|vw}}\right)\star K_{NM}+\log\frac{1-\frac{1}{Y^{(a)}_-}}{1-\frac{1}{Y^{(a)}_+}}\,\hat\star\,K_M-\log(1+Y_Q)\star K_{xv}^{QM}\,,\\
\log Y^{(a)}_{M|w}=&-\mu_{M|w}^{(a)}+\log\left(1+\frac{1}{Y^{(a)}_{N|w}}\right)\star K_{NM}+\log\frac{1-\frac{1}{Y^{(a)}_-}}{1-\frac{1}{Y^{(a)}_+}}\,\hat\star\, K_M\,,\\
\log Y^{(a)}_{\pm}=&-\mu_{\pm}^{(a)}-\log\left(1+Y_Q\right)\star K^{Qy}_\pm+\log \frac{1+\frac{1}{Y^{(a)}_{M|vw}}}{1+\frac{1}{Y^{(a)}_{M|w}}}\star K_M\,.\label{TBAy}
\end{align}
We employed the notation of \cite{vanTongeren:2013gva}, where the various kernels $K^{A,B}$ relate to the previously discussed functions \eqref{kernel} through fusing the constituent particles of bound states together. Explicit formulas are given in the appendix of \cite{vanTongeren:2013gva}. In the detailed rapidity analysis, it turns out that the kernels for the two roots of $y$-particles are related via analytic continuation. To avoid ``double counting", one only needs to integrated them over $u\in(-2g,2g)$ which we denoted by $\hat{\star}$ in the convolution. Later we will also come across the complement $\check{\star}$, which denotes integration over $(-\infty,-2g)\cup(2g,\infty)$.

We can use a solution for the full set of $Y$ functions to compute the free energy
\begin{equation}\label{TBE}
F=-\frac{\beta}{L}\int\dd u\sum_{Q=1}^{\infty}\frac{1}{2\pi}\frac{\dd \tilde p^Q}{\dd u}\log(1+Y_Q)\,.
\end{equation}
The ground state energy of the original model is then given by the limit 
\begin{equation}\label{goal}
E_0=\lim_{\beta\to\infty}\left(\frac{L}{\beta }F\right)=-\int\dd u\sum_{Q=1}^{\infty}\frac{1}{2\pi}\frac{\dd \tilde p^Q}{\dd u}\log(1+Y_Q)\,.
\end{equation}

Higher excitations can be described by analytic continuation, which introduces certain driving terms in the TBA equations \cite{Dorey:1996re, Bajnok:2008bm, Gromov:2009tv, deLeeuw:2011rw}. However, this is only partially known and the quantum spectral curve fixes this issue quite elegantly. 

Solving the TBA equations is quite challenging, so instead of tackling them directly, a few simplifications can be made. However, these simplifications drop some of the information contained in the full (canonical) TBA equations. Especially, we will find that the chemical potentials drop out, so the twisting we introduced does not influence the simplifications of TBA. To restore this information, one has to impose conditions on the asymptotics of the solutions and there we will see the twists re-emerge.

\subsection{Twisted $Y$- and $T$-system}

\begin{figure}
	\centering
	\includegraphics[width=0.8\textwidth]{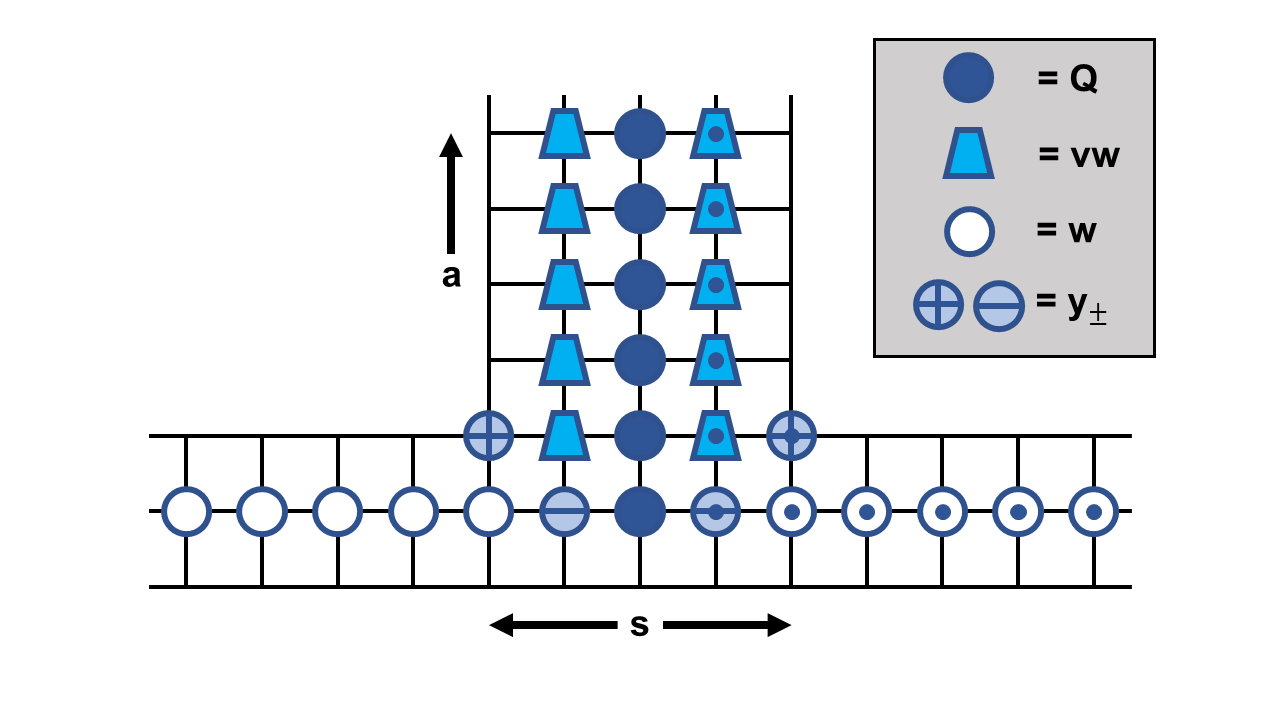}
	\caption{\small{The ``T-hook" lattice spanned by the $T$-functions \cite{Gromov:2009tv}. Outside the lattice we set $T=0$ and then require the Hirota equation \eqref{Tsys} to hold on every node of the lattice. Additionally, we can relate the various $Y$-functions to ratios \eqref{Yfunc} of $T$-functions surrounding the nodes we placed them at. The $Y$-system then also only includes next neighbour interaction. We denoted the right wing of functions with a dot to connect to our naming convention for angles. One can observe how the two $PSU(2|2)$ sectors are held together only by the fundamental $Q$-strings, which again mirrors the mostly independent treatment they get in the ABA.}}\label{figure2}
\end{figure}

The TBA equations in the form given above are highly interconnected and involve sums over all constituent numbers of bound states. To decouple the equations one can make use of the fact that bound states just behave as the sum of their constituents. We can therefore rearrange particles of two bound states. Take for example two strings of length $Q$ and shift their rapidities by $\pm \frac{ i}{2}$. This configuration is equivalent to one string of length $Q+1$ and one of length $Q-1$. This is captured by relations of the form
\begin{equation}\label{reshuffle}
K^{\chi Q}\left(v,u+\frac{i}{2}\right)+K^{\chi Q}\left(v,u-\frac{i}{2}\right)=K^{\chi Q+1}(v,u)+K^{\chi Q-1}(v,u)\,.
\end{equation} 
More accurately, one introduces the kernel  
\begin{equation}
s(u)=\frac{1}{2 \cosh(\pi u)}
\end{equation}
and operator $s^{-1}$ such that 
\begin{equation}
f\star s^{-1}(u)=\lim_{\epsilon\to 0}\left[f\left(u+\frac{i}{2}-i\epsilon\right)+f\left(u-\frac{i}{2}+i\epsilon\right)\right] \quad \Rightarrow \quad s\star s^{-1}(u)=\delta(u)\,.
\end{equation}
This allows us to identify the inverse 
\begin{equation}
(K+1)^{-1}_{PQ}=\delta_{P,Q}-\underbrace{(\delta_{P,Q+1}+\delta_{P,Q-1})}_{=:I_{PQ}}s\,,
\end{equation}
which captures the reshuffling characteristics \eqref{reshuffle} and can be proven rigorously in Fourier space. Application of this operator to the TBA equations leads to ``localised" equations that each depend only on a few $Y$-functions. These are called the simplified TBA equations which we shall just list without going into the details, which can be found in \cite{vanTongeren:2013gva}
\begin{align}
\log Y_1&=\sum_{(a)}\log\left(1-\frac{1}{Y_-^{(a)}}\right)\,\hat{\star}\,s-\log(1+\frac{1}{Y_2})\star s-\check\Delta\,\check\star\, s\,, \\
\log Y_Q&=\log\frac{Y_{Q+1}Y_{Q-1}}{(1+Y_{Q-1})(1+Y_{Q+1})}\star s + \sum_{(a)}\left(1+\frac{1}{Y^{(a)}_{Q-1|vw}}\right)\star s  \qquad Q>1\,,\\
\log Y_{M|vw}^{(a)}&=\log\left(1+Y_{M+1|vw}^{(a)}\right)\left(1+Y_{M-1|vw}^{(a)}\right)\star s-\log(1+Y_{M+1})\star s+\delta_{M,1}\log\frac{1-Y_-^{(a)}}{1-Y_+^{(a)}}\,\hat{\star}\,s\,,
\\
\log Y_{M|w}^{(a)}&=\log \left(1+Y_{M+1|w}^{(a)}\right)\left(1+Y_{M-1|w}^{(a)}\right)\star s +\delta_{M,1}\log\frac{1-\frac{1}{Y_-^{(a)}}}{1-\frac{1}{Y_+^{(a)}}}\,\hat{\star}\,s\,,\\
\log\frac{Y_+^{(a)}}{Y_-^{(a)}}&=\log(1+Y_Q)\star K_{Qy}\,,\\
\log Y_-^{(a)}Y_+^{(a)}&=-\log(1+Y_Q)\star K_{Q}+2\log(1+Y_Q)\star K_{xv}^{Q1}\star s+ 2\log \frac{1+Y_{1|vw}^{(a)}}{1+Y_{1|w}^{(a)}}\star s\,.
\end{align} 
In the first equation we abbreviated terms convoluted on $\abs{u}>2g$ as
\begin{equation}\begin{split}
\check{\Delta}=&L\check{\epsilon}+\sum_{(a)}\log\left(1-\frac{1}{Y_-^{(a)}}\right)\left(1-\frac{1}{Y_+^{(a)}}\right)\,\hat\star\,\check K+2\log(1+Y_Q)\star\check K^\Sigma_Q\\
&+\sum_{(a)}\log\left(1-\frac{1}{Y_{M|vw}^{(a)}}\right)\star \check K_M+\sum_{(a)}\log \left(1-Y_{k-1|vw}^{(a)}\right)\star\check K_{k-1}\,.
\end{split}\end{equation}
If we now act with $s^{-1}$ we find purely algebraic relations called the ``$Y$-system"
\begin{align}
\frac{Y_1^+Y_1^-}{Y_2}&=\frac{\prod_{(a)}\left(1-\frac{1}{Y_-^{(a)}}\right)}{1+Y_2}\,, \qquad \abs{u}<2g\,,\\
\frac{Y_Q^+Y_Q^-}{Y_{Q-1}Y_{Q+1}}&=\frac{\prod_{(a)}\left(1-\frac{1}{Y_{Q-1|vw}^{(a)}}\right)}{(1+Y_{Q-1})(1+Y_{Q+1})}\,,\\
Y^+_{1|vw}Y^-_{1|vw}&=\frac{1+Y_{2|vw}}{1+Y_2}\left(\frac{1-Y_-}{1-Y_+}\right)^{\theta(2g-\abs{u})}\,,\\
Y^+_{M|vw}Y^-_{M|vw}&=\frac{(1+Y_{M-1|vw})(1+Y_{M+1|vw})}{1+Y_{M+1}}\,,\\
Y^+_{1|w}Y^-_{1|w}&=(1+Y_{2|w})\left(\frac{1-Y_-^{-1}}{1-Y_+^{-1}}\right)^{\theta(2g-\abs{u})}\,,\\
Y^+_{M|w}Y^-_{M|w}&=(1+Y_{M-1|w})(1+Y_{M+1|w})\,,\\
Y_-^+Y_-^-&=\frac{1+Y_{1|vw}}{1+Y_{1|w}}\frac{1}{1+Y_1}\,,
\end{align}
where the index $(a)$ has been suppressed since both wings obey the same equations and $Y_+$ can be derived as analytic continuation of $Y_-$. The upper index $\pm$ corresponds to shifts of the rapidity by $\pm \frac{i}{2}$. 
We see that the various $Y$ functions only depend on a number of maximally 4 ``neighbours", which means we can arrange them pictorially in a square lattice. This results in the famous ``T-hook" (see figure \ref{figure2}). The lattice structure of these relations can be used to index all $Y$-functions by two integers 
\begin{equation}
\begin{split}
Y_Q=Y_{Q,0}\,,\quad Y_{M|vw}^{(r)}=Y_{M+1,1}^{-1}\,,\quad Y_{M|w}^{(r)}=Y_{1,M+1}\,,\quad Y_-^{(r)}=-Y_{1,1}^{-1}\,,\quad Y_+^{(r)}=-Y_{2,2}
\end{split}
\end{equation}
and similar for the left wing, where the second index is to be taken negative. With this set-up the entire $Y$-system (for  $\abs{u}<2g$)\footnote{The domain of validity can be extended from $\abs{u}<2g$ to a domain with ``long" branch cuts along $(-\infty,-2g)\cup(2g,\infty)$, which we will meet again when discussing the QSC.} can be written in a uniform way
\begin{equation}\label{Y-system}
Y^-_{a,s}Y^+_{a,s}=\frac{(1+Y_{a,s-1})(1+Y_{a,s+1})}{(1+Y_{a-1,s}^{-1})(1+Y_{a+1,s}^{-1})}\,,
\end{equation}
This system can be related to the Hirota equation \cite{doi:10.1143/JPSJ.50.3785}, which appeared in the discussion of canonical examples of integrable systems, such as the KdV equation or the sine-Gordon model. If we represent the $Y$-functions as
\begin{equation}\label{Yfunc}
Y_{a,s}=\frac{T_{a,s+1}T_{a,s-1}}{T_{a+1,s}T_{a-1,s}}\,,
\end{equation}
the Y-system \eqref{Y-system} reduces to the Hirota equation
\begin{equation}\label{Tsys}
T^+_{a,s}T^-_{a,s}=T_{a+1,s}T_{a-1,s}+T_{a,s+1}T_{a,s-1}\,,
\end{equation}
which is also referred to as the $T$-system. Note, however, that the identification $\eqref{Yfunc}$ is not unique, since there is a fourfold ``gauge freedom" with respect to transformations 
\begin{equation}\label{gauge}
T_{a,s}\to g_1^{[a+s]}g_2^{[a-s]}g_3^{[-a+s]}g_4^{[-a-s]}T_{a,s}\,,
\end{equation}
where $f^{[a]}(u)=f(u+{ia\over2} )$. One can check that this leaves the $Y$-functions invariant. 

The $T$-system \eqref{Tsys} is the most compact form to encode the algebraic structures underlying the spectral problem of $\mathcal{N}=4$ SYM. In fact the $T$-system encodes certain characters of the $PSU(2,2|4)$ symmetry group \cite{Gromov:2010vb}, so it was already postulated before the connection to the TBA was made.

In this chain of simplifications we acted with projection operators (namely $(K+1)^{-1}$ and $s^{-1}$) which both drop some of the information specified in the canonical TBA equations \cite{Cavaglia:2010nm,Balog:2011nm,Gromov:2011cx}. We note especially that any constant terms of the canonical TBA equations have dropped out. As previously mentioned, this means that as far as the $Y$-system is concerned, orbifolds are treated on an equal footing to the untwisted case. The $Y$-system is thus more generic than the TBA, so we have to select the physical solution out of the set of all possible solutions to the $Y$-system. This selection can be made by analysing the large-$u$ asymptotics. Instead of first generating a set of solutions to select from, we can turn this reasoning around and first specify the appropriate asymptotics $Y^0$ explicitly and then only search for solutions of the $Y$-system with said asymptotics.\\

If we take a closer look at the canonical TBA equations we realise that for large $L$ the first term in \eqref{can1} dominates and the functions $Y_Q$ are exponentially suppressed
\begin{equation}
Y_Q\sim Y^0_Q\sim e^{-\tilde{\epsilon}_QL}\,.
\end{equation}
Therefore we can drop all $\log(1+Y_Q)$ terms. This in turn results in 
\begin{equation}
Y_+^{0}=Y_-^{0}\,,
\end{equation}
which decouples $y$-particles from $vw$-strings and $w$-strings. Then, their TBA equations can be solved by expressions constant in $u$, utilising the q-numbers
\begin{equation}\label{q-number}
[n]_q=\frac{q^n-q^{-n}}{q^1-q^{-1}}\,.
\end{equation}
The large-$L$ asymptotic solutions are given by \cite{vanTongeren:2013gva}
\begin{equation}\label{asymp}
Y_{M|vw}^{0,(r)}=[M]_{e^{i\dot\alpha}}[M+2]_{e^{i\dot\alpha}}\,,\quad Y_{M|w}^{0,(r)}=[M]_{e^{i\dot\vartheta}}[M+2]_{e^{i\dot\vartheta}}\,,\quad Y_\pm^{0,(r)}=\frac{[2]_{e^{i\dot\alpha}}}{[2]_{e^{i\dot\vartheta}}}\,
\end{equation}
and the left wing follows analogously. The main $Y$-functions are
\begin{equation}\label{YQ}
Y_Q^0=([2]_{e^{i\alpha}}-[2]_{e^{i\vartheta}})([2]_{e^{i\dot\alpha}}-[2]_{e^{i\dot\vartheta}})[Q]_{e^{i\alpha}}[Q]_{e^{i\dot\alpha}}e^{-\tilde{\epsilon}_QL}=:A_Qe^{-\tilde{\epsilon}_QL}\,.
\end{equation}
$Y_Q^0$ is exponentially suppressed but only vanishes if supersymmetry is preserved. The full solution is then given by an expansion
\begin{equation}
Y=Y^0\,(1+y+\dots)\,.
\end{equation}
To find the leading order (LO) energy correction for large $L$ we need to expand $\log(1+Y_Q)$ in \eqref{goal} 
\begin{equation}
E_0^{LO}=-\sum_{Q=1}^{\infty}\int\frac{\dd\tilde p}{2\pi}~Y^0_Q(\tilde p)=-\sum_{Q=1}^{\infty}A_Q\int\frac{\dd\tilde p}{2\pi} ~e^{-\tilde{\epsilon}_Q(\tilde p)L}\,.
\end{equation}
The next-to-leading order (NLO) includes both a $(Y^0)^2$ and a $Y_0\cdot y$ term.\\

Such expressions have been derived much earlier by a simple argument due to L\"uscher \cite{Luscher:1985dn}. Starting from the mirror-model partition function with the defect operator $A=e^{iD}$, we can expand in large $L$ and find that single particle states contribute at leading order 
\begin{equation}
Z=\lim_{\beta\to\infty}\Tr(e^{-\tilde H (\beta)L}e^{iD })=1+\sum_{k,\alpha}e^{iD_{\alpha}-\tilde\epsilon(\tilde p_k)L}+\sum_{k,l,(\alpha,\beta)}e^{i D_{(\alpha,\beta)}-\left(\tilde\epsilon(\tilde p_k)+\epsilon(\tilde p_l)\right)L}+\dots\,,\label{lusch}
\end{equation}
where we assumed that $D$ acts diagonally on Fock space. 
The ground state energy is just the logarithm of the partition function 
\begin{equation}
E_0(L)=-\lim_{\beta\to\infty}\frac{1}{\beta}\log\left[\Tr(e^{-\tilde H (\beta)L}e^{iD})\right]\,.
\end{equation}
The leading order contribution can then be phrased as
\begin{equation}\label{LO}
E_0^{LO}(L)=-\sum_{Q=1}^{\infty}\Str_Q(e^{iD}) \int\frac{\dd\tilde p}{2\pi}e^{-\tilde \epsilon_Q(\tilde{p})L}\,,
\end{equation}
which agrees with the TBA result when we identify the prefactors $A_Q=\Str_Q(e^{iD})$. As a simple sanity check, we observe that a trivial $D$ as in the supersymmetric case results in a vanishing energy contribution as $\Str_Q(\mathbbm{1})=0$. The logarithm expansion works similar to the TBA expression, so we have a nice interpretation of the TBA expansion in terms of particles moving around the circle. \\

Let us conclude this summary of orbifolds in the TBA framework by commenting on (the asymptotics of) the $T$-system. It has been shown \cite{Gromov:2010vb}, that the constant T-hook can be solved in terms of characters of the underlying $PSU(2,2|4)$, which pointed the way towards the quantum spectral curve. In our orbifolded model, we expect the asymptotics of the $T$-functions to reproduce the asymptotic $Y$-functions we found earlier. 

In the large-$u$ asymptotic case the shifts by $\frac{i}{2}$ become negligible. We can then try to solve the simplified Hirota equations
\begin{equation}
(T^0_{a,s})^2=T^0_{a+1,s}T^0_{a-1,s}+T^0_{a,s+1}T^0_{a,s-1}
\end{equation}
by constants independent of $u$. We set all $T$-functions outside the T-hook to $0$, which determines the boundary conditions for this discretised PDE. Of the gauge freedom \eqref{gauge} we only recover exponential dependence on the $(a,s)$ indices
\begin{equation}
T^0_{a,s}\to \exp(c_0+c_1 a+ c_2 s+c_{3} as)T^0_{a,s}\,,
\end{equation}
which we use to set $T^0_{1,0}=1$ and $T^0_{0,s}=1$ for all $s$. The latter condition uses two gauge parameters to fix $T^0_{0,0}=T^0_{0,1}=1$, the other $T^0_{0,s}$ follow due to the Hirota equations and the boundary conditions. The last gauge parameter will be used later.

Now we can compare with the asymptotic $Y$-functions and see whether we find an appropriate set of $T$-functions. Take for example the $Y_{M|w}$-asymptotics \eqref{asymp}
\begin{equation}
Y^{0,(r)}_{M|w}=[M]_{e^{i\dot \vartheta}}[M+2]_{e^{i\dot \vartheta}}= \frac{T_{1,M}^0T_{1,M+2}^0}{T_{2,M+1}^0}
\end{equation}
and similar for the left wing. We therefore expect that ($M>0$)
\begin{equation}
T_{1, M}^{0}\sim[M]_{e^{i\dot \vartheta}}\,,\qquad T_{1, -M}^{0}\sim[M]_{e^{i\vartheta}}\,.
\end{equation}
This is indeed the case, but to arrive at this conclusion and fix the prefactors, we first need to solve a system of algebraic equations at the middle of the T-hook. If we impose the Hirota equations for $T_{1,1}$, $T_{1,2}$, $T_{2,1}$ and $T_{2,2}$ and the correct asymptotics for the $Y$ functions on these nodes, we can determine the surrounding $T$-functions. The top and right wing can be constructed iteratively. The left wing side follows analogously.
Let us introduce the abbreviations 
\begin{equation}
 \xi=([2]_{e^{i \vartheta}}-[2]_{e^{i \alpha}})\,,\qquad \dot\xi=([2]_{e^{i\dot \vartheta}}-[2]_{e^{i\dot \alpha}})\,.
\end{equation}
Then the asymptotic $T$-functions take the values ($M>0$)
\begin{equation}
\begin{split}
&T_{1,-M}^0=[M]_{e^{i\vartheta}}\xi \epsilon^{M}\,,\quad T_{M,0}=1\,,\quad T_{1,M}^0=[M]_{e^{i\dot\vartheta}}\dot\xi \dot{\epsilon}^{M}\,,\\
&T_{M,-1}=(-1)^{M+1}[M]_{e^{i\alpha}}\xi \epsilon^{M}\,,\quad T_{M,-2}=\xi^2\epsilon^{2M}\,, \\ &T_{M,1}=(-1)^{M+1}[M]_{e^{i\dot\alpha}}\dot\xi \dot{\epsilon}^{M}\,,\quad T_{M,2}=\dot\xi^2\dot{\epsilon}^{2M}\,.
\end{split}
\end{equation}
These parameters $\epsilon$ and $\dot{\epsilon}$ are yet to be determined and we have not yet imposed the middle node Hirota equations for $T^0_{a,0}$
\begin{equation}
1=T_{a,0}^2=T^0_{a+1,0}T^0_{a-1,0}+T^0_{a,+1}T^0_{a,-1}=1+[M]_{e^{i\alpha}}[M]_{e^{i\dot\alpha}}\xi \dot\xi (\epsilon\dot{\epsilon})^{M}\,.
\end{equation} 
Clearly, this equation can only be solved for infinitesimal $\epsilon\dot{\epsilon}$, but since this is an asymptotic $T$-system, we have a good explanation for this: the exponential $e^{-\tilde{\epsilon}_QL}$ in the middle node $Y_Q^0$-functions is responsible for this suppression. Indeed, we can approximate $\tilde{\epsilon}_Q\sim Q$ and find that
\begin{equation}\label{delta}
\epsilon\dot{\epsilon}\sim e^{-L}\,.
\end{equation}

To completely fix the individual factors $\epsilon$ and $\dot{\epsilon}$, we remind ourselves of the last remaining gauge parameter $c_{as}$ and see that we can choose $\epsilon=1$ or any other combination that yields the asymptotic \eqref{delta}. Indeed, with this set of $T$-functions we immediately find
\begin{equation}
Y_Q^0\sim\xi\dot{\xi}[Q]_{e^{i\alpha}}[Q]_{e^{i\dot\alpha}}\dot{\epsilon}^Q=\xi\dot{\xi}[Q]_{e^{i\alpha}}[Q]_{e^{i\dot\alpha}}e^{-QL}\,,
\end{equation}
which is a good approximation to the $Y$-system result \eqref{YQ}.

\section{Orbifolding the quantum spectral curve}\label{QSC}

The most recent progress in the integrability program was the formulation of the quantum spectral curve \cite{Gromov:2014caa} (see \cite{Gromov:2017blm,Levkovich-Maslyuk:2019awk} for a more pedagogical review). To derive its structure, one can reorder the $T$-system in into certain $\mathbf{P}_a$ and $\mathbf{Q}_i$ functions, whose algebraic relations, asymptotics and analytic properties capture the structure of the TBA and fully determine the spectrum. Moreover, the QSC can be related both to the spin-chain and the sigma model in a very intuitive way. We will present the QSC from this more natural angle, keeping in mind that the proper derivation is built on the TBA \cite{Gromov:2014caa}. We will address this connection to the $T$-system in appendix \ref{appendix}.

As in the case of the previously described $Y$- and $T$ system we can deform the QSC by changing the large-$u$ asymptotics of the $Q_i$ functions. This twisting has been suggested in \cite{Kazakov:2015efa} and performed for the case of $\gamma$-deformations. The case of orbifolds can be treated in a similar manner and we develop the appropriate twisting in \ref{new}.

The QSC makes use of the dualities that arise from different choices of hierarchies in the nested Bethe\,-Yang equations. Remember that for the TBA we introduced a hierarchy starting from $w_1$ particles. However, we could have chosen any other ground state and in fact even our splitting of $PSU(2,2|4)$ was already dependant on an arbitrary light-cone gauge. To explain how to avoid these arbitrary choices, we need to further analyse the structure of the Bethe\,-Yang equations. 

\subsection{From spin-chain to QSC}

The S-matrices in the Bethe\,-Yang equations \eqref{BY1}-\eqref{BY3} take the generic form 
\begin{equation}
S^{A,B}(u_{A,k},u_{B,j})\sim \frac{u_{A,k}-u_{B,j}+n_{A,B}\frac{i}{2}}{u_{A,k}-u_{B,j}-n_{A,B}\frac{i}{2}}\,,
\end{equation}
where $u_{A,k}$ are the Bethe roots of type $A$ in some appropriate rapidity parametrisation and $n_{A,B}$ are some fixed integers. If we define $Q$-functions
\begin{equation}
Q_A(u)\sim\prod_{j=1}^{K^A}(u-u_{A,j})\,,
\end{equation}
we can rewrite the Bethe\,-Yang equations entirely in terms of $Q$-functions. As an example, \eqref{BY3} can be rewritten as
\begin{align}\label{QBY}
1=\prod_{l=1}^{K^{II}}\frac{w_k-y_l+\frac{i}{2}}{w_k-y_l-\frac{i}{2}}\prod_{l\neq k}^{K^{III}}\frac{w_k-w_l-i}{w_k-w_l+i}\quad \Leftrightarrow \quad
-1=\frac{Q_{II}^{+}Q_{III}^{[-2]}}{Q_{II}^{-}Q_{III}^{[+2]}}\,,\,\, u=u_{III,k}\,.
\end{align}
We notice that numerator and denominator look very similar and  we can instead package this information in a ``$QQ$-relation"
\begin{equation}\label{QBYq}
Q_{\emptyset}Q_{II}=Q_{III}^+ \hat Q _{III}^--Q_{III}^-\hat Q _{III}^+\,,
\end{equation}
where $Q_{\emptyset}=1$ has been introduced for later convenience. Indeed, if we shift this equation by $\pm\frac{i}{2}$, evaluate at $u_{III,k}$ and take the ratio, we get back to \eqref{QBY}. The function $\hat Q_{III}$ is implicitly defined via \eqref{QBYq} in terms of the other $Q$-functions and cancels out in the Bethe\,-Yang equations. Nevertheless, we can try to interpret it. Note that it should be proportional to a polynomial of order $K^{II}-K^{III}$. As mentioned earlier, we chose a specific hierarchy of excitations on the spin-chain, but similarly, we could have chosen other ground states and orderings. In a $SU(2)$ spin-chain, we can think of up-states $\uparrow$ excited on an all-down ground state $\ket{\downarrow\downarrow\dots\downarrow}$ just as well as of down-states $\downarrow$ excited on an all-up ground state $\ket{\uparrow\uparrow\dots\uparrow}$. A similar logic applies here. With our arbitrary choice of perspective, we lost sight of this fundamental symmetry of the spin-chain, but the ``dual" $ \hat Q_{III}$-functions encode the ``hole"-excitations and restore the symmetry. 

We could simply write the alternative Bethe equation 
\begin{align}
-1=\frac{Q_{II}^{+}\hat Q_{III}^{[-2]}}{Q_{II}^{-}\hat Q_{III}^{[+2]}}\,,\quad u=\hat u_{III,k}\,,
\end{align}
which captures the same physics, encoded over a different hierarchy.

For the other Bethe\,-Yang equations, a similar description via $Q$-functions is possible and when we combine the two $\mathfrak{psu}(2|2)_{c.e.}$ factors we end up with a system of $2^8=256$ $Q$-functions. Every equivalent set of Bethe\,-Yang equations then relates $8$ of these in a hierarchy. However, due to the intricate net of dualities, it is actually enough to specify any $8$ independent $Q$-functions. \\

\begin{figure}
	\centering
	\begin{minipage}{0.60\textwidth}%
		\includegraphics[width=\textwidth]{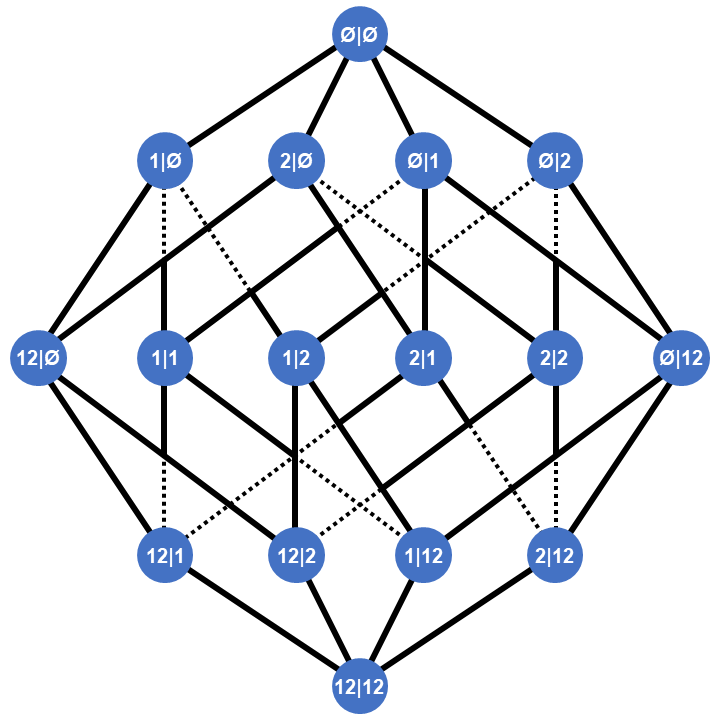}
	\end{minipage}

	\caption{\small We can think of the $Q$-functions as vertices of a hypercube, the faces of which represent the $QQ$-relations. This diagram is known as the Hasse diagram of the Index-set. For simplicity, we only present a 4D-hypercube, but this can be thought of as a $PSU(2|2)$ subsystem of $PSU(2,2|4)$.}\label{figure3}
\end{figure}

A useful ordering prescription for the various $Q$-functions is to introduce ordered multiindices of the form $a...d|i...j$, where a distinction has been made with respect to the $SU(4)$ $R$-symmetry ($S^5$) and the $SU(2,2)$ Poincare-symmetry ($AdS_5$). Starting with $Q_{\emptyset|\emptyset}=1$, we successively add indices to get a valid set of Bethe\,-Yang equations. For example the standard asymptotic Bethe Ansatz appearing in \cite{Beisert:2005fw} corresponds to
\begin{equation}\label{path}
Q_{\emptyset|\emptyset}\to Q_{\emptyset|1}\to Q_{1|1}\to Q_{12|1}\to Q_{12|12}\to Q_{12|123}\to Q_{123|123}\to Q_{1234|123}\to Q_{1234|1234}\,.
\end{equation}
This can be visualised as climbing up the Hasse diagram or the Dynkin diagram of $SU(4|4)$, as shown in figure \ref{figure3}. 
The dualities then correspond to exchanging the order in which these indices are added. Succinctly, we have the following $QQ$-relations
\begin{align}
Q_{A|I}Q_{Aab|I}&\propto Q_{Aa|I}^+Q_{Ab|I}^--Q_{Ab|I}^+Q_{Aa|I}^-\,,\label{QQb1}\\
Q_{A|I}Q_{A|Iij}&\propto Q_{A|Ii}^+Q_{A|Ij}^--Q_{A|Ij}^+Q_{A|Ii}^-\,,\label{QQb2}\\
Q_{Aa|I}Q_{A|Ii}&\propto Q_{Aa|Ii}^+Q_{A|I}^--Q_{A|I}^+Q_{Aa|Ii}^-\,,\label{QQf}
\end{align}
where $A$ and $I$ summarise any uninvolved ``spectator" indices. The first two sets of $QQ$-relations are called bosonic, while the last set is called fermionic. 

Our previous example \eqref{QBYq} was linked to $w$-particles, which sit on-top of all other excitations or in other words their Bethe\,-Yang equation is the first to be solved in the hierarchy of Bethe\,-Yang equations. It belongs to the bosonic $QQ$-relations and can be represented by 
\begin{equation}\label{protoQQ}
Q_{\emptyset|\emptyset}Q_{12|\emptyset}\propto Q_{1|\emptyset}^+ Q_{2|\emptyset}^-- Q_{2|\emptyset}^+ Q_{1|\emptyset}^-\,.
\end{equation}

We can also define a ``Hodge dual" set of $Q$-functions as
\begin{equation}
Q^{a_1\dots a_n|i_1\dots i_m}:=(-1)^{nm}\epsilon^{a_1\dots a_nb_1\dots b_{4-n}}\epsilon^{i_1\dots i_mi_1\dots i_{4-m}}Q_{b_1\dotsb_{4-n}|i_1\dots i_{4-m}}\,,
\end{equation}
which satisfy the same $QQ$-relations. We also adopt a slightly different naming convention
\begin{equation}
\mathbf{P}_a:=Q_{a|\emptyset}\,,\quad \mathbf{P}^a:=Q^{a|\emptyset}\,, \quad \mathbf{Q}_i:=Q_{\emptyset|i}\,,\quad \mathbf{Q}^i:=Q^{\emptyset|i}\,.
\end{equation}

Departing from the spin-chain where $Q_{1234|1234}=u^L$ , for $\mathcal{N}=4$ SYM we need to make the Hodge duality manifest and set $Q_{1234|1234}=1$. This is possible if an overall scale is factored out, which is implied by the $P$ in $PSU(2,2|4)$. This leads to the identities 
\begin{equation}\label{PQ}
\mathbf{Q}_i\mathbf{P}_aQ^{a|i,\pm}=0\,, \quad \mathbf{P}^a=Q^{a|i,\pm}\mathbf{Q}_i\,, \quad \mathbf{Q}^i=Q^{a|i,\pm}\mathbf{P}_a\,, \quad \mathbf{P}_a\mathbf{P}^a=0\,,\quad \mathbf{Q}_i\mathbf{Q}^i=0\,.
\end{equation} 
In a sense $Q_{a|i}$ acts as a metric for the $\mathbf{P}_a$ and $\mathbf{Q}_i$-functions, and indeed $Q_{a|i}Q^{b|i}=-\delta_a^b$.

At this point the $Q$-functions are not polynomials anymore. In fact, as they are related to the $T$-functions (see appendix \ref{appendix}), whose dependence on $u$ is governed by the Zhukovski variables $x(u)$ \eqref{Zhukovski}, they $Q$-functions inherit their non-analyticity. Therefore, the $Q$-functions are in general multi-valued functions with branch cuts along the real interval $(-2g,2g)$ (short cut), its complement $(-\infty,-2g)\cup(2g,\infty)$ (long cut) or $\frac{i}{2}$-shifted versions thereof.
It is believed that $Q$-function asymptotics, glueing conditions along the cuts and the $QQ$-relations determine the full spectrum of single-trace operators of $\mathcal{N}=4$ SYM to a priori any order in $g$. 

Let us mention, that the subset of $\mathbf{P}_a$, $\mathbf{Q}_i$ and $Q_{a|i}$ already encodes all information necessary to solve the spectral problem. $\mathbf{P}_a$ and $\mathbf{Q}_i$ have to be specified asymptotically while $Q_{a|i}$ is formally given by 
\begin{equation}\label{primQQ}
Q_{a|i}=-\sum_{n=0}^\infty \mathbf{P}_a^{[2n+1]} \mathbf{Q}_i^{[2n+1]}\,.
\end{equation}

The asymptotics of the $Q$-functions can be deduced by comparison to the sigma-model, which we shall demonstrate in the next section.

\subsection{Relating the QSC to the sigma-model} 

To interpret the quantum spectral curve, we can draw parallels to the standard WKB approximation of quantum mechanics, where we solve the Schr\"odinger equation by the exponential Ansatz for the wave function
\begin{equation} \label{wavefunction}
\Psi(x)\sim C_+e^{ +\frac{i}{\hbar}\int \dd x~ p(x)}+ C_-e^{ -\frac{i}{\hbar}\int \dd x~ p(x)}\,,\qquad p(x)=\sqrt{2m(E-V(x))}\,,
\end{equation}
which has a branch cut along the classical solution between the two turning points, where $V(x)=E$. The Born-Sommerfeld quantisation condition, which specifies that the integral of the pseudomomentum $p(x)$ around the branch cut can only take discrete values $2\pi\hbar(n+\frac{1}{2})$, derives from the condition that the wave function glues nicely around these branch points. From far away in the $x$-plane the branch cut shrinks to a point and we expect the pseudomomentum to behave like a meromorphic function with a single pole at $x=0$ and residue $2\pi\hbar(n+\frac{1}{2})$.\\

A very similar behaviour is presented by the $Q$-functions and indeed we can think of them as wave functions (for a pedagogical comparison to the harmonic oscillator, see \cite{Gromov:2017blm}) and the sigma-model as the appropriate classical model. Remember that in section \ref{sigma} we defined a bunch of pseudomomenta via the eigenvalues of the monodromy matrix \eqref{monodromy}. These had branch cuts between $\pm1$ but after a proper redefinition 
\begin{equation}
z=x_{ph}(u)\,,
\end{equation}
we can match their analyticity properties to the $Q$ functions and recover the asymptotics
\begin{equation}\label{relation}
\begin{split}
\mathbf{P}_a\sim\exp(-\int^u\dd v~ p_a(v)),\quad &\mathbf{P}^a\sim\exp(+\int^u\dd v~ p_a(v)),\\
\mathbf{Q}^i\sim\exp(-\int^u\dd v~ p_{\hat i}(v)),\quad &\mathbf{Q}_i\sim\exp(+\int^u\dd v~ p_{\hat i}(v))\,.
\end{split}
\end{equation} 
With this knowledge, we can deduce the large $u$ asymptotics of the $Q$-functions. The asymptotics of the pseudomomenta depend on the charges/quantum numbers $(\Delta,S_1,S_2|J_3,J_1,J_2)$ of the state at hand , as was spelled out in \eqref{charges}. Since this is a classical matching of charges, there may be quantum corrections to the asymptotic charges (similar to the $1/2$ contribution to the energy of the harmonic oscillator) and we will derive these from the twisted case in the next section. 

The glueing conditions can be derived (heuristically) by analysing the glueing conditions of the Lax connection. It turns out that moving through the short branch cut $(-2g,2g)$, one has to identify
\begin{equation}
\tilde p_{\hat 1}(u)=-p_{\hat 2}(u)\,,\quad\tilde p_{\hat 2}(u)=-p_{\hat 1}(u)\,,\quad\tilde p_{\hat 3}(u)=-p_{\hat 4}(u)\,,\quad\tilde p_{\hat 4}(u)=-p_{\hat 3}(u)\,,
\end{equation}
where the tilde denotes analytical continuation through the branch cut. These translate to
\begin{equation}\label{glue}
\tilde{\mathbf{Q}}_1=\bar{\mathbf{Q}^2}\,,\quad \tilde{\mathbf{Q}}_2=\bar{\mathbf{Q}}^1\,,\quad \tilde{\mathbf{Q}}_3=\bar{\mathbf{Q}}^4\,,\quad \tilde{\mathbf{Q}}_4=\bar{\mathbf{Q}}^3\,,
\end{equation}
where we took into account the complex nature of the $Q$-functions in the quantum model.
The spectral problem thus reduces to:
\begin{itemize}
	\item Specifying asymptotics for $\mathbf{Q}_i$ and $\mathbf{P}_a$ as we had to do in the $Y$- and $T$-system. These are given in terms of fixed charges $(S_1,S_2|J_3,J_1,J_2)$ as well as the undetermined anomalous dimension $\Delta$.
	\item Solving $QQ$-relations like \eqref{primQQ} to derive the other $Q$-functions.
	\item Imposing gluing conditions of the form $\tilde{\mathbf{Q}}_1=\bar{\mathbf{Q}}^2$ etc.
	\item Reading off the anomalous dimension $\Delta$ from the solution.	
\end{itemize}
Although this problem can, in principle, be solved exactly, in practice one usually employs numerical methods (see e.g. the Mathematica code in \cite{Gromov:2017blm}).

\

Let us end this exposition by remarking that a very simple subsector of states is the ``left-right" (LR) symmetric subsector that includes the $\mathfrak{su}(2)$ and $\mathfrak{sl}(2)$ subsectors. It is specified by $J_2=S_2=0$.
In that case 
\begin{equation} \label{LR}
\mathbf{P}_a=\chi_{ab}\mathbf{P}^b\,,\quad \mathbf{Q}_i=\chi_{ij}\mathbf{Q}^j\,,\quad \chi=\begin{pmatrix}
0&0&0&1\\ 0&0&-1&0\\0&1&0&0\\-1&0&0&0
\end{pmatrix}.
\end{equation}

\subsection{Implementing twist}\label{new}
For the orbifold case, we can now use inspiration from both ABA and sigma-model considerations. Consistency with both models will teach us exactly how to implement twists in the quantum spectral curve. Indeed, twisted QSC has been discussed previously, since it is a way to derive the quantum shifts of charges we saw earlier \cite{Kazakov:2015efa}. A number of deformations have been discussed there, including $\gamma$ and $\beta$-deformations \cite{Kazakov:2015efa,Levkovich-Maslyuk:2020rlp,Marboe:2019wyc}. Here, we present the case of orbifolds which so far have not been treated in the QSC literature but give rise to interesting models as for example the $\mathcal{N}=2$ $\mathbb{Z}_2$-orbifold or type 0 string theory which we will discuss in the next section. \\

In the $T$-system \eqref{Tsys} the algebra was unchanged by the twisting, but the asymptotics had to be modified. In the QSC we expect this to be true, too (see appendix \ref{appendix} for the comparison). From the sigma-model discussion we expect a constant shift of the pseudomomenta, as given in \eqref{tcharges} so we should get asymptotics of the schematic form
\begin{equation}
\mathbf{P}_a\sim x_a^{iu}u^{-M_a}\,,\qquad  \mathbf{Q}_i\sim y_i^{-iu}u^{\hat M_i}\,,
\end{equation}
where the twist factors are abbreviated as
\begin{equation}
(x_1,x_2,x_3,x_4,y_1,y_2,y_3,y_4)=(e^{-i\dot{\vartheta}},e^{i\dot{\vartheta}},e^{-i\vartheta},e^{i\vartheta},e^{i\dot\alpha},e^{-i\dot{\alpha}},e^{i\alpha},e^{-i\alpha})
\end{equation} 
and the asymptotic charges $(M_a,\hat M_i)$ depend on the classical charges given in \eqref{tcharges} and quantum shifts. We will not care about the constant prefactors of these leading asymptotics here, but move their discussion to the appendix \ref{appendix2}.

If we turn to the Bethe\,-Yang equations, we see that the twist operators we insert in \eqref{tBY1}-\eqref{tBY3} will have to be matched by similar twists in the the $QQ$-relations. E.g. our prototype \eqref{QBY} should be changed to 
\begin{equation}
Q_{\emptyset}Q_{II}=e^{i\dot\vartheta }Q_{III}^+ \hat Q _{III}^--e^{-i\dot\vartheta }Q_{III}^-\hat Q _{III}^+\,.
\end{equation}
Alternatively, we can absorb the twists into the $Q$-functions. If we define 
\begin{equation}
Q_{III}'= e^{\dot\vartheta u}Q_{III}\,, \qquad{\hat{Q}}'_{III}=e^{-\dot\vartheta  u}\hat Q_{III}\,,
\end{equation} 
we find that they obey the standard untwisted $QQ$-relations. As ${Q}'_{III}$ and $\hat{Q}'_{III}$ are related to $\mathbf{P}_1$ and $\mathbf{P}_2$ \eqref{protoQQ}, we immediately recognise the twisted asymptotics we found in the sigma-model approach, which is demonstrating consistency of our formalism. We will henceforth only use twisted $Q$-functions and untwisted $QQ$-relations, therefore dropping the prime.\\

Now the only subtlety we have to take care of are the potential (quantum) shifts in charges $(\hat M_i,M_a)$. 
These arise as follows. Whenever two twists factors coincide, the leading asymptotic cancels in the $QQ$-relations. Let us demonstrate this in our prototype equation. If we plug in the assumed asymptotics we find
\begin{equation}\label{protocancel}
Q_{\emptyset}Q_{II}\sim e^{i\dot\vartheta } u^{ M_1}u^{ M_2}- e^{-i\dot\vartheta }u^{ M_1}u^{ M_2}+\order{u^{ M_1+M_2-1}}\,.
\end{equation}
One realises that for $\dot\vartheta=0$ or $\pi$ the leading asymptotics cancel. This reduces the effective charge of the $Q_{II}$-function and all further $Q$-functions we build from this.\footnote{A further subtlety arises if in addition $M_1+M_2=0$. Then we find that we have to set either $Q_\emptyset$ or $Q_{II}$ to $0$. This is related to the appearance of short multiplets and has been discussed in section 3.3.4 of \cite{Kazakov:2015efa}. We will avoid this case for now.} To have a consistent net of $QQ$-relations we therefore need to correct the asymptotic charges of all related $Q$-functions. Since this only happens when two twist factors coincide, the easiest scenario to study is the one with all twist angles different and unequal to $0$ and $\pi$. In that case no cancellations happen and all charges can be taken at face value. On the other extreme, for the completely untwisted case, there is a group-theoretic discussion leading to the right corrections \cite{Gromov:2014caa}. For a general twisted case one has to introduce some hierarchy of $Q$-functions to determine the appropriate shifts \cite{Kazakov:2015efa}, as we will explain below. \\

\begin{figure}
	\centering
	\begin{minipage}{0.49\textwidth}
		\includegraphics[width=0.95\textwidth]{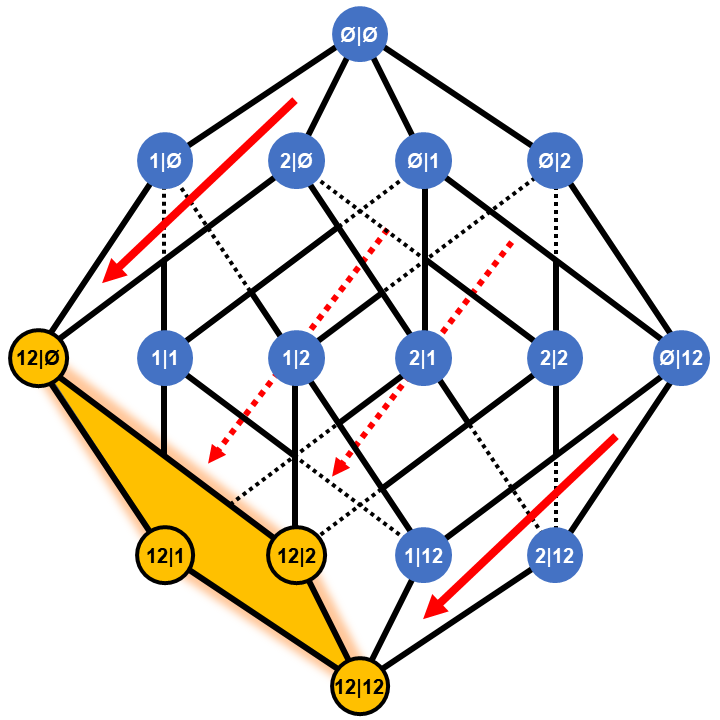}
		\caption*{\textbf{a}}
	\end{minipage}
	\begin{minipage}{0.49\textwidth}
		\includegraphics[width=0.95\textwidth]{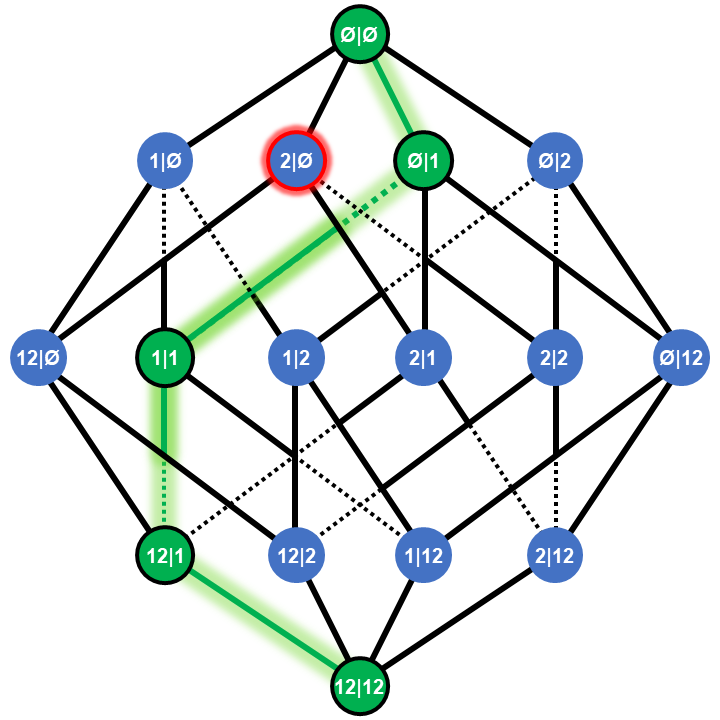}
		\caption*{\textbf{b}}
	\end{minipage}
\caption{\small{\textbf{a)} When two twist angles coincide (here $x_1$ and $x_2$) a sub-cube of codimension 2 picks up a relative shift in asymptotic charge. We have also drawn arrows along the $QQ$-relations that lead to cancellations. \textbf{b)} Finally, to link to the actual physical charges, a distinguished spin-chain has to be traced out, along which no overall shifts should appear. This means all other $Q$ functions have to be appropriately chosen to conspire to this end. In the case considered in b, we would need to raise the charge of $Q_{2|\emptyset}$, which we marked in red, by 1.}}\label{figure4}
\end{figure}
Systematically, we can think of the $QQ$-relations spanning the 2D faces of an 8D Hypercube, while the $Q$-functions live at the vertices. Pictorially, this is just the Hasse diagram of the set of indices (see fig. \ref{figure3}). Every time two twist factors $x_a$, $x_b$ coincide, the bosonic $QQ$-relations \eqref{QQb1} involving $a$ and $b$ and arbitrary spectator indices exhibit a cancellation. The $Q$-functions of the form $Q_{Aab|I}$ form a 6D sub-cube of the original Hasse diagram and their asymptotic charge is reduced by $1$. A similar effect arises when two $y_i$, $y_j$ coincide. Opposed to this, coinciding $x_a$ and $y_b$ factors yield a cancellation in the fermionic $QQ$-relations \eqref{QQf}, and due to their inverted structure this raises the charge of $Q_{Aa,Ii}$ by 1. 

We see that in the general case a number of integer shifts of charges occur in various 6D sub-cubes (see fig. \ref{figure4}a). The shifts add up linearly in the intersections of these sub-cubes and since the ``lowest-weight" state $Q_{1234|1234}$ sits in the intersection of all shifted sub-cubes it picks up the total of all shifts
\begin{equation}
Q_{1234|1234}\sim u^{\left(-\sum_{a} M_a+\sum_{i} \hat M_i-\sum_{a<b}\delta_{x_a,x_b}-\sum_{i<j}\delta_{y_i,y_j}+\sum_{a,i}\delta_{x_a,y_i}\right)}\,.
\end{equation}
However, we require $Q_{1234|1234}=1$, so we have to impose the condition 
\begin{equation}
-\sum_{a} M_a+\sum_{i} \hat M_i\underbrace{-\sum_{a<b}\delta_{x_a,x_b}-\sum_{i<j}\delta_{y_i,y_j}+\sum_{a,i}\delta_{x_a,y_i}}_{:=S_{total}}=0\,.
\end{equation}
We can also analyse the Hodge dual functions
\begin{equation}\label{dual}
\mathbf{P}^a\sim x_a^{-iu}u^{M^a}\,,\qquad  \mathbf{Q}^i\sim y_i^{iu}u^{-\hat M^i}\,,
\end{equation}
where we anticipated the asymptotic charges to be inverse to the original functions. However the shifts change this behaviour, so we introduced the dual charges with raised indices. They satisfy the relationship
\begin{equation}
M^a=M_a\underbrace{+\sum_{b\neq a}\delta_{x_a,x_b}-\sum_{i}\delta_{x_a,y_i}}_{:=-S_a}\,,\qquad \hat M^i=\hat M_i\underbrace{-\sum_{j\neq i}\delta_{y_i,y_j}+\sum_{a}\delta_{x_a,y_i}}_{:=S_i}\,.
\end{equation}
This sets up the overall tally of shifts but the questions remains, how to distribute them. To this end, we need to specify, where in the Hasse diagram we want the classical charges to appear. 

In our discussion we build upon the BMN ground state \eqref{BMN} so we can set up the QSC such that when we run through the Hasse diagram according to \eqref{path} we precisely recover the ABA equations in the form specified in \cite{Beisert:2005fw}. This path maps out a particularly nice form of Bethe\,-Yang equations which has been studied a lot in the literature (see e.g. \cite{Beisert:2005he} for an application to orbifolds). This means we have to introduce a partial ordering \footnote{In general one has to be careful how to choose this ordering. Depending on the operator of interest, there may be ``shortening" effects where some leading asymptotics vanish. There is an elegant way to organise operators in terms of $\mathcal{N}=4$ multiplets at $g=0$ which then mix at finite $g$. When we break the supersymmetry, the $\mathcal{N}=4$ multiplets break down into smaller multiplets and to access the entire spectrum, one has to choose different gradings of $Q$-functions for different states. We refer to \cite{Volin:2010xz, Marboe:2017dmb,Marboe:2019wyc} for a detailed discussion. Our grading is also referred to as ``0224" in that context.}
\begin{equation}
\hat 1\prec 1\prec 2\prec \hat 2\prec \hat 3 \prec 3\prec 4\prec \hat 4\,, 
\end{equation}
and twist all $\mathbf{P}_a$ and $\mathbf{Q}_i$ such that they conspire in vanishing shifts along the main path of $Q$-functions (see fig. \ref{figure4}b). A little algebra results in the correct asymptotics
\begin{equation}
\mathbf{P}_a\sim x_a^{iu}u^{-\mathcal{M}_a-S_{\prec a}}\,,\qquad \mathbf{Q}_i\sim y_i^{-iu}u^{\hat{\mathcal{M}}_i-\hat S_{\prec i}}\,,
\end{equation}
 where we introduced the notation $(\mathcal{M}_a,\hat{\mathcal{M}}_i)$ for the classical charges given in \eqref{tcharges} and 
 \begin{equation}
 S_{\prec a}:=-\sum_{b\prec a}\delta_{x_a,x_b}+\sum_{i\prec a}\delta_{x_a,y_i}\,,\qquad \hat S_{\prec i}:=-\sum_{i\prec j}\delta_{y_i,y_j}+\sum_{a\prec i}\delta_{x_a,y_i}\,,
 \end{equation}
 which denotes the shifts ``up to the position in the hierarchy". The asymptotics of the Hodge dual functions can now be read off from \eqref{dual} as
 \begin{equation}
 \mathbf{P}^a\sim x_a^{-iu}u^{\mathcal{M}_a-S_{\succ a}}\,,\qquad \mathbf{Q}^i\sim y_i^{iu}u^{-\hat{\mathcal{M}}_i-\hat S_{\succ i}}\,,
\end{equation}
with the obvious extension of the definition\footnote{To avoid any confusion, here are the relations between the various shift numbers we introduced;
\begin{equation*}
S_{total}=\frac{1}{2}\left(\sum_a S_a+\sum_i \hat S_i\right),\qquad S_a=S_{\prec a}+S_{\succ a}\,,\qquad \hat S_{i}=\hat S_{\prec i}+\hat S_{\succ i}\,.
\end{equation*}
In the untwisted case we have $S_{total}=4$ and $S_a=\hat S_i=1$ for any $a$ and $i$. In the type 0B case (see section \ref{0B}) we have $S_{total}=-12$ and $S_a=\hat S_i=-3$ for any $a$ and $i$. These are actually the extremal values of $S_{total}$.}
 \begin{equation}
S_{\succ a}:=S_a-S_{\prec a}=-\sum_{b\succ a}\delta_{x_a,x_b}+\sum_{i\succ a}\delta_{x_a,y_i}\,,\qquad \hat S_{\succ i}:=\hat S_i-\hat S_{\prec i}=-\sum_{i\succ j}\delta_{y_i,y_j}+\sum_{a\succ i}\delta_{x_a,y_i}\,.
\end{equation}

In the completely untwisted case this reproduces the shifts given in \cite{Gromov:2014caa,Gromov:2017blm}
\begin{equation}
M_a=\left\{\mathcal{M}_1+1,~\mathcal{M}_2,~\mathcal{M}_3+1,~\mathcal{M}_4\right\}\,,\qquad M_i=\left\{\hat{\mathcal{M}}_1,~\hat{\mathcal{M}}_2-1,~\hat{\mathcal{M}}_3,~\hat{\mathcal{M}}_4-1\right\}.
\end{equation}
Of course, this is only one path to go through the Hasse diagram and in principle all other systems of Bethe\,-Yang equations are equivalent. However, transitioning between different paths we need to keep track of the new shifts that then modify the charges and have to be subtracted in the end.\\

We have seen that orbifolds of $AdS_5\cross S^5$ can easily be implemented in the framework of the quantum spectral curve. The only influence the twisting has is (as expected) captured in the asymptotics of the $Q$ functions. An exponential factor $e^{\phi u}$ appears in the asymptotics and whenever two twist factors coincide the charges are shifted. This makes the fully twisted case the easiest to deal with, all charges being directly related to the classical ones via \eqref{relation}. 

In our previous parametrisation there are a few different cases to discuss. First of all the twist angles $0$ and $\pi$ (modulo $2\pi$) are special as they lead to two coinciding twist factors and an overall $-1$ shift. For example, we get $x_1=x_2$ for $\dot{\vartheta}=0$ and a cancellation in the $QQ$-relations for $Q_{A|I}Q_{A12|I}$ leading to a $-1$ shift of all $Q$-functions of the form $Q_{A12|I}$. The same amount of degeneracy occurs when $\alpha=\dot{\alpha}$ or $\vartheta=\dot{\vartheta}$. On the other hand any cross-coincidence like $\alpha=\vartheta$ or any dotted combination thereof leads to a fermionic degeneracy of two pairs of twist factors, so an overall shift of $S_{total}=+2$. 

Apart from these simplest cases of two coincident angles, there are of course cases of multiple angles coinciding, the completely untwisted case being the most degenerate case of total shift $S_{total}=4$. This is the highest value $S_{total}$ can take, although there are other twisted cases with this number. In the next section we will analyse the ``opposite" corner of special symmetry: The case of $\vartheta=\dot{\vartheta}=\pi$ and $\alpha=\dot{\alpha}=0$,
which has the most negative total shift of $S_{total}=-12$. This exactly corresponds to the twisted sector of type 0B string theory on $AdS_5\cross S^5$.

\section{$\mathbf{\mathbb{Z}_2}$-examples}\label{examples}

So far we have only recapitulated the available techniques for a general orbifold of $AdS_5\cross S^5$. In this section, we will apply them to the simplest examples we can consider: $\mathbb{Z}_2$-orbifolds of the $S^5$. The natural choices for twist angles are $\vartheta=\pi$ (or equivalently $\dot\vartheta=\pi$) as well as the combination of both angles $\vartheta=\dot{\vartheta}=\pi$. In the first case we retain half of the supersymmetry of the untwisted $\mathcal{N}=4$ SYM, while in the second case we break all of the supersymmetry. 

\subsection{$\mathcal{N}=2$ supersymmetric orbifold}\label{N=2}

If we only break the $R$-symmetry of one of the $\mathfrak{psu}(2|2)_{c.e.}$ factors, the theory retains half the supersymmetry ($\mathcal{N}=2$) \cite{Kachru:1998ys}. This happens for example in the case where only one twist angle gets turned on, either $\vartheta$ or $\dot{\vartheta}$.\footnote{Other $\mathcal{N}=2$ supersymmetric orbifolds can be constructed by twisting the $AdS_5$ angles $\alpha$ or $\dot{\alpha}$.} We want to discuss the $\mathbb{Z}_2$-orbifold which arises from the $\mathbb{Z}_2$-transformation that takes two orthogonal planes in the embedding space $\mathbb{R}^6$ of the $S^5$ and rotates them by $\pi$. This orbifold of $AdS_5\cross S^5$ was first constructed as near-horizon limit of a stack of D3-branes located at the orbifold locus of $\mathbb{C}^2/\mathbb{Z}_2$ \cite{Douglas:1996sw,Kachru:1998ys,Lawrence:1998ja}. 

String theory on this background splits into two sectors. The untwisted sector consists of the $\mathbb{Z}_2$-invariant states of the original theory. The twisted sector arises from strings that only close up to a $\mathbb{Z}_2$-transformation. 

On the gauge theory side, one is to start with $\mathcal{N}=4$ SYM with gauge group $SU(2N)$. The orbifold action splits the gauge group into $SU(N)\cross SU(N)$ and the $\mathcal{N}=4$ multiplet decomposes into two $\mathcal{N}=2$ vector multiplets and two hypermultiplets, which are in the bifundamental representation of the two gauge groups. The $\mathbb{Z}_2$-symmetry acts by exchanging the two $SU(N)$ factors. We can therefore relate the untwisted and twisted sectors of the string theory to single-trace operators that are either symmetric or antisymmetric under the exchange of the gauge groups. 

\

The spectrum of untwisted states can be deduced by projecting the spectrum of the untwisted $\mathcal{N}=4$ theory to its $\mathbb{Z}_2$-invariant subset. The twisted sector on the other hand has to be treated more carefully, so we employ the previously described integrability toolbox. We trace the effect of the $\vartheta=\pi$ twist throughout all stages of our previous journey:
\begin{itemize}
	\item In the left-wing Bethe\,-Yang equations \eqref{tBY2} we need to insert a $(-1)$-factor for the $y$-particles. 
	\item In the mirror theory the $\mathbb{Z}_2$-twist corresponds to a defect operator on the timelike circle. Therefore we can just introduce an operator $\exp(i\pi N_{y^{(l)}_{\pm}})$ in the partition function. This is precisely the kind of operator introduced in \eqref{lusch}.
	\item In the free energy \eqref{free} we now need a chemical potential $\mu_\pm^{(l)}=i\pi$ for the left-wing y-particles. This results in a constant term $-i\pi$ on the right-hand side of their canonical TBA equation \eqref{TBAy}.
	\item The chemical potential cancels in the simplified TBA equations, so simplified TBA, $Y$- and $T$-system remain unchanged.
	\item The asymptotics of the $Y$-functions depend on q-numbers \eqref{q-number}, which for our values of $q$ become $[n]_{e^{i\alpha}}=[n]_{e^{i\dot\alpha}}=[n]_{e^{i\dot\vartheta}}=[n]_1=n$ and $[n]_{e^{i\vartheta}}=[n]_{-1}=(-1)^{n+1}n$. The asymptotic solution for $Y_\pm^{0,(l)}$ becomes $-1$. The fundamental $Y_Q$ function \eqref{YQ} vanishes because
	\begin{equation}
	Y_Q^0\sim([2]_{e^{i\dot\alpha}}-[2]_{e^{i\dot\vartheta}})=0\,.
	\end{equation}
	\item In the $T$-system, we can similarly rewrite asymptotics in simpler terms, using that $\xi=-4$. For example, we find $T_{1,-M}^0=T_{M,-1}^0=(-1)^M 4M$. 
	\item The (large-$u$) asymptotic $\mathbf{P}_a$ functions in the QSC receive factors $\left\{1,1,e^{\pi u},e^{-\pi u} \right\}$.
	\item The asymptotic charges get shifted by $S_{\prec a}=\{1,0,0,-1\}$ and $\hat S_{\prec i}=\{0,1,0,-1\}$. The total shift $S_{total}$ vanishes. To summarise, we present the correct asymptotics in the following tables \ref{tab_asymp} and \ref{tab_tasymp}.
\end{itemize}
\begin{table}[h!]
	\centering
	\renewcommand{\arraystretch}{1.5}
	\begin{tabular}{c||c|c|c|c}
		$a/i$ & 1 & 2 & 3 & 4 \\ \hhline{=#=|=|=|=}
		$\mathbf{P}_a$ & $ u^{-\frac{J_3+J_1-J_2}{2}-1}$&$u^{-\frac{J_3-J_1+J_2}{2}}$&$u^{-\frac{-J_3+J_1+J_2}{2}-1}$&$u^{-\frac{-J_3-J_1-J_2}{2}}$\\ \hline
		$\mathbf{P}^a$&$u^{\frac{J_3+J_1-J_2}{2}}$&$u^{\frac{J_3-J_1+J_2}{2}-1}$&$u^{\frac{-J_3+J_1+J_2}{2}}$&$u^{\frac{-J_3-J_1-J_2}{2}-1}$\\ \hline
		$\mathbf{Q}_i$&$u^{\frac{\Delta -S_1+S_2}{2}}$&$u^{\frac{\Delta +S_1-S_2}{2}-1}$&$u^{\frac{-\Delta -S_1-S_2}{2}}$&$u^{\frac{-\Delta +S_1+S_2}{2}-1}$\\ \hline
		$\mathbf{Q}^i$&$u^{-\frac{\Delta -S_1+S_2}{2}-1}$&$u^{-\frac{\Delta +S_1-S_2}{2}}$&$u^{-\frac{-\Delta -S_1-S_2}{2}-1}$&$u^{-\frac{-\Delta +S_1+S_2}{2}}$
	\end{tabular}
	\caption{Asymptotics of $Q$-functions in the untwisted sector of the $\mathcal{N}=2$ orbifold  $AdS_5\cross S^5/\mathbb{Z}_2$ \protect\phantom{Table 1: }(same as type IIB or the untwisted sector of type 0B string theory on $AdS_5\cross S^5$).}
	\label{tab_asymp}
\end{table}
\begin{table}[h!]
	\centering
	\renewcommand{\arraystretch}{1.5}
	\begin{tabular}{c||c|c|c|c}
		$a/i$ & 1 & 2 & 3 & 4 \\ \hhline{=#=|=|=|=}
		$\mathbf{P}_a$ & $u^{-\frac{J_3+J_1-J_2}{2}-1}$&$ u^{-\frac{J_3-J_1+J_2}{2}}$&$e^{\pi u}u^{-\frac{-J_3+J_1+J_2}{2}}$&$e^{-\pi u}u^{-\frac{-J_3-J_1-J_2}{2}+1}$\\ \hline
		$\mathbf{P}^a$&$u^{\frac{J_3+J_1-J_2}{2}-2}$&$u^{\frac{J_3-J_1+J_2}{2}-3}$&$e^{-\pi u}u^{\frac{-J_3+J_1+J_2}{2}+1}$&$e^{+\pi u}u^{\frac{-J_3-J_1-J_2}{2}}$\\ \hline
		$\mathbf{Q}_i$&$u^{\frac{\Delta -S_1+S_2}{2}}$&$u^{\frac{\Delta +S_1-S_2}{2}-1}$&$u^{\frac{-\Delta -S_1-S_2}{2}}$&$u^{\frac{-\Delta +S_1+S_2}{2}+1}$\\ \hline
		$\mathbf{Q}^i$&$u^{-\frac{\Delta -S_1+S_2}{2}+1}$&$u^{-\frac{\Delta +S_1-S_2}{2}+2}$&$u^{-\frac{-\Delta -S_1-S_2}{2}+1}$&$u^{-\frac{-\Delta +S_1+S_2}{2}}$
	\end{tabular}
	\caption{Asymptotics of $Q$-functions in the twisted sector of the $\mathcal{N}=2$ orbifold  $AdS_5\cross S^5/\mathbb{Z}_2$.}
	\label{tab_tasymp}
\end{table}
Since supersymmetry is preserved, we retain a number of BPS operators, whose quantum numbers are protected. Indeed the single-trace operators 
\begin{equation}
\mathcal{O}_{\pm J}=\Tr\,(Z^J)\pm\Tr\,(\tilde Z^J)\,,
\end{equation}
which correspond to the BMN ground states in both untwisted and twisted sector are such protected operators. The TBA gives their energy \eqref{goal} and since $Y_Q=0$, we find that the ground state energy vanishes
\begin{equation}
E_0=0\,.
\end{equation}

This is precisely what we expected. The calculation of other conformal dimensions is complicated by the fact that the orbifolding pattern is left-right asymmetric. Therefore one cannot access the simple set of operators governed by \eqref{LR}. This makes the calculation of non-trivial operator dimensions like the twisted sector Konishi-operator \eqref{twistedKonishi} much harder than in the symmetric case \cite{deLeeuw:2011rw}.

 Nevertheless, the QSC is capable of describing the Konishi-operator and since this operator has been studied extensively in the $\mathcal{N}=4$ theory \cite{Bajnok:2008bm,Frolov:2010wt,Roiban:2011fe} and some deformed theories \cite{deLeeuw:2011rw,Marboe:2019wyc}, the outcome can be compared to known results. Furthermore, one could investigate potential relationships to semi-classical string states along the lines of \cite{Beccaria:2012xm}. Therefore, the study of QSC for the $\mathcal{N}=2$ orbifolded Konishi-state is an interesting question for future research. 

\subsection{Type 0B string on $AdS_5 \cross S^5$}\label{0B}

Let us move on to our second example: Type 0B string theory on $AdS_5\cross S^5$. In the RNS formalism for superstring theory on flat space, type 0A and type 0B string theory are results of a modified GSO projection. They are constructed by a ``diagonal" choice of sectors, which guarantees modular invariance and results in a purely bosonic spectrum. The lowest-energy mode is a tachyon of mass
\begin{equation}\label{tachyon}
m_T^2=-\frac{2}{\alpha'}\,.
\end{equation}
At the massless level we find the same bosonic modes we would find in type IIA or IIB, respectively, and an additional copy of the RR-fields. These type 0 theories are in principle valid string theories apart from the appearance of the tachyon, which suggests that flat space is not a stable background for type 0 string theory. 

The type 0 string theory on flat space can also be constructed in the Green-Schwarz formalism as a $\mathbb{Z}_2$-orbifold of type II string theory \cite{Dixon:1986iz}. The symmetry we want to quotient by is the $\mathbb{Z}_2$-subgroup of the $\mathbb{Z}_4$-symmetry described in \eqref{Z4} that switches the sign of all spacetime fermions. An explicit spacetime implementation of this symmetry transformation is the rotation of a plane in the target space by $2\pi$, which only affects the fermions. 

Quotienting by this $\mathbb{Z}_2$ symmetry results in two sectors. In the untwisted sector we proceed just as in type II theory but project out all spacetime fermions. In the twisted sector we impose anti-periodic boundary conditions on the worldsheet fermionic fields. This results in the appearance of the tachyonic ground state and the other copy of RR-fields on the first (massless) level. Thus we reproduce the spectrum found in the RNS formalism.

We can also study type 0 string theory on non-trivial curved and/or fluxed Type II backgrounds, if we find a way to perform a similar orbifolding. In \cite{Takayanagi:2001jj, Skrzypek:2021eue} an explicit construction using a continuous deformation called Melvin-twist was investigated and applied to pp-wave backgrounds. For $AdS_5\cross S^5$ we can use the large amount of symmetry and perform the necessary orbifolding using the techniques we presented above.

\

Starting from type IIB string theory on $AdS_5\cross S^5$ we choose the twist angle $\phi_{J_2}=2\pi$, which should result in type 0B string theory on the same background. In the parametrisation defined above, the twisted sector is characterised as $(\alpha,\dot{\alpha},\vartheta,\dot{\vartheta})=(0,0,\pi,\pi)$. Again we perform a straight-forward analysis of the integrability framework. Compared to the $\mathcal{N}=2$ supersymmetric orbifold case we need to adjust not only the left wing of \eqref{tBY2}, but both wings simultaneously: 

\begin{itemize}
	\item In the Bethe\,-Yang equation \eqref{tBY2} we need to insert a $(-1)$-factor for $y$-particles of both wings. This has essentially the same effect as raising the winding number $m$ by one. We see that only fermions are affected, which is precisely what we want.
	\item We introduce the twist operator $e^{i\pi N_f}$ in the partition function of the mirror theory. This is precisely the kind of operator introduced in \eqref{lusch}.
	\item in the free energy \eqref{free} we now need chemical potentials $\mu_\pm^{(l)}=\mu_\pm^{(r)}=i\pi$ for all y-particles. This results in a constant term $-i\pi$ on the right hand side of the canonical TBA equation \eqref{TBAy}.
	\item The relevant q-numbers \eqref{q-number} for the $Y$-function asymptotics are $[n]_{e^{i\alpha}}=[n]_{e^{i\dot\alpha}}=[n]_1=n$ and $[n]_{e^{i\vartheta}}=[n]_{e^{i\dot\vartheta}}=[n]_{-1}=(-1)^{n+1}n$. We find $Y_\pm^{0,(a)}=-1$. The fundamental $Y$-function for $Q$-strings becomes
	\begin{equation}
	Y_Q=(4Q)^2e^{-\tilde\epsilon_QL}\,.
	\end{equation}
	This is non-vanishing because of completely broken supersymmetry and agrees completely with the L\"uscher prediction since $\Str_Q(e^{i\pi N_f})=(4Q)^2$.
	\item In the $T$-system, we can again rewrite asymptotics in simpler terms, using that $\xi=\dot{\xi}=-4$. Now also the right wing gets adjusted as e.g. $T_{1,M}^0=T_{M,1}^0=(-1)^M 4M\dot{\epsilon}^Q$. 
	\item The (large-$u$) asymptotic $\mathbf{P}_a$ functions in the QSC receive factors $\left\{e^{\pi u},e^{-\pi u},e^{\pi u},e^{-\pi u} \right\}$.
	\item The asymptotic charges get shifted by $S_{\prec a}=\{0,-1,-2,-3\}$ and $\hat S_{\prec i}=\{0,-1,-2,-3\}$. To summarise, we present the twisted asymptotics in the following table \ref{tab_tasymp0} (the untwisted sector is the same as in the previous example).
\end{itemize}
\begin{table}[h!]
	\centering
	\renewcommand{\arraystretch}{1.5}
	\begin{tabular}{c||c|c|c|c}
		$a/i$ & 1 & 2 & 3 & 4 \\ \hhline{=#=|=|=|=}
		$\mathbf{P}_a$ & $ e^{\pi u}u^{-\frac{J_3+J_1-J_2}{2}}$&$e^{-\pi u}u^{-\frac{J_3-J_1+J_2}{2}+1}$&$e^{\pi u}u^{-\frac{-J_3+J_1+J_2}{2}+2}$&$e^{-\pi u}u^{-\frac{-J_3-J_1-J_2}{2}+3}$\\ \hline
		$\mathbf{P}^a$&$e^{-\pi u}u^{\frac{J_3+J_1-J_2}{2}+3}$&$e^{\pi u}u^{\frac{J_3-J_1+J_2}{2}+2}$&$e^{-\pi u}u^{\frac{-J_3+J_1+J_2}{2}+1}$&$e^{+\pi u}u^{\frac{-J_3-J_1-J_2}{2}}$\\ \hline
		$\mathbf{Q}_i$&$u^{\frac{\Delta -S_1+S_2}{2}}$&$u^{\frac{\Delta +S_1-S_2}{2}+1}$&$u^{\frac{-\Delta -S_1-S_2}{2}+2}$&$u^{\frac{-\Delta +S_1+S_2}{2}+3}$\\ \hline
		$\mathbf{Q}^i$&$u^{-\frac{\Delta -S_1+S_2}{2}+3}$&$u^{-\frac{\Delta +S_1-S_2}{2}+2}$&$u^{-\frac{-\Delta -S_1-S_2}{2}+1}$&$u^{-\frac{-\Delta +S_1+S_2}{2}}$
	\end{tabular}
	\caption{Asymptotics of $Q$-functions in the twisted sector of type 0B theory on $AdS_5\cross S^5$.}
\label{tab_tasymp0}
\end{table}

Now that we have modified the integrability machinery for type 0B, we can immediately extract some physical results. 

We can start with the ground state \eqref{BMN} at finite coupling. This is the ground state that is accessible to the TBA without introducing further driving terms. The asymptotic Bethe Ansatz requires an infinite angular momentum $J$ such that winding corrections are suppressed but the TBA can capture these and gives results at finite $J$. At weak coupling we can follow \cite{vanTongeren:2013gva} and immediately compute the leading-order (LO) winding correction. Remembering the mirror dispersion relation \eqref{disp} for bound states of $Q$ particles, we see that 
\begin{equation}
e^{-\tilde \epsilon_Q(\tilde{p})J}=\frac{(2g)^{2J}}{( \tilde p^2+Q^2)^J}+\order{g^{2L+2}}\,.
\end{equation} 
Using the residue theorem we can compute integrals of the form
\begin{equation}
\int\frac{\dd\tilde p}{2\pi}\frac{1}{(\tilde p^2+Q^2)^k}=\frac{\Gamma({2k-1})}{\Gamma({k})^2}(2Q)^{1-2k}
\end{equation}
and thus we get
\begin{equation}\label{series}
E_0^{LO}(J)=-\frac{\Gamma({2J-1})}{\Gamma({J})^2}(2g)^{2J}\sum_{Q=1}^{\infty}(4Q)^2  (2Q)^{1-2J}+\order{g^{2J+2}}\,.
\end{equation}
The sum yields the $\zeta$-function and we arrive at
\begin{equation}
E_0^{LO}(J)=-\frac{\Gamma({2J-1})}{\Gamma({J})^2}g^{2J}2^{5}\zeta(2J-3)+\order{g^{2J+2}}\,,
\end{equation}
which by using $\Gamma$-function identities can be put in the form 
\begin{equation}\label{energy}
E_0^{LO}(J)=-\frac{8\Gamma\left(J-\frac{1}{2}\right)\zeta(2J-3)}{\sqrt{\pi}\Gamma(J)}(2g)^{2J}+\order{g^{2J+2}}\,.
\end{equation}
The next-to-leading-order (NLO) corrections come in at $g^{4J}$.

\begin{center}
	\includegraphics[width=10cm]{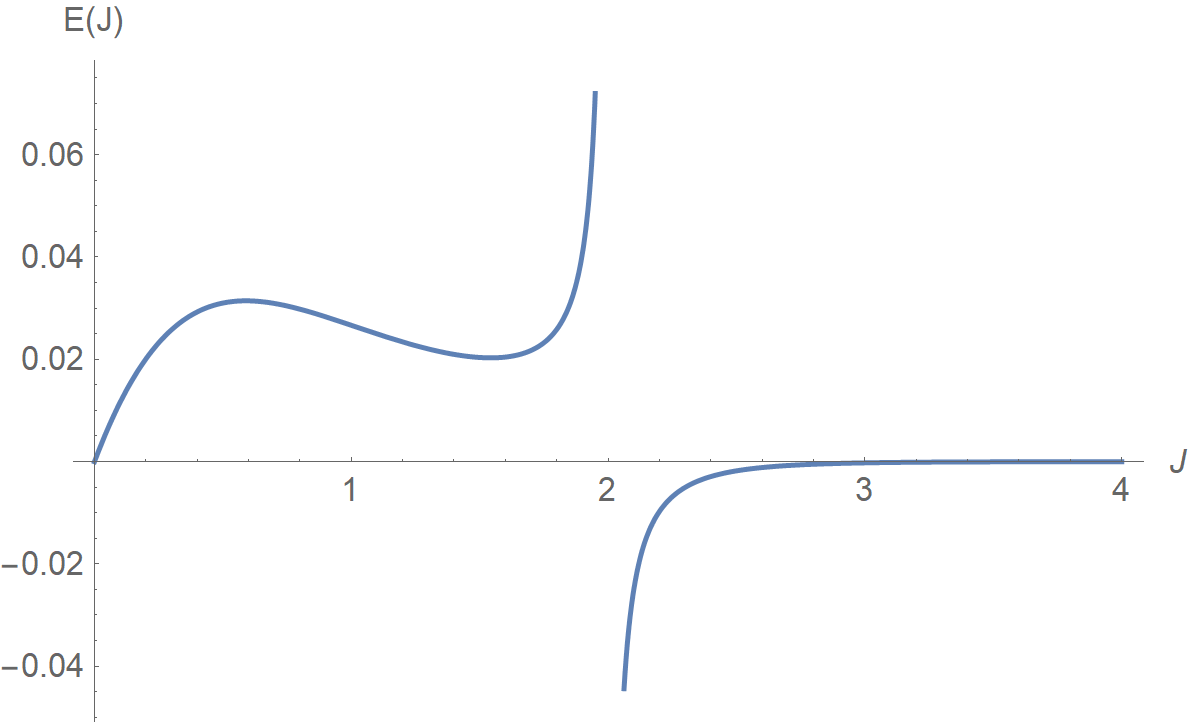}
	\captionof{figure}{\small {\small Leading order energy correction $E_0^{LO}(J)$ at weak coupling $g=0.1$ . Note the pole at $J=2$.}}
	\label{pic1}
\end{center}

 \eqref{energy} has a pole at $J=2$ which already appears in discussions of the ground state of the untwisted model \cite{Frolov:2009in}. There, however, one can introduce an infinitesimal twist to regularise the pole. In the twisted model, the pole at $J=2$ seems to be yet unresolved in the TBA framework. An interesting open question is what the QSC would predict for this state.\footnote{The validity of the TBA result for $J=1$ may be questioned here, too. \eqref{energy} yields a finite result, which is the zeta-regularised value of the divergent series in \eqref{series}. However, as the pole at $J=2$ already signalises a break-down that needs to be treated, this value ``behind" the pole might change, too.} Comparing to \cite{Levkovich-Maslyuk:2020rlp}, where a similar calculation has been performed for the $\gamma$-deformation we expect an imaginary but finite correction at weak coupling. Extrapolating from the result found in (4.14) of \cite{Levkovich-Maslyuk:2020rlp}, we conjecture a weak coupling expansion of the form
 \begin{equation}\label{conjecture}
 \Delta_\pm=2\pm i\left[8g^2-64(3 \zeta_3+1)g^6 +\order{g^{8}}\right]\,,
 \end{equation}
 
which would signal an instability in the spectrum that a priori is not related to the type 0 string theory tachyon in the flat space limit.\footnote{The two choices of sign correspond to the two possible conformal fixed points, as we will discuss below.}

\begin{center}
	\includegraphics[width=10cm]{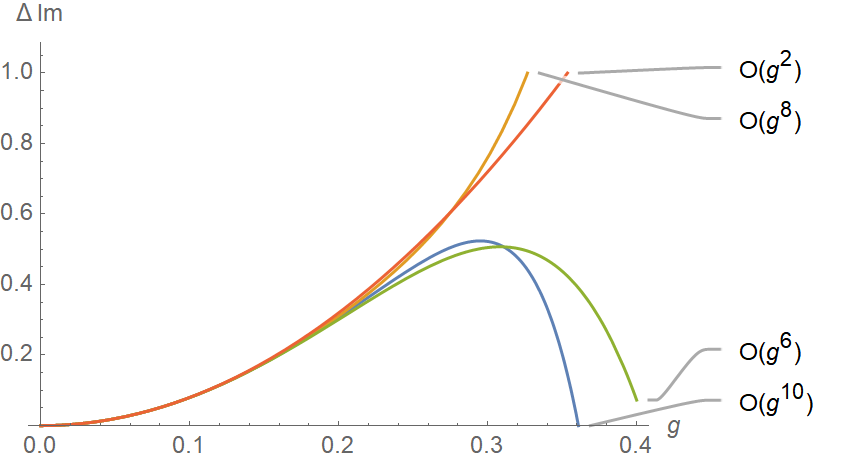}
	\captionof{figure}{\small {\small Perturbative QSC prediction for the imaginary part of $\Delta_+$ for $J=2$, extrapolating from the result in \cite{Levkovich-Maslyuk:2020rlp}.}}
	\label{pic2}
\end{center}

The $\gamma$-deformation discussed in \cite{Levkovich-Maslyuk:2020rlp} is a three-parameter deformation of $\mathcal{N}=4$ SYM, which introduces non-commutativity into the $\mathcal{N}=4$ SYM Lagrangian, governed by the parameters $(\gamma_1,\gamma_2, \gamma_3)$. On the gravity-dual side, this deformation is connected to the maximally symmetric $AdS_5\cross S^5$ background via three TsT-transformations with respect to the three Cartan elements $J_1$, $J_2$ and $J_3$. In terms of the integrability machinery, this results in similar twists to the orbifolds we discussed, but these are now dependent on the quantum numbers of the state in question. However, the structure of the QSC calculations is quite similar when we restrict our attention to a specific state and indeed the state discussed in \cite{Levkovich-Maslyuk:2020rlp} results in twists similar to the orbifold groundstate we want to discuss here. Specifically, we can look at a ``partially $\gamma$-deformed" state as discussed around equation (4.16) in \cite{Levkovich-Maslyuk:2020rlp}. In terms of the conventions there, the correct extrapolation is reached when we set 
\begin{equation}
\gamma_1\to 0\,,\quad \gamma_2\to 0\,,\quad\gamma_3 \to \pi\,,\quad S_\pm \to 1\,, \quad \kappa\to i\,,\quad \hat\kappa\to i\,.
\end{equation}

Given these values we can read off \eqref{conjecture} from (4.16) in \cite{Levkovich-Maslyuk:2020rlp}. We note, however, that this extrapolation is only partially justified, as the parametrisation used in \cite{Levkovich-Maslyuk:2020rlp} becomes singular at these values, and the asymptotic charges we discussed in table \ref{tab_tasymp0} are different. We were able to use the rescaling invariance of the $Q$-functions to absorb these singularities, and confirmed that the leading order solutions $\mathbf{Q}_i^{(0)}$ agree with the ones found in \cite{Levkovich-Maslyuk:2020rlp}. A brief outline of this perturbative calculation is presented in appendix \ref{appendix3}. To further back up our conjecture, more extensive perturbative and numerical studies should be performed, which turn out to be rather time-consuming and are left to future work.
 
The ground state energy of the twisted sector gives the conformal dimension of the corresponding single-trace operator of R-charge $J$ in the field theory via \eqref{conformal}
\begin{equation}
E_0=\Delta-J\,.
\end{equation}
The holographic dual is a twisted-sector point-like string moving around the $S^5$ with angular momentum $J$. We can use the holographic dictionary to relate the conformal dimension to the mass of this string state. A massive scalar field in $AdS_5$ obeys the relation
\begin{equation}
\Delta(\Delta-4)=m^2R^2\,,
\end{equation}
We can compute this mass naively, but since we are working at small $g$, we expect the dual string theory to be strongly coupled. This means there is no valid perturbative description of string states but instead a highly non-trivial world-sheet theory would have to be discussed. If we remain oblivious to this difficulty for the moment and rely on supergravity language, we find real valued conformal dimensions for all $J\neq2$. These states satisfy the Breitenloher-Freedman bound ($m^2R^2>-4$) and are physically unproblematic. If, however, \eqref{conjecture} indeed holds for $J=2$, this state would be of tachyonic nature.

\section{Discussion and outlook}\label{Discussion}

In this paper we discussed the integrability treatment of orbifolds of $\mathcal{N}=4$ SYM and the dual string theories. We explained how to describe the twisted sector in asymptotic Bethe Ansatz (ABA) and thermodynamic Bethe Ansatz (TBA). Furthermore, we specialised the twisted quantum spectral curve (QSC) of \cite{Kazakov:2015efa} to orbifolds.

With this set-up, it is possible to expand our knowledge about the spectra of orbifold theories, which provide a rich testing ground for the AdS/CFT conjecture away from the maximally symmetric case. We presented a simple example for this with the $\mathcal{N}=2$ orbifold and suggested the twisted-sector Konishi-operator \eqref{twistedKonishi} as interesting object to study. Despite not being left-right symmetric \eqref{LR}, this operator can be studied using the QSC formalism we described. Perturbative methods and numerical methods for this problem have been developed, for example in \cite{Gromov:2009zb, Gromov:2015vua,Gromov:2017blm}. 

Other more complicated orbifolds, such as $\mathbb{Z}_{n>2}$ orbifolds, orbifolds along the $AdS_5$ angles or orbifolds by non-abelian groups can be implemented analogously. They generically introduce multiple twisted sectors, characterised by the various group elements. Nonetheless, the spectrum of these sectors can be accessed by the presented techniques, only requiring more book-keeping. In fact, the TsT-transformations that lead to the $\beta$- and $\gamma$-deformed models fit into this framework, too. One only needs to make the twist angles state-dependent, e.g. $\vartheta\to \vartheta(J_3,J_1,J_2)$. 

Computing the spectra of orbifolds of $\mathcal{N}=4$ SYM opens up an entire landscape of theories we can apply the AdS/CFT dictionary to and test the conjecture against. Most interestingly, non-supersymmetric cases can be studied. SUSY-breaking is a pressing issue in general and more specifically for a dual string description of QCD. A major hurdle for many instances of non-supersymmetric string theories is the appearance of tachyons, which has partially motivated this work. Let us look at the type 0B case we studied above and outline what we know and would like to understand about the tachyons appearing there. 

\

As mentioned previously, type 0B string theory on flat space has a tachyon of mass-squared
\begin{equation}
m_T^2=-\frac{2}{\alpha'}\,,\tag{\ref{tachyon}}
\end{equation}
which is the lowest energy mode of the twisted sector in the GS string construction \cite{Dixon:1986iz,Seiberg:1986by}. 
In \cite{Klebanov:1999ch} it was suggested that on the $AdS_5\cross S^5$ background with 5-form flux $F_5$ the tachyon may get stabilised, with some further evidence provided in \cite{Skrzypek:2021eue}. There, we analysed the type 0B string on the pp-wave background that arises in the BMN-limit ($J$, $g\to \infty$ with $\mathcal{J}=\frac{J}{g}$ fixed) of $AdS_5\cross S^5$. The mass is corrected by a light-cone momentum $p^u$-dependant term
\begin{equation}
(m_{T,\text{pp}})^2=-\frac{2}{\alpha'}+8p^uf+\order{\alpha'}\,,
\end{equation}
where $f$ is a constant that determines the field strength of the RR 5-form flux $F_5$. For small $p^u$ this still results in a tachyon but for large $p^u$ we find that the tachyon gets stabilised. It is now an interesting question to ask what happens if one returns to the full $AdS_5\cross S^5$ geometry. 

When we describe a classical string with large angular momentum $J$ but with fixed $g$, we can expand the string action in terms of $\frac{1}{\mathcal{J}}$. The leading order term in this action is precisely the action we get in the pp-wave background. The sub-leading corrections become significant when $\mathcal{J}$ becomes small. Therefore, the BMN-limit is only a good approximation to finite size $AdS_5\cross S^5$ when $\mathcal{J}$ is large. 

Since $p^u$ roughly corresponds to $\mathcal{J}$ we find that the domain of instability for small $p^u$ overlaps with the parameter region where the BMN-limit breaks down. Therefore the instability we found in the pp-wave background is not necessarily an instability in the full $AdS_5\cross S^5$ background. Indeed, we expect the curvature of the $AdS_5$ space to lift the instability at least partially, given that slightly tachyonic fields are allowed by the Breitenloher-Freedman bound. However, it is still unclear whether a small number of tachyons remains at low $g$ or whether all states get stabilised at some $g_{crit.}$.  

To address this question, we studied the twisted sector operators $\mathcal{O}_{-J}$ of the form \eqref{groundstates}. For $J>2$ we computed the leading-order correction to the conformal dimension \eqref{energy}. It turned out as real and finite, so no instability is indicated. However, we found evidence for a break-down of the perturbative TBA description for the operator 
\begin{equation}\label{problem}
\mathcal{O}_{{-2}}=\Tr\big(Z Z\big)-\Tr\big(\tilde Z\tilde Z\big)\,.
\end{equation}
It is not assumed that this is the only operator that faces this issue, other states of the same quantum numbers might mix in, too. Nevertheless, resolving the break-down for the simplest operator $\mathcal{O}_{-2}$ might already provide us with a better understanding of instabilities in AdS/CFT.\footnote{Another interesting operator to study would be the one that couples to the tachyon field $T$ in the low-energy effective action of type 0B string theory.  In \cite{Klebanov:1999ch,Klebanov:1999um} this operator was identified as
	\begin{equation*}
	\mathcal{O}_T=\frac{1}{4}\Tr~{F^2_{\alpha\beta}}-\frac{1}{4}\Tr~{\tilde F^2_{\alpha\beta}}+\dots\,,
	\end{equation*}
	where $F_{\alpha\beta}$ and $\tilde F_{\alpha\beta}$ are the field strength tensors for the two $SU(N)$ gauge groups.	To study this operator using the QSC, one would have to use a different path in the Hasse diagram, as opposed to \eqref{path} \cite{Marboe:2017dmb,Marboe:2019wyc}. Note that in $\mathcal{N}=4$ SYM the two operators 
	\begin{equation*}
	\Tr\left(Z Z\right) \quad \text{and} \quad  \Tr\,F^2_{\alpha\beta}
	\end{equation*}
	are in the same supermultiplet, so naively one might hope for some kind or similar relationship between $\mathcal{O}_{-2}$ and $\mathcal{O}_{T}$. However, in this non-supersymmetric scenario there is no multiplet structure relating them immediately. It is therefore an interesting open question, whether the algebraic relationships that underpin the QSC make any connection between these operators possible.}

A similar breakdown of the TBA appears in the $\gamma$-deformation of $\mathcal{N}=4$ SYM \cite{Frolov:2005dj}. However, in this case the QSC has successfully produced the anomalous dimension of the twisted sector $J=2$ state \cite{Levkovich-Maslyuk:2020rlp} and gave a finite but imaginary answer. We therefore conjectured a similar behaviour in the type 0B case, as given in \eqref{conjecture}. A closer analysis of the details, continuing from the preliminary calculations presented in appendix \ref{appendix3}, remains to be carried out.

Interpreting this breakdown of the TBA or the conjectured imaginary anomalous dimension from the QSC for $\mathcal{O}_{-2}$ is difficult. As discussed above, a naive supergravity interpretation of $\mathcal{O}_{-2}$ would attribute the imaginary anomalous dimension to the appearance of a tachyon at weak coupling. In the gauge theory on the other hand, the operator $\mathcal{O}_{-2}$ has been linked to a breakdown of conformality, which the TBA breakdown might be related to.

In non-supersymmetric orbifold gauge theories the interactions are not protected by supersymmetry, so we have to renormalise the fields and couplings. Quantum corrections will generally turn on all possible terms in the action, but since we are only analysing the large-$N$ limit, most of them drop out. However, there are double-trace operators \cite{Tseytlin:1999ii,Klebanov:1999um,Adams:2001jb} which contribute at leading order in large-$N$ and therefore have to be added to the effective action, such as
\begin{equation}
f\int\dd x^4 ~\mathcal{O}_{{-2}}\bar{\mathcal{O}}_{{-2}}\,.
\end{equation}
The single trace operators inherit their vanishing $\beta$-function from the original theory but the double-trace coupling $f$ has a non-vanishing $\beta$-function. We may want to tune the double-trace coupling $f$ to a fixed point, but it turns out that for type 0B no real-valued fixed point exists and thus conformal symmetry is broken \cite{Dymarsky:2005nc,Dymarsky:2005uh, Pomoni:2009joh,Fokken:2013aea}. We can sacrifice unitarity and formally tune $f$ to a complex fixed point, but this causes the conformal dimension of the operator $\mathcal{O}_{{-2}}$ to pick up an imaginary anomalous dimension $\Delta_{\mathcal{O}_{{-2}}}=2\pm i \gamma$, $\gamma\in\mathbb{R}$ \cite{Pomoni:2009joh}. 

We see that the double-trace induced breakdown of conformality results in a similar imaginary anomalous dimension for $\mathcal{O}_{{-2}}$ as we expect from the QSC. This suggests that the QSC indeed takes into account all quantum corrections and naturally describes the theory at the complex fixed point in $f$. Such behaviour was found in the fishnet model which also has a non-unitary fixed point \cite{Gurdogan:2015csr, Gromov:2017cja}. Note that we can tune $f$ to two separate conformal fixed points related by complex conjugation. This results in two complex conjugate results for the anomalous dimensions in agreement with \eqref{conjecture}.

The parallels we just drew lead us to conjecture that the TBA breaks down precisely due to non-conformality of the gauge theory at weak coupling. Although we can formally remedy this issue by going to a complex fixed point, we then find its traces in the appearance of a tachyon at weak coupling, which may be confirmed by the QSC.

We now have discussed two separate notions of tachyons appearing at strong and weak coupling, respectively. They can be described in terms of the two sides of the AdS/CFT duality: the strong coupling tachyon is related to the standard type 0B string tachyon, while the weak coupling tachyon can be related in the gauge theory to the breakdown of conformality and a complex fixed point. The obvious question to ask is whether the tachyonic instabilities extend to all finite couplings or whether they get stabilised at some critical values $g_{crit.}$.\footnote{Transitions at intermediate values of $g$ are known to appear in the TBA for excited states. At critical values $g_{crit.}$, zeros of $Y$-functions merge \cite{Arutyunov:2009ax}, making it necessary to modify the TBA at every transition.} A priori there is no reason why these two tachyons should be related so we can think of four different scenarios
\begin{enumerate}
	\item The two tachyons are related and remain tachyonic for all values of $g$.
	\item The two tachyons are unrelated and get stabilised but overlap in an intermediate regime of $g$ where both modes are tachyonic.
	\item The two tachyons get stabilised with no overlap, thus resulting in a domain of stability. 
	\item Other modes we did not discuss here become tachyonic at intermediate couplings, resulting in a plethora of possible domains of overlap or stability.
\end{enumerate}
Of these scenarios, confirming the first scenario would give us an interesting example of matching conformal dimensions from weak to strong coupling. However, a precise relationship between the two tachyons is so far unknown. The most exciting scenario would be number 3, as it would give us access to a stable background of type 0B string theory and a dual strongly coupled non-supersymmetric gauge theory. This would be the first rigorous example for non-supersymmetric holography and a contender for describing QCD by a dual string theory.

To decide which scenario is true, one needs to probe the intermediate coupling regime, which despite some success with numerical and tentative approaches \cite{Gromov:2009zb,Frolov:2010wt,Hegedus:2016eop, Marboe:2018ugv} is still a hard problem, at least for deformations of $AdS_5\cross S^5$. This ties into the general issue of finding a strong coupling description of the QSC, which would be an important step towards a full understanding of the AdS/CFT conjecture.
Studying a comparably accessible operator like $\mathcal{O}_{{-2}}$, which may actually change its behaviour qualitatively should be an interesting starting point to this broader endeavour and to the question whether non-supersymmetric holography is justifiable. 

\section*{Acknowledgements}
I would like to thank my supervisor Arkady A. Tseytlin for his continuous support and guidance. I would further like to thank Nikolay Gromov, Andrea Cavagli\`a, Michelangelo Preti, Julius Julius, Fedor Levkovich-Maslyuk, Stijn J. van Tongeren, Matthias Wilhelm, Simon Eckhammar, Dennis le Plat, Elli Pomoni and Francesco Galvagno for useful discussions and/or correspondence. I acknowledge funding by
the President's PhD Scholarship of Imperial College London.

\appendix

\section{Derivation of QSC asymptotics from the $T$-system}\label{appendix}

To derive the twisted QSC from the $T$-system, it is useful to go back to the roots of the QSC and its original derivation \cite{Gromov:2014caa,Kazakov:2015efa}. Reminding ourselves of the $T$-system \eqref{Tsys}
\begin{equation}
T^+_{a,s}T^-_{a,s}=T_{a+1,s}T_{a-1,s}+T_{a,s+1}T_{a,s-1}
\end{equation}
living on the extended T-hook (fig. \ref{figure2}), we previously realised that there is a certain amount of ``gauge"-freedom. Indeed transforming
\begin{equation}
T_{a,s}\to g_1^{[a+s]}g_2^{[a-s]}g_3^{[-a+s]}g_4^{[-a-s]}T_{a,s}\,,
\end{equation}
where $f^{[a]}(u)=f(u+{ia\over2} )$, leaves \eqref{Tsys} and the $Y$-functions
\begin{equation}
Y_{a,s}=\frac{T_{a,s+1}T_{a,s-1}}{T_{a+1,s}T_{a-1,s}}
\end{equation}
invariant. 
This allows for a particularly nice gauge choice that makes the analytical properties of the $T$-functions as simple as possible.

In general, the $Y$- and $T$-functions have branch cuts. This is due to the Zhukovski-map
\eqref{Zhukovski} appearing in the S-matrices of the asymptotic Bethe Ansatz.
In combinations of S-matrices into kernels appearing in the TBA, we expect the appearance of an a priori infinite number of branch cuts shifted by integer multiples of $\frac{i}{2}$. This should then be reflected by the $Y$- and $T$-functions. However, we can use the gauge freedom of the $T$-system to reduce the number of branch cuts for a subset of $T$-functions.

Indeed, it is possible to set $T_{0,s}=1$ and reduce $T_{1,s}$ $\abs{s}>0$ to only having two short cuts. Then, we can introduce functions $\mathbf{P}_a$, $\mathbf{P}^a$ with only a single short cut along the real axis and parametrise
\begin{align}
T_{1,s}&=\mathbf{P}_1^{[+s]}\mathbf{P}_2^{[-s]}-\mathbf{P}_2^{[+s]}\mathbf{P}_1^{[-s]}\,\, \qquad s>0\,,\\
T_{1,s}&=\mathbf{P}^{4[+s]}\mathbf{P}^{3[-s]}-\mathbf{P}^{3[+s]}\mathbf{P}^{4[-s]} \quad s<0\,.
\end{align}
One can show that together with the prescription
\begin{equation}
T_{2,\pm s}=T_{1,\pm 1}^{[+s]}T_{1,\pm 1}^{[-s]} \quad s>1\,,
\end{equation}
the Hirota $T$-system is satisfied for the left and right wings of the T-hook. The remaining $T$-functions can then be reconstructed step by step, utilising previous knowledge of some group-theoretical relationships. It turns out that by specifying a few auxiliary single-cut functions $\mathbf{P}_3,\mathbf{P}_4$ and $\mathbf{P}^1,\mathbf{P}^2$ as well as an infinitely cut, but $i$-periodic matrix $\mu_{ab}$, one can encapsulate the entire $T$-system by the following $\mathbf{P}\mu$-system \footnote{By specifying the analyticity properties of the $\mathbf{P}_a$-functions, we actually capture more than only the algebraic properties of the $T$-system, but also some of the additional data that is contained within the TBA, see \cite{Cavaglia:2010nm,Balog:2011nm,Gromov:2011cx}.}
\begin{equation}\label{pmu}
\tilde \mu_{ab}-\mu_{ab}=\mathbf{P}_a\tilde{\mathbf{P}}_b-\mathbf{P}_b\tilde{\mathbf{P}}_a\,, \quad \tilde{\mathbf{P}}_a=\mu_{ab}\mathbf{P}^b\,,\quad \mathbf{P}_a\mathbf{P}^a=0\,,\quad \Pf(\mu)=1\,,
\end{equation} 
where the tilde denotes analytic continuation through the branch cut and $\Pf(\mu)=\mu_{12}\mu_{34}+\mu_{23}\mu_{14}-\mu_{13}\mu_{24}$ denotes the Pfaffian of the matrix $\mu$. 

This system is very useful to describe the left and right wings of the $T$-system, but deriving the middle part of the T-hook involves a tedious iterative process. However, the heart of this process can be captured by formally summing up finite differences of the form
\begin{equation}
Q_{a|i}=(U^{[+1]}U^{[+3]}....)_a^j M_{ji}\,, \qquad U_{a}^b=\delta_a^b + \mathbf{P}_a\mathbf{P}^b\,,
\end{equation}
for some constant matrix $M_{ij}$. If we further define auxiliary functions 
\begin{equation}
\mathbf{Q}_i=-\mathbf{P}^aQ_{a|i}^+ 
\end{equation}
above the real axis, we find that these $\mathbf{Q}_i$ functions only have a single long cut in the mirror kinematics. Instead of solving $\mathbf{Q}_i$ as function of $\mathbf{P}_a$'s, we regard them as an alternative basis, related to the $\mathbf{P}_a$ by 
\begin{equation}\label{Qai}
Q_{a|i}^+-Q_{a|i}^-=\mathbf{P}_a\mathbf{Q}_i\,.
\end{equation} 
An analogous treatment of the $\mathbf{Q}_i$-functions shows, that one can introduce $\mathbf{Q}^i$ and $\omega_{ij}$ functions that lead to a fully consistent $\mathbf{Q}\omega$-system
\begin{equation}
\tilde \omega_{ij}-\omega_{ij}=\mathbf{Q}_i\tilde{\mathbf{Q}}_j-\mathbf{Q}_j\tilde{\mathbf{Q}}_i\,, \quad \tilde{\mathbf{Q}}_i=\omega_{ij}\mathbf{Q}^j\,,\quad \mathbf{Q}_i\mathbf{Q}^i=0\,,\quad \Pf(\omega)=1\,,
\end{equation} 
which again sums up the entire $T$-system. \\

As the naming suggests, the $\mathbf{Q}$- and $\mathbf{P}$-functions we defined here are the ones appearing in the modern formulation of the QSC. The net of dualities between $Q$-functions turns out as much more efficient then the ``bases" $\mathbf{P}\mu$ and $\mathbf{Q}\omega$.  Indeed, it was later noted that the matrix $\omega_{ij}$ could be reduced to the glueing conditions given in \eqref{glue}. All other relationships in this set-up can be understood as being encoded in the $QQ$-relations and the Hodge duality structure of the QSC. Therefore the information of the $T$-system is retained in the QSC. \\

In both approaches one has to fix asymptotics of the respective functions and this is the only place where the twists we introduced show up. Having understood the relationship of $T$-system and QSC, we can now simply match these asymptotics.

The easiest example is the asymptotic solution of $T_{1,s}$ which we found as the constant expression 
\begin{equation}\label{match}
T^0_{1,s}=[s]_{e^{i\dot\vartheta}}\dot\xi=\frac{e^{is\dot{\vartheta}}-e^{-is\dot{\vartheta}}}{e^{i\dot{\vartheta}}-e^{-i\dot{\vartheta}}}\dot{\xi}\overset{!}{=}\mathbf{P}_1^{0,[+s]}\mathbf{P}_2^{0,[-s]}-\mathbf{P}_2^{0,[+s]}\mathbf{P}_1^{0,[-s]}\,.
\end{equation}
The $s$-dependence of the $\mathbf{P}$-functions relies only on shifts of $u$, so we find that we have to introduce an exponential factor

\begin{equation}
\mathbf{P}^0_1\sim e^{u\dot{\vartheta}}\,, \qquad \mathbf{P}^0_2\sim e^{-u\dot{\vartheta}}\,,
\end{equation}
which is precisely the exponential prefactor we find in comparing to the sigma-model. The other factors have to be distributed between $\mathbf{P}_1$ and $\mathbf{P}_2$ such that the product of both matches the previous formula \eqref{match}. We note that for $\dot\vartheta=0,\pi$ the denominator diverges, which is linked to the cancellations in the $QQ$-system (see discussion after \eqref{protocancel}).

The left wing works similarly, the difference being the raised indices which for the asymptotic solution merely result in a changed sign of $\vartheta$. lowering the indices we find similar exponential prefactors
\begin{equation}
\mathbf{P}^0_3\sim e^{u{\vartheta}}\,, \qquad \mathbf{P}^0_4\sim e^{-u{\vartheta}}\,.
\end{equation}
Interestingly, the infinitesimal factor $\dot\epsilon$ that appeared in the $T$-system seems to be related to the $\mu_{ab}$ matrices that connect the two wings in the $\mathbf{P}\mu$-system. 

The other $T$-functions are connected much less directly to the $Q$-functions but appendix B of \cite{Gromov:2014caa} gives a full dictionary. 

\section{Leading coefficients of QSC functions}\label{appendix2}

The charges of the state we want to describe only provide us with the asymptotics of the $Q$-functions. However, the $QQ$-relations and glueing conditions are in principle restrictive enough to derive the full $Q$-functions. As a first step we can derive the leading coefficients.

Let us first introduce generic constant prefactors 
\begin{equation}\begin{split}
&\mathbf{P}_a\sim A_a \cdot x_a^{iu}u^{-\mathcal{M}_a-S_{\prec a}}\,\,\,,\quad \mathbf{P}^a\sim A^a \cdot x_a^{-iu}u^{\mathcal{M}_a-S_{\succ a}}\,,\\
&\mathbf{Q}_i\sim B_i \cdot y_i^{-iu}u^{\hat{\mathcal{M}}_i-\hat S_{\prec i}}\quad,\,\quad \mathbf{Q}^i\sim B^i \cdot y_i^{iu}u^{-\hat{\mathcal{M}}_i-\hat S_{\succ i}}\,\,.
\end{split}\end{equation}
We can then use the QQ relations to run through the prefactors of all leading order asymptotic $Q$-functions. At every step we have to discern whether cancellations occur or not
\begin{align}
Q_{Aab|I}&\sim\frac{Q_{Aa|I}Q_{Ab|I}}{Q_{A|I}}\cross\begin{cases}\frac{x_b-x_a}{\sqrt{x_ax_b}} & \quad\, x_a\neq x_b\\
i\frac{M_b-M_a}{u} &\quad\, x_a=x_b\end{cases}\,,\\
Q_{A|Iij}&\sim\frac{Q_{A|Iij}Q_{A|Iij}}{Q_{A|I}}\cross\begin{cases}\frac{y_i-y_j}{\sqrt{y_iy_j}} & \quad y_i\neq y_j\\
i\frac{\hat M_i-\hat M_j}{u} &\quad y_i=y_j \end{cases}\,,\\
Q_{Aa|Ii}&\sim\frac{Q_{Aa|I}Q_{A|Ii}}{Q_{A|I}}\cross\begin{cases}\frac{\sqrt{x_ay_i}}{y_i-x_a} & x_a\neq y_i\\
i\frac{u}{-M_a+\hat M_i+1} & x_a=y_i \end{cases}\,.
\end{align}
Here we returned to the ``naive" asymptotic charges $M_a=\mathcal{M}_a+S_{\prec a}$ and $\hat M_i=\hat{\mathcal{M}}_i-\hat{S}_{\prec i}$ for notational simplicity. The multiplication/division by $u$ is of course exactly the cancellation effect we described earlier. 

As a simple example we can determine the asymptotics of $Q_{a|i}$. We have to distinguish two cases: 
\begin{itemize}
	\item If $x_a\neq y_i$ no cancellation occurs and we find that 
	\begin{equation}\label{Qai1}
	Q_{a|i}\sim A_aB_i \frac{\sqrt{x_ay_i}}{x_a-y_i}\left(\frac{x_a}{y_i}\right)^{iu}u^{-M_a+\hat M_i}\,.
	\end{equation}
	\item If $x_a=y_i$ we have to take into account a cancellation and we find
	\begin{equation}\label{Qai2}
	Q_{a|i}\sim i \frac{A_aB_i}{-M_a+\hat M_i+1}u^{-M_a+\hat M_i+1}\,.
	\end{equation}
	In this case we can again run into the additional problem when $M_a-\hat M_i=1$. In this case either $A_a$ or $B_i$ have to vanish, symbolising the appearance of a short representation, which needs to be handled more carefully.
\end{itemize}
A lengthier exercise in accounting leads to the prefactors $A^a$ and $B^i$, as these are related to 7-index $Q$-functions. One finds \cite{Kazakov:2015efa}
\begin{equation}
A^a=\frac{1}{A_a x_a}\frac{\prod_i z_{a,i}}{\prod_{b\neq a}z_{b,a}}\,,\qquad B^i=\frac{1}{B_i y_i}\frac{\prod_a z_{a,i}}{\prod_{i\neq j}z_{j,i}}\,,
\end{equation} 
where the $z$-factors denote the numerical factors appearing in the asymptotic QQ-relations,
namely 
\begin{equation}
\begin{split}
z_{a,b}&=\begin{cases}
x_b-x_a &\quad\quad\,\,\, x_a\neq x_b\\ i x_a (M_b-M_a) &\quad\quad\,\,\, x_a=x_b
\end{cases}\,,\\
z_{i,j}&=\begin{cases}
y_i-y_j &\qquad\quad y_i\neq y_j\\ i y_i (\hat M_i-\hat M_j) &\qquad\quad y_i=y_j
\end{cases}\,,\\
z_{a,i}&=\begin{cases}
y_i-x_a & x_a\neq y_i\\ i x_a (-M_a+\hat M_i+1) &x_a=x_b
\end{cases}\,.
\end{split}
\end{equation}

In the completely untwisted model we always need \eqref{Qai2} for $Q_{a|i}$ and we find the relations (no summation convention here)
\begin{equation}
A^aA_a=i \frac{\prod_i (-M_a+\hat M_i+1)}{\prod_{b\neq a} (M_a-M_b)}\,,\qquad B^iB_i=i \frac{\prod_a (-M_a+\hat M_i+1)}{\prod_{j\neq i} (M_i-M_j)}\,.
\end{equation}
In the twisted sector of type 0B we find that we always need $\eqref{Qai1}$ for $Q_{a|i}$ giving the simple asymptotic
\begin{equation}
Q_{a|i}\sim \, \frac{A_a B_i}{2}x_a^{iu+\frac{1}{2}}u^{-M_a+\hat{M}_i}\,.
\end{equation}
The prefactors $A^a$ and $B^i$ satisfy the relationships. 
\begin{equation}
A^aA_a=i \frac{16}{\prod_{b\neq a} (M_a-M_b)}\,,\qquad B^iB_i=-i \frac{16}{\prod_{j\neq i} (M_i-M_j)}\,.
\end{equation}

\section{Outline of perturbative algorithm for type 0B twisted sector}\label{appendix3}

Now that we know the leading asymptotics exactly from the previous appendix \ref{appendix2}, we can take a look at the sub-leading terms. We restrict our attention to the ground state of the type 0B twisted sector, which is right-left symmetric, i.e. follows \eqref{LR}, which greatly simplifies calculations. These discussions follow the recipe of \cite{Levkovich-Maslyuk:2020rlp}, but we use slightly modified parametrisations to avoid singular values.

\subsubsection*{Ansatz for $\mathbf{P}_a$}
The $\mathbf{P}$-functions should only have one short branch-cut, so they should depend on $u$ only through $x_{ph}(u)$ \eqref{Zhukovski}, which we shall call $X$ in this section.
For small $g$ we can expand
\begin{equation}
X(u):=x_{ph}(u)\sim \frac{u}{g}-\frac{g}{u}-\frac{g^3}{u^3}-2\frac{g^5}{u^5}+\order{g^7}\,,
\end{equation}
so instead of $u$ we should use $gX$ for the leading term. Then we introduce the normalised functions $p_a(X)$ 
\begin{equation}\label{defp}
\mathbf{P}_a(X)= A_a x_a^{iu} (g X)^{-M_{\text{max}}}\cdot p_a(X)\,,
\end{equation}
where we divided out the highest order in $1/g X$. To this end we introduced $M_{\text{max}}:=\max\{M_a\}_a$ which for our case is $M_{\text{max}}=M_1$. The factor $ (g X)^{-M_{\text{max}}}=(gX)^{-M_1}=(gX)^{-J/2}$ makes sure that all $p_a(X)$ start at non-negative orders in $gX$. We choose the Ansatz
\begin{equation}
p_a(X)=(g X)^{M_{\text{max}}-M_a}+\sum_{n=0}^{M_{\text{max}}-M_a-1} c_{a,-n}(gX)^n + \sum_{n=1}^\infty \frac{c_{a,n}g^{2n}}{(gX)^n}\,.
\end{equation}
The coefficients $c_{a,n}$ can be expanded in a Taylor series around $g=0$
\begin{equation}
c_{a,n}=\sum_{k=0}^\infty c_{a,n}^{(k)}~g^{2k}\,.
\end{equation}
Note that the coefficients $c_{a,n>0}$ of the infinite sum are already suppressed by $g^{2n}$. This is due to the expected scaling of $\mathbf{P}_a\sim\order{g^0}$ and $\tilde{\mathbf{P}}_a\sim\order{g^{-J}}$ with $g$. To verify the later scaling condition, we remember that analytic continuations can be performed by substituting the Zhukovski variable $X$ with $X^{-1}$ and therefore we get
\begin{equation}
\tilde p_a(x)=\left(\frac{g^2}{g X}\right)^{M_{\text{max}}-M_a}+\sum_{n=0}^{M_{\text{max}}-M_a-1} c_{a,-n}\left(\frac{g^2}{g X}\right)^n + \sum_{n=1}^\infty c_{a,n}(gX)^n \sim\order{g^0}\,,
\end{equation}
which indeed leads to $\tilde{\mathbf{P}}_a\sim\order{g^{-J}}$ via the prefactor in \eqref{defp}.

\subsection*{Fixing $H$-symmetry}
The choice of $\mathbf{P}$ functions is not unique as there is a symmetry with respect to linear transformations $\mathbf{P}_a=H_a^b\mathbf{P}_b$ called the $H$-symmetry, which leaves the $QQ$-relations invariant. However, by fixing the asymptotics of the $\mathbf{P}_a$ functions and the left-right-symmetry matrix $\chi$, we have already broken parts of this symmetry. Indeed, one can work out that the remaining symmetry transformations are
\begin{equation}\begin{split}
\mathbf{P}_1\quad&\rightarrow\quad \alpha \mathbf{P}_1\,,\\
\mathbf{P}_2\quad&\rightarrow\quad \beta \mathbf{P}_2\,, \\
\mathbf{P}_3\quad&\rightarrow\quad \alpha^{-1} \mathbf{P}_3 + \gamma \mathbf{P}_1\,, \\
\mathbf{P}_4\quad&\rightarrow\quad \beta^{-1}\mathbf{P}_4 + \delta\mathbf{P}_2\,.
\end{split}\end{equation}
This allows us to fix 
\begin{equation}
A_1=-i \frac{16}{(J+2)(J+3)}\,,\quad A_2=-i \frac{16}{(J+1)(J+2)}\,,\quad A_3=1\,,\quad A_4=1\,,
\end{equation}
as well as to set to 0 two more sub-leading factors.\\

Given this Ansatz we first use the $QQ$-relations to derive the $\mathbf{Q}_i$ functions order by order in perturbation theory. Then the glueing conditions and asymptotics of the $\mathbf{Q}_i$ will constrain the remaining $c_{a,n}$ parameters, including $\Delta=J+\order{g^{2}}$ which is the quantity we want to compute. 

Let us repeat the $QQ$-relations we will use in the following:
\begin{equation}\label{keyQQ}
Q_{a|i}^+-Q_{a|i}^-=\mathbf{P}_a\mathbf{Q}_i\,,\qquad \mathbf{Q}_i=-\mathbf{P}^aQ_{a|i}^\pm\,.
\end{equation}
There are two approaches that are commonly used in the literature. 

\subsection*{4th-order Baxter equation}
The first approach, first demonstrated in \cite{Alfimov:2014bwa}, starts by shifting the second equation in \eqref{keyQQ} by $\pm i$ and $\pm 2i$ and using the first equation to rewrite everything in terms of $Q_{a|i}^{+}$. This gives a set of 16 equations, completely determining $Q_{a|i}^{+}$. Then one can shift the second order equation by $+3i$ and reinsert $Q_{a|i}$ to receive a 4th-order difference equation for $\mathbf{Q}_i$ in terms of shifted $\mathbf{P}_a$ functions:
\begin{equation}\label{Baxter}
f_4\mathbf{Q}_i^{[+4]}+f_2\mathbf{Q}_i^{[+2]}+f_0\mathbf{Q}_i+f_{-2}\mathbf{Q}_i^{[-2]}+f_{-4}\mathbf{Q}_i^{[-4]}=0,
\end{equation}
where the factors $f_i(\mathbf{P})$ encode the $\mathbf{P}_a$-dependence via 
\begin{equation}\begin{split}
f_4(\mathbf{P})=&\epsilon^{abcd}\mathbf{P}_a^{[+2]}\mathbf{P}_b^{[0]}\mathbf{P}_c^{[-2]}\mathbf{P}_d^{[-4]}\,, \\
f_2(\mathbf{P})=&-\epsilon^{abcd}\mathbf{P}_a^{[+4]}\mathbf{P}_b^{[0]}\mathbf{P}_c^{[-2]}\mathbf{P}_d^{[-4]} +\mathbf{P}_e^{[+2]}\mathbf{P}^{e[+4]}\epsilon^{abcd}\mathbf{P}_a^{[+2]}\mathbf{P}_b^{[0]}\mathbf{P}_c^{[-2]}\mathbf{P}_d^{[-4]}\,, \\
f_0(\mathbf{P})=&\epsilon^{abcd}\mathbf{P}_a^{[+4]}\mathbf{P}_b^{[+2]}\mathbf{P}_c^{[-2]}\mathbf{P}_d^{[-4]}
-\mathbf{P}_e^{[0]}\mathbf{P}^{e[+2]}\epsilon^{abcd}\mathbf{P}_a^{[+4]}\mathbf{P}_b^{[0]}\mathbf{\mathbf{P}}_c^{[-2]}\mathbf{P}_d^{[-4]}\,, \\
&+\mathbf{P}_e^{[0]}\mathbf{P}^{e[+4]}\epsilon^{abcd}\mathbf{P}_a^{[+2]}\mathbf{P}_b^{[0]}\mathbf{P}_c^{[-2]}\mathbf{P}_d^{[-4]}\,, \\
f_{-2}(\mathbf{P})=&-\epsilon^{abcd}\mathbf{P}_a^{[-4]}\mathbf{P}_b^{[0]}\mathbf{P}_c^{[+2]}\mathbf{P}_d^{[+4]} -\mathbf{P}_e^{[-2]}\mathbf{P}^{e[-4]}\epsilon^{abcd}\mathbf{P}_a^{[-2]}\mathbf{P}_b^{[0]}\mathbf{P}_c^{[+2]}\mathbf{P}_d^{[+4]}\,, \\
f_{-4}(\mathbf{P})=&\epsilon^{abcd}\mathbf{P}_a^{[-2]}\mathbf{P}_b^{[0]}\mathbf{P}_c^{[+2]}\mathbf{P}_d^{[+4]}\,.
\end{split}\end{equation}

\subsection*{Perturbative solution of $Q_{a|i}$ relations}
The second approach pioneered in \cite{Gromov:2015vua} starts from a zeroth-order solution $\mathbf{Q}_i^{(0)}$for $\mathbf{Q}_i$ and then derives $Q_{a|i}^{(0)}$ from it. Then order by order one solves the equation
\begin{equation}
Q_{a|i}^{(n)+}-Q_{a|i}^{(n)-}+\mathbf{P}_a\mathbf{P}^bQ_{b|i}^{(n)-}=\order{g^{2n+2}}.
\end{equation}
to get to the desired order of $Q_{a|i}^{(n)}$.
Plugging in the Ansatz
\begin{equation}
Q_{a|i}^{(n+1)}=Q_{a|i}^{(n)}+(b^{(n)})_i^jQ_{a|i}^{(n)},
\end{equation}
where $(b^{(n)})_i^j$ is of order $g^2$, we only have to solve a number of finite-difference equations. 
As a final step one can then use the second equation in \eqref{keyQQ} to find $\mathbf{Q}_i^{(n)}$.

\subsection*{Hybrid approach}
The second approach is more amenable to perturbation theory but requires an initial guess at $\mathbf{Q}_i^{(0)}$. We therefore follow \cite{Levkovich-Maslyuk:2020rlp} and use a hybrid of both approaches, namely solve the Baxter equation at leading order and then use this as the initial solution $\mathbf{Q}_i^{(0)}$ to run the perturbative machinery. 

Before we do so, however, we use the information we have about the asymptotics of the $\mathbf{Q}_i$-functions to fix a few parameters. When we plug the $\mathbf{Q}_i$ asymptotics into the 4th-order Baxter equation \eqref{Baxter} we expect that it should be satisfied for large $u$ at leading order in $g$. This already fixes a number of $c_{a,n}^{(0)}$. For example, for $J=2$ we find 
\begin{equation}\begin{split}
c_{3,-3}^{(0)}&=0\,,\quad c_{3,-2}^{(0)}=2\,,\quad c_{3,-1}^{(0)}=-\frac{2}{5c_{2,0}^{(0)}}\,,\quad c_{3,0}=0\,,\\
c_{4,-4}^{(0)}&=\frac{5}{3}c_{2,0}^{(0)}\,, \quad c_{4,-3}^{(0)}=-\frac{10}{3}\,,\quad c_{4,-2}^{(0)}=-\frac{10}{3}c_{2,0}^{(0)}\,,\quad c_{4,-1}=0\,,
\end{split}\end{equation}
where we also fixed the residual $H$-symmetry to set $c_{3,0}=c_{4,-1}=0$. The only zeroth-order factors in $\mathbf{P}_a$ remaining unfixed are $c_{2,0}$ and $c_{4,0}$. 

\subsection*{$\eta$-functions and leading order solutions}

To solve the leading order Baxter equation we need to introduce $\eta$-functions, which are defined as
\begin{equation}
\eta_{s_1,\dots,s_k}^{z_1,\dots,z_k}(u)=\sum_{n_1>n_2>\dots>n_k\geq0}\frac{z_1^{n_1}z_2^{n_2}\dots z_k^{n_k}}{(u+i n_1)^{s_1}(u+i n_2)^{s_2}\dots(u+i n_k)^{s_k}}.
\end{equation}
and arise (together with polynomials and exponentials) as common solutions to finite difference equations, since e.g.
\begin{equation}
\eta_s^1(u+i)-\eta_s^1(u)=-\frac{1}{u^s}.
\end{equation}
Furthermore they have an infinite number of poles in the upper half-plane, which is precisely the behaviour we expect from the $\mathbf{Q}$-functions after the branch-cuts (of length $4 g$) have reduced to poles in the $g\to 0$ limit. A Mathematica package dealing with these functions can be taken from the appendix to \cite{Gromov:2015dfa}, but in our case the orbifolding introduced various $z$-factors, which forced us to slightly extend the package for the special case $z=-1$.

In the example of $J=2$, we find 4 solutions to the Baxter equations given by
\begin{equation}\begin{split}
\mathbf{Q}_I^{(0)}=&u^2\\
\mathbf{Q}_{II}^{(0)}=&-u^2 ~\eta_2^1(u)+ u~\eta_1^1(u)+\frac{i}{4 u}\\
\mathbf{Q}_{III}^{(0)}=&u\\
\mathbf{Q}_{IV}^{(0)}=&-u^2 \eta_1^1(u)+\frac{i}{4}\\
\end{split}\end{equation}
which serves as the zeroth-order basis from which to later build the full $\mathbf{Q}_i$ functions. We note that this precisely matches the zeroth-order result of \cite{Levkovich-Maslyuk:2020rlp} in the questionable case $\kappa=\hat{\kappa}=i$. In \cite{Levkovich-Maslyuk:2020rlp}, the $A_i$ were chosen in a way that would be highly divergent for these values. Here, we chose a different parametrisation and much to our relief no previously unknown cancellations occurred, leading to essentially the same functions. This leads us to believe that the physical (i.e. parametrisation independent) result of \cite{Levkovich-Maslyuk:2020rlp} for $\Delta$ should be readily extendable to this special case, resulting in our conjecture \eqref{conjecture}. 

\bibliographystyle{JHEP}
\bibliography{Integrable0.bib}

\providecommand{\href}[2]{#2}\begingroup\raggedright\begin{thebibliography}{10}

\bibitem{Maldacena:1997re}
J.M.~Maldacena, \emph{{The Large N limit of superconformal field theories and
  supergravity}}, \href{https://doi.org/10.1023/A:1026654312961}{\emph{Adv.
  Theor. Math. Phys.} {\bfseries 2} (1998) 231}
  [\href{https://arxiv.org/abs/hep-th/9711200}{{\ttfamily hep-th/9711200}}].

\bibitem{tHooft:1973alw}
G.~'t~Hooft, \emph{{A Planar Diagram Theory for Strong Interactions}},
  \href{https://doi.org/10.1016/0550-3213(74)90154-0}{\emph{Nucl. Phys. B}
  {\bfseries 72} (1974) 461}.

\bibitem{Bena:2003wd}
I.~Bena, J.~Polchinski and R.~Roiban, \emph{{Hidden symmetries of the AdS(5) x
  S**5 superstring}},
  \href{https://doi.org/10.1103/PhysRevD.69.046002}{\emph{Phys. Rev. D}
  {\bfseries 69} (2004) 046002}
  [\href{https://arxiv.org/abs/hep-th/0305116}{{\ttfamily hep-th/0305116}}].

\bibitem{Berenstein:2002jq}
D.E.~Berenstein, J.M.~Maldacena and H.S.~Nastase, \emph{{Strings in flat space
  and pp waves from N=4 superYang-Mills}},
  \href{https://doi.org/10.1088/1126-6708/2002/04/013}{\emph{JHEP} {\bfseries
  04} (2002) 013} [\href{https://arxiv.org/abs/hep-th/0202021}{{\ttfamily
  hep-th/0202021}}].

\bibitem{Minahan:2002ve}
J.A.~Minahan and K.~Zarembo, \emph{{The Bethe ansatz for N=4 superYang-Mills}},
  \href{https://doi.org/10.1088/1126-6708/2003/03/013}{\emph{JHEP} {\bfseries
  03} (2003) 013} [\href{https://arxiv.org/abs/hep-th/0212208}{{\ttfamily
  hep-th/0212208}}].

\bibitem{Beisert:2003yb}
N.~Beisert and M.~Staudacher, \emph{{The N=4 SYM integrable super spin chain}},
  \href{https://doi.org/10.1016/j.nuclphysb.2003.08.015}{\emph{Nucl. Phys. B}
  {\bfseries 670} (2003) 439}
  [\href{https://arxiv.org/abs/hep-th/0307042}{{\ttfamily hep-th/0307042}}].

\bibitem{Beisert:2006ez}
N.~Beisert, B.~Eden and M.~Staudacher, \emph{{Transcendentality and Crossing}},
  \href{https://doi.org/10.1088/1742-5468/2007/01/P01021}{\emph{J. Stat. Mech.}
  {\bfseries 0701} (2007) P01021}
  [\href{https://arxiv.org/abs/hep-th/0610251}{{\ttfamily hep-th/0610251}}].

\bibitem{Beisert:2005fw}
N.~Beisert and M.~Staudacher, \emph{{Long-range psu(2,2$|$4) Bethe Ansatze for
  gauge theory and strings}},
  \href{https://doi.org/10.1016/j.nuclphysb.2005.06.038}{\emph{Nucl. Phys. B}
  {\bfseries 727} (2005) 1}
  [\href{https://arxiv.org/abs/hep-th/0504190}{{\ttfamily hep-th/0504190}}].

\bibitem{Luscher:1985dn}
M.~Luscher, \emph{{Volume Dependence of the Energy Spectrum in Massive Quantum
  Field Theories. 1. Stable Particle States}},
  \href{https://doi.org/10.1007/BF01211589}{\emph{Commun. Math. Phys.}
  {\bfseries 104} (1986) 177}.

\bibitem{Zamolodchikov:1989cf}
A.B.~Zamolodchikov, \emph{{Thermodynamic Bethe Ansatz in Relativistic Models.
  Scaling Three State Potts and Lee-yang Models}},
  \href{https://doi.org/10.1016/0550-3213(90)90333-9}{\emph{Nucl. Phys. B}
  {\bfseries 342} (1990) 695}.

\bibitem{Arutyunov:2007tc}
G.~Arutyunov and S.~Frolov, \emph{{On String S-matrix, Bound States and TBA}},
  \href{https://doi.org/10.1088/1126-6708/2007/12/024}{\emph{JHEP} {\bfseries
  12} (2007) 024} [\href{https://arxiv.org/abs/0710.1568}{{\ttfamily
  0710.1568}}].

\bibitem{Bombardelli:2009ns}
D.~Bombardelli, D.~Fioravanti and R.~Tateo, \emph{{Thermodynamic Bethe Ansatz
  for planar AdS/CFT: A Proposal}},
  \href{https://doi.org/10.1088/1751-8113/42/37/375401}{\emph{J. Phys. A}
  {\bfseries 42} (2009) 375401}
  [\href{https://arxiv.org/abs/0902.3930}{{\ttfamily 0902.3930}}].

\bibitem{Arutyunov:2009ur}
G.~Arutyunov and S.~Frolov, \emph{{Thermodynamic Bethe Ansatz for the AdS(5) x
  S(5) Mirror Model}},
  \href{https://doi.org/10.1088/1126-6708/2009/05/068}{\emph{JHEP} {\bfseries
  05} (2009) 068} [\href{https://arxiv.org/abs/0903.0141}{{\ttfamily
  0903.0141}}].

\bibitem{Gromov:2009tv}
N.~Gromov, V.~Kazakov and P.~Vieira, \emph{{Exact Spectrum of Anomalous
  Dimensions of Planar N=4 Supersymmetric Yang-Mills Theory}},
  \href{https://doi.org/10.1103/PhysRevLett.103.131601}{\emph{Phys. Rev. Lett.}
  {\bfseries 103} (2009) 131601}
  [\href{https://arxiv.org/abs/0901.3753}{{\ttfamily 0901.3753}}].

\bibitem{Gromov:2014caa}
N.~Gromov, V.~Kazakov, S.~Leurent and D.~Volin, \emph{{Quantum spectral curve
  for arbitrary state/operator in AdS$_{5}$/CFT$_{4}$}},
  \href{https://doi.org/10.1007/JHEP09(2015)187}{\emph{JHEP} {\bfseries 09}
  (2015) 187} [\href{https://arxiv.org/abs/1405.4857}{{\ttfamily 1405.4857}}].

\bibitem{Marboe:2017dmb}
C.~Marboe and D.~Volin, \emph{{The full spectrum of AdS5/CFT4 I: Representation
  theory and one-loop Q-system}},
  \href{https://doi.org/10.1088/1751-8121/aab34a}{\emph{J. Phys. A} {\bfseries
  51} (2018) 165401} [\href{https://arxiv.org/abs/1701.03704}{{\ttfamily
  1701.03704}}].

\bibitem{Gromov:2007aq}
N.~Gromov and P.~Vieira, \emph{{The AdS(5) x S**5 superstring quantum spectrum
  from the algebraic curve}},
  \href{https://doi.org/10.1016/j.nuclphysb.2007.07.032}{\emph{Nucl. Phys. B}
  {\bfseries 789} (2008) 175}
  [\href{https://arxiv.org/abs/hep-th/0703191}{{\ttfamily hep-th/0703191}}].

\bibitem{Schafer-Nameki:2005yyn}
S.~Schafer-Nameki, M.~Zamaklar and K.~Zarembo, \emph{{Quantum corrections to
  spinning strings in AdS(5) x S(5) and Bethe ansatz: A Comparative study}},
  \href{https://doi.org/10.1088/1126-6708/2005/09/051}{\emph{JHEP} {\bfseries
  09} (2005) 051} [\href{https://arxiv.org/abs/hep-th/0507189}{{\ttfamily
  hep-th/0507189}}].

\bibitem{Roiban:2007jf}
R.~Roiban, A.~Tirziu and A.A.~Tseytlin, \emph{{Two-loop world-sheet corrections
  in AdS(5) x S**5 superstring}},
  \href{https://doi.org/10.1088/1126-6708/2007/07/056}{\emph{JHEP} {\bfseries
  07} (2007) 056} [\href{https://arxiv.org/abs/0704.3638}{{\ttfamily
  0704.3638}}].

\bibitem{Lunin:2005jy}
O.~Lunin and J.M.~Maldacena, \emph{{Deforming field theories with U(1) x U(1)
  global symmetry and their gravity duals}},
  \href{https://doi.org/10.1088/1126-6708/2005/05/033}{\emph{JHEP} {\bfseries
  05} (2005) 033} [\href{https://arxiv.org/abs/hep-th/0502086}{{\ttfamily
  hep-th/0502086}}].

\bibitem{Frolov:2005dj}
S.~Frolov, \emph{{Lax pair for strings in Lunin-Maldacena background}},
  \href{https://doi.org/10.1088/1126-6708/2005/05/069}{\emph{JHEP} {\bfseries
  05} (2005) 069} [\href{https://arxiv.org/abs/hep-th/0503201}{{\ttfamily
  hep-th/0503201}}].

\bibitem{Frolov:2005iq}
S.A.~Frolov, R.~Roiban and A.A.~Tseytlin, \emph{{Gauge-string duality for
  (non)supersymmetric deformations of N=4 super Yang-Mills theory}},
  \href{https://doi.org/10.1016/j.nuclphysb.2005.10.004}{\emph{Nucl. Phys. B}
  {\bfseries 731} (2005) 1}
  [\href{https://arxiv.org/abs/hep-th/0507021}{{\ttfamily hep-th/0507021}}].

\bibitem{Alday:2005ww}
L.F.~Alday, G.~Arutyunov and S.~Frolov, \emph{{Green-Schwarz strings in
  TsT-transformed backgrounds}},
  \href{https://doi.org/10.1088/1126-6708/2006/06/018}{\emph{JHEP} {\bfseries
  06} (2006) 018} [\href{https://arxiv.org/abs/hep-th/0512253}{{\ttfamily
  hep-th/0512253}}].

\bibitem{Hanany:1999sp}
A.~Hanany and Y.-H.~He, \emph{{A Monograph on the classification of the
  discrete subgroups of SU(4)}},
  \href{https://doi.org/10.1088/1126-6708/2001/02/027}{\emph{JHEP} {\bfseries
  02} (2001) 027} [\href{https://arxiv.org/abs/hep-th/9905212}{{\ttfamily
  hep-th/9905212}}].

\bibitem{Solovyov:2007pw}
A.~Solovyov, \emph{{Bethe Ansatz Equations for General Orbifolds of N=4 SYM}},
  \href{https://doi.org/10.1088/1126-6708/2008/04/013}{\emph{JHEP} {\bfseries
  04} (2008) 013} [\href{https://arxiv.org/abs/0711.1697}{{\ttfamily
  0711.1697}}].

\bibitem{Beisert:2005he}
N.~Beisert and R.~Roiban, \emph{{The Bethe ansatz for Z(S) orbifolds of N=4
  super Yang-Mills theory}},
  \href{https://doi.org/10.1088/1126-6708/2005/11/037}{\emph{JHEP} {\bfseries
  11} (2005) 037} [\href{https://arxiv.org/abs/hep-th/0510209}{{\ttfamily
  hep-th/0510209}}].

\bibitem{Beccaria:2011qd}
M.~Beccaria and G.~Macorini, \emph{{Y-system for $Z_S$ Orbifolds of N=4 SYM}},
  \href{https://doi.org/10.1007/JHEP01(2012)112}{\emph{JHEP} {\bfseries 06}
  (2011) 004} [\href{https://arxiv.org/abs/1104.0883}{{\ttfamily 1104.0883}}].

\bibitem{Berenstein:2004ys}
D.~Berenstein and S.A.~Cherkis, \emph{{Deformations of N=4 SYM and integrable
  spin chain models}},
  \href{https://doi.org/10.1016/j.nuclphysb.2004.09.005}{\emph{Nucl. Phys. B}
  {\bfseries 702} (2004) 49}
  [\href{https://arxiv.org/abs/hep-th/0405215}{{\ttfamily hep-th/0405215}}].

\bibitem{vanTongeren:2013gva}
S.J.~van Tongeren, \emph{{Integrability of the ${\rm Ad}{{{\rm S}}_{5}}\times
  {{{\rm S}}^{5}}$ superstring and its deformations}},
  \href{https://doi.org/10.1088/1751-8113/47/43/433001}{\emph{J. Phys. A}
  {\bfseries 47} (2014) 433001}
  [\href{https://arxiv.org/abs/1310.4854}{{\ttfamily 1310.4854}}].

\bibitem{Douglas:1996sw}
M.R.~Douglas and G.W.~Moore, \emph{{D-branes, quivers, and ALE instantons}},
  \href{https://arxiv.org/abs/hep-th/9603167}{{\ttfamily hep-th/9603167}}.

\bibitem{Kachru:1998ys}
S.~Kachru and E.~Silverstein, \emph{{4-D conformal theories and strings on
  orbifolds}}, \href{https://doi.org/10.1103/PhysRevLett.80.4855}{\emph{Phys.
  Rev. Lett.} {\bfseries 80} (1998) 4855}
  [\href{https://arxiv.org/abs/hep-th/9802183}{{\ttfamily hep-th/9802183}}].

\bibitem{Lawrence:1998ja}
A.E.~Lawrence, N.~Nekrasov and C.~Vafa, \emph{{On conformal field theories in
  four-dimensions}},
  \href{https://doi.org/10.1016/S0550-3213(98)00495-7}{\emph{Nucl. Phys. B}
  {\bfseries 533} (1998) 199}
  [\href{https://arxiv.org/abs/hep-th/9803015}{{\ttfamily hep-th/9803015}}].

\bibitem{Gadde:2010zi}
A.~Gadde, E.~Pomoni and L.~Rastelli, \emph{{Spin Chains in $\mathcal{N}$=2
  Superconformal Theories: From the $\mathbb{Z}_{2}$ Quiver to Superconformal
  QCD}}, \href{https://doi.org/10.1007/JHEP06(2012)107}{\emph{JHEP} {\bfseries
  06} (2012) 107} [\href{https://arxiv.org/abs/1006.0015}{{\ttfamily
  1006.0015}}].

\bibitem{deLeeuw:2011rw}
M.~de~Leeuw and S.J.~van Tongeren, \emph{{Orbifolded Konishi from the Mirror
  TBA}}, \href{https://doi.org/10.1088/1751-8113/44/32/325404}{\emph{J. Phys.
  A} {\bfseries 44} (2011) 325404}
  [\href{https://arxiv.org/abs/1103.5853}{{\ttfamily 1103.5853}}].

\bibitem{Beccaria:2012xm}
M.~Beccaria, S.~Giombi, G.~Macorini, R.~Roiban and A.A.~Tseytlin,
  \emph{{'Short' spinning strings and structure of quantum $AdS_5 \times S^5$
  spectrum}}, \href{https://doi.org/10.1103/PhysRevD.86.066006}{\emph{Phys.
  Rev. D} {\bfseries 86} (2012) 066006}
  [\href{https://arxiv.org/abs/1203.5710}{{\ttfamily 1203.5710}}].

\bibitem{Pomoni:2019oib}
E.~Pomoni, \emph{{4D $\mathcal{N}=2$ SCFTs and spin chains}},
  \href{https://doi.org/10.1088/1751-8121/ab7f66}{\emph{J. Phys. A} {\bfseries
  53} (2020) 283005} [\href{https://arxiv.org/abs/1912.00870}{{\ttfamily
  1912.00870}}].

\bibitem{Pomoni:2021pbj}
E.~Pomoni, R.~Rabe and K.~Zoubos, \emph{{Dynamical spin chains in 4D $
  \mathcal{N} $ = 2 SCFTs}},
  \href{https://doi.org/10.1007/JHEP08(2021)127}{\emph{JHEP} {\bfseries 08}
  (2021) 127} [\href{https://arxiv.org/abs/2106.08449}{{\ttfamily
  2106.08449}}].

\bibitem{Pestun:2007rz}
V.~Pestun, \emph{{Localization of gauge theory on a four-sphere and
  supersymmetric Wilson loops}},
  \href{https://doi.org/10.1007/s00220-012-1485-0}{\emph{Commun. Math. Phys.}
  {\bfseries 313} (2012) 71} [\href{https://arxiv.org/abs/0712.2824}{{\ttfamily
  0712.2824}}].

\bibitem{Mitev:2014yba}
V.~Mitev and E.~Pomoni, \emph{{Exact effective couplings of four dimensional
  gauge theories with $\mathcal N=$ 2 supersymmetry}},
  \href{https://doi.org/10.1103/PhysRevD.92.125034}{\emph{Phys. Rev. D}
  {\bfseries 92} (2015) 125034}
  [\href{https://arxiv.org/abs/1406.3629}{{\ttfamily 1406.3629}}].

\bibitem{Mitev:2015oty}
V.~Mitev and E.~Pomoni, \emph{{Exact Bremsstrahlung and Effective Couplings}},
  \href{https://doi.org/10.1007/JHEP06(2016)078}{\emph{JHEP} {\bfseries 06}
  (2016) 078} [\href{https://arxiv.org/abs/1511.02217}{{\ttfamily
  1511.02217}}].

\bibitem{Pestun:2016zxk}
V.~Pestun et~al., \emph{{Localization techniques in quantum field theories}},
  \href{https://doi.org/10.1088/1751-8121/aa63c1}{\emph{J. Phys. A} {\bfseries
  50} (2017) 440301} [\href{https://arxiv.org/abs/1608.02952}{{\ttfamily
  1608.02952}}].

\bibitem{Niarchos:2019onf}
V.~Niarchos, C.~Papageorgakis and E.~Pomoni, \emph{{Type-B Anomaly Matching and
  the 6D (2,0) Theory}},
  \href{https://doi.org/10.1007/JHEP04(2020)048}{\emph{JHEP} {\bfseries 04}
  (2020) 048} [\href{https://arxiv.org/abs/1911.05827}{{\ttfamily
  1911.05827}}].

\bibitem{Galvagno:2020cgq}
F.~Galvagno and M.~Preti, \emph{{Chiral correlators in $ \mathcal{N} $ = 2
  superconformal quivers}},
  \href{https://doi.org/10.1007/JHEP05(2021)201}{\emph{JHEP} {\bfseries 05}
  (2021) 201} [\href{https://arxiv.org/abs/2012.15792}{{\ttfamily
  2012.15792}}].

\bibitem{Niarchos:2020nxk}
V.~Niarchos, C.~Papageorgakis, A.~Pini and E.~Pomoni, \emph{{(Mis-)Matching
  Type-B Anomalies on the Higgs Branch}},
  \href{https://doi.org/10.1007/JHEP01(2021)106}{\emph{JHEP} {\bfseries 01}
  (2021) 106} [\href{https://arxiv.org/abs/2009.08375}{{\ttfamily
  2009.08375}}].

\bibitem{Billo:2021rdb}
M.~Billo, M.~Frau, F.~Galvagno, A.~Lerda and A.~Pini, \emph{{Strong-coupling
  results for $ \mathcal{N} $ = 2 superconformal quivers and holography}},
  \href{https://doi.org/10.1007/JHEP10(2021)161}{\emph{JHEP} {\bfseries 10}
  (2021) 161} [\href{https://arxiv.org/abs/2109.00559}{{\ttfamily
  2109.00559}}].

\bibitem{Beccaria:2022ypy}
M.~Beccaria, G.P.~Korchemsky and A.A.~Tseytlin, \emph{{Strong coupling
  expansion in \ensuremath{\mathscr{N}} = 2 superconformal theories and the
  Bessel kernel}}, \href{https://doi.org/10.1007/JHEP09(2022)226}{\emph{JHEP}
  {\bfseries 09} (2022) 226}
  [\href{https://arxiv.org/abs/2207.11475}{{\ttfamily 2207.11475}}].

\bibitem{Beccaria:2022kxy}
M.~Beccaria, G.P.~Korchemsky and A.A.~Tseytlin, \emph{{Exact strong coupling
  results in $ \mathcal{N} $ = 2 Sp(2N) superconformal gauge theory from
  localization}}, \href{https://doi.org/10.1007/JHEP01(2023)037}{\emph{JHEP}
  {\bfseries 01} (2023) 037}
  [\href{https://arxiv.org/abs/2210.13871}{{\ttfamily 2210.13871}}].

\bibitem{Dixon:1986iz}
L.J.~Dixon and J.A.~Harvey, \emph{{String Theories in Ten-Dimensions Without
  Space-Time Supersymmetry}},
  \href{https://doi.org/10.1016/0550-3213(86)90619-X}{\emph{Nucl. Phys. B}
  {\bfseries 274} (1986) 93}.

\bibitem{Seiberg:1986by}
N.~Seiberg and E.~Witten, \emph{{Spin Structures in String Theory}},
  \href{https://doi.org/10.1016/0550-3213(86)90297-X}{\emph{Nucl. Phys. B}
  {\bfseries 276} (1986) 272}.

\bibitem{Takayanagi:2001jj}
T.~Takayanagi and T.~Uesugi, \emph{{Orbifolds as Melvin geometry}},
  \href{https://doi.org/10.1088/1126-6708/2001/12/004}{\emph{JHEP} {\bfseries
  12} (2001) 004} [\href{https://arxiv.org/abs/hep-th/0110099}{{\ttfamily
  hep-th/0110099}}].

\bibitem{Skrzypek:2021eue}
T.~Skrzypek and A.A.~Tseytlin, \emph{{On type 0 string theory in solvable RR
  backgrounds}}, \href{https://doi.org/10.1007/JHEP03(2022)173}{\emph{JHEP}
  {\bfseries 03} (2022) 173}
  [\href{https://arxiv.org/abs/2110.14683}{{\ttfamily 2110.14683}}].

\bibitem{Nekrasov:1999mn}
N.~Nekrasov and S.L.~Shatashvili, \emph{{On nonsupersymmetric CFT in
  four-dimensions}},
  \href{https://doi.org/10.1016/S0370-1573(99)00059-9}{\emph{Phys. Rept.}
  {\bfseries 320} (1999) 127}
  [\href{https://arxiv.org/abs/hep-th/9902110}{{\ttfamily hep-th/9902110}}].

\bibitem{Klebanov:1999ch}
I.R.~Klebanov and A.A.~Tseytlin, \emph{{A Nonsupersymmetric large N CFT from
  type 0 string theory}},
  \href{https://doi.org/10.1088/1126-6708/1999/03/015}{\emph{JHEP} {\bfseries
  03} (1999) 015} [\href{https://arxiv.org/abs/hep-th/9901101}{{\ttfamily
  hep-th/9901101}}].

\bibitem{Tseytlin:1999ii}
A.A.~Tseytlin and K.~Zarembo, \emph{{Effective potential in nonsupersymmetric
  SU(N) x SU(N) gauge theory and interactions of type 0 D3-branes}},
  \href{https://doi.org/10.1016/S0370-2693(99)00471-2}{\emph{Phys. Lett. B}
  {\bfseries 457} (1999) 77}
  [\href{https://arxiv.org/abs/hep-th/9902095}{{\ttfamily hep-th/9902095}}].

\bibitem{Klebanov:1999um}
I.R.~Klebanov, \emph{{Tachyon stabilization in the AdS / CFT correspondence}},
  \href{https://doi.org/10.1016/S0370-2693(99)01084-9}{\emph{Phys. Lett. B}
  {\bfseries 466} (1999) 166}
  [\href{https://arxiv.org/abs/hep-th/9906220}{{\ttfamily hep-th/9906220}}].

\bibitem{Arutyunov:2010gu}
G.~Arutyunov, M.~de~Leeuw and S.J.~van Tongeren, \emph{{Twisting the Mirror
  TBA}}, \href{https://doi.org/10.1007/JHEP02(2011)025}{\emph{JHEP} {\bfseries
  02} (2011) 025} [\href{https://arxiv.org/abs/1009.4118}{{\ttfamily
  1009.4118}}].

\bibitem{Adams:2001jb}
A.~Adams and E.~Silverstein, \emph{{Closed string tachyons, AdS / CFT, and
  large N QCD}}, \href{https://doi.org/10.1103/PhysRevD.64.086001}{\emph{Phys.
  Rev. D} {\bfseries 64} (2001) 086001}
  [\href{https://arxiv.org/abs/hep-th/0103220}{{\ttfamily hep-th/0103220}}].

\bibitem{Dymarsky:2005nc}
A.~Dymarsky, I.R.~Klebanov and R.~Roiban, \emph{{Perturbative gauge theory and
  closed string tachyons}},
  \href{https://doi.org/10.1088/1126-6708/2005/11/038}{\emph{JHEP} {\bfseries
  11} (2005) 038} [\href{https://arxiv.org/abs/hep-th/0509132}{{\ttfamily
  hep-th/0509132}}].

\bibitem{Dymarsky:2005uh}
A.~Dymarsky, I.R.~Klebanov and R.~Roiban, \emph{{Perturbative search for fixed
  lines in large N gauge theories}},
  \href{https://doi.org/10.1088/1126-6708/2005/08/011}{\emph{JHEP} {\bfseries
  08} (2005) 011} [\href{https://arxiv.org/abs/hep-th/0505099}{{\ttfamily
  hep-th/0505099}}].

\bibitem{Pomoni:2009joh}
E.~Pomoni and L.~Rastelli, \emph{{Large N Field Theory and AdS Tachyons}},
  \href{https://doi.org/10.1088/1126-6708/2009/04/020}{\emph{JHEP} {\bfseries
  04} (2009) 020} [\href{https://arxiv.org/abs/0805.2261}{{\ttfamily
  0805.2261}}].

\bibitem{Fokken:2013aea}
J.~Fokken, C.~Sieg and M.~Wilhelm, \emph{{Non-conformality of ${{\gamma
  }_{i}}$-deformed N = 4 SYM theory}},
  \href{https://doi.org/10.1088/1751-8113/47/45/455401}{\emph{J. Phys. A}
  {\bfseries 47} (2014) 455401}
  [\href{https://arxiv.org/abs/1308.4420}{{\ttfamily 1308.4420}}].

\bibitem{Gurdogan:2015csr}
O.~G\"urdo\u{g}an and V.~Kazakov, \emph{{New Integrable 4D Quantum Field
  Theories from Strongly Deformed Planar $\mathcal N = $ 4 Supersymmetric
  Yang-Mills Theory}},
  \href{https://doi.org/10.1103/PhysRevLett.117.201602}{\emph{Phys. Rev. Lett.}
  {\bfseries 117} (2016) 201602}
  [\href{https://arxiv.org/abs/1512.06704}{{\ttfamily 1512.06704}}].

\bibitem{Gromov:2017cja}
N.~Gromov, V.~Kazakov, G.~Korchemsky, S.~Negro and G.~Sizov,
  \emph{{Integrability of Conformal Fishnet Theory}},
  \href{https://doi.org/10.1007/JHEP01(2018)095}{\emph{JHEP} {\bfseries 01}
  (2018) 095} [\href{https://arxiv.org/abs/1706.04167}{{\ttfamily
  1706.04167}}].

\bibitem{Levkovich-Maslyuk:2020rlp}
F.~Levkovich-Maslyuk and M.~Preti, \emph{{Exploring the ground state spectrum
  of \ensuremath{\gamma}-deformed N = 4 SYM}},
  \href{https://doi.org/10.1007/JHEP06(2022)146}{\emph{JHEP} {\bfseries 06}
  (2022) 146} [\href{https://arxiv.org/abs/2003.05811}{{\ttfamily
  2003.05811}}].

\bibitem{Metsaev:1998it}
R.R.~Metsaev and A.A.~Tseytlin, \emph{{Type IIB superstring action in AdS(5) x
  S**5 background}},
  \href{https://doi.org/10.1016/S0550-3213(98)00570-7}{\emph{Nucl. Phys. B}
  {\bfseries 533} (1998) 109}
  [\href{https://arxiv.org/abs/hep-th/9805028}{{\ttfamily hep-th/9805028}}].

\bibitem{Zarembo:2017muf}
K.~Zarembo, \emph{{Integrability in Sigma-Models}},
  \href{https://arxiv.org/abs/1712.07725}{{\ttfamily 1712.07725}}.

\bibitem{Arutyunov:2009ga}
G.~Arutyunov and S.~Frolov, \emph{{Foundations of the AdS$_{5} \times S^{5}$
  Superstring. Part I}},
  \href{https://doi.org/10.1088/1751-8113/42/25/254003}{\emph{J. Phys. A}
  {\bfseries 42} (2009) 254003}
  [\href{https://arxiv.org/abs/0901.4937}{{\ttfamily 0901.4937}}].

\bibitem{deLeeuw:2007akd}
M.~de~Leeuw, \emph{{Coordinate Bethe Ansatz for the String S-Matrix}},
  \href{https://doi.org/10.1088/1751-8113/40/48/008}{\emph{J. Phys. A}
  {\bfseries 40} (2007) 14413}
  [\href{https://arxiv.org/abs/0705.2369}{{\ttfamily 0705.2369}}].

\bibitem{Mitev:2012vt}
V.~Mitev, M.~Staudacher and Z.~Tsuboi, \emph{{The Tetrahedral Zamolodchikov
  Algebra and the ${AdS_5\times S^5}$ S-matrix}},
  \href{https://doi.org/10.1007/s00220-017-2905-y}{\emph{Commun. Math. Phys.}
  {\bfseries 354} (2017) 1} [\href{https://arxiv.org/abs/1210.2172}{{\ttfamily
  1210.2172}}].

\bibitem{Cavaglia:2020hdb}
A.~Cavaglia, D.~Grabner, N.~Gromov and A.~Sever, \emph{{Colour-twist operators.
  Part I. Spectrum and wave functions}},
  \href{https://doi.org/10.1007/JHEP06(2020)092}{\emph{JHEP} {\bfseries 06}
  (2020) 092} [\href{https://arxiv.org/abs/2001.07259}{{\ttfamily
  2001.07259}}].

\bibitem{Fiamberti:2008sh}
F.~Fiamberti, A.~Santambrogio, C.~Sieg and D.~Zanon, \emph{{Anomalous dimension
  with wrapping at four loops in N=4 SYM}},
  \href{https://doi.org/10.1016/j.nuclphysb.2008.07.014}{\emph{Nucl. Phys. B}
  {\bfseries 805} (2008) 231}
  [\href{https://arxiv.org/abs/0806.2095}{{\ttfamily 0806.2095}}].

\bibitem{Bajnok:2008bm}
Z.~Bajnok and R.A.~Janik, \emph{{Four-loop perturbative Konishi from strings
  and finite size effects for multiparticle states}},
  \href{https://doi.org/10.1016/j.nuclphysb.2008.08.020}{\emph{Nucl. Phys. B}
  {\bfseries 807} (2009) 625}
  [\href{https://arxiv.org/abs/0807.0399}{{\ttfamily 0807.0399}}].

\bibitem{Ahn:2011xq}
C.~Ahn, Z.~Bajnok, D.~Bombardelli and R.I.~Nepomechie, \emph{{TBA, NLO Luscher
  correction, and double wrapping in twisted AdS/CFT}},
  \href{https://doi.org/10.1007/JHEP12(2011)059}{\emph{JHEP} {\bfseries 12}
  (2011) 059} [\href{https://arxiv.org/abs/1108.4914}{{\ttfamily 1108.4914}}].

\bibitem{Arutyunov:2009zu}
G.~Arutyunov and S.~Frolov, \emph{{String hypothesis for the AdS(5) x S**5
  mirror}}, \href{https://doi.org/10.1088/1126-6708/2009/03/152}{\emph{JHEP}
  {\bfseries 03} (2009) 152} [\href{https://arxiv.org/abs/0901.1417}{{\ttfamily
  0901.1417}}].

\bibitem{Dorey:1996re}
P.~Dorey and R.~Tateo, \emph{{Excited states by analytic continuation of TBA
  equations}}, \href{https://doi.org/10.1016/S0550-3213(96)00516-0}{\emph{Nucl.
  Phys. B} {\bfseries 482} (1996) 639}
  [\href{https://arxiv.org/abs/hep-th/9607167}{{\ttfamily hep-th/9607167}}].

\bibitem{doi:10.1143/JPSJ.50.3785}
R.~Hirota, \emph{Discrete analogue of a generalized toda equation},
  \href{https://doi.org/10.1143/JPSJ.50.3785}{\emph{Journal of the Physical
  Society of Japan} {\bfseries 50} (1981) 3785}
  [\href{https://arxiv.org/abs/https://doi.org/10.1143/JPSJ.50.3785}{{\ttfamily
  https://doi.org/10.1143/JPSJ.50.3785}}].

\bibitem{Gromov:2010vb}
N.~Gromov, V.~Kazakov and Z.~Tsuboi, \emph{{PSU(2,2$|$4) Character of
  Quasiclassical AdS/CFT}},
  \href{https://doi.org/10.1007/JHEP07(2010)097}{\emph{JHEP} {\bfseries 07}
  (2010) 097} [\href{https://arxiv.org/abs/1002.3981}{{\ttfamily 1002.3981}}].

\bibitem{Cavaglia:2010nm}
A.~Cavaglia, D.~Fioravanti and R.~Tateo, \emph{{Extended Y-system for the
  $AdS_5/CFT_4$ correspondence}},
  \href{https://doi.org/10.1016/j.nuclphysb.2010.09.015}{\emph{Nucl. Phys. B}
  {\bfseries 843} (2011) 302}
  [\href{https://arxiv.org/abs/1005.3016}{{\ttfamily 1005.3016}}].

\bibitem{Balog:2011nm}
J.~Balog and A.~Hegedus, \emph{{$AdS_5\times S^5$ mirror TBA equations from
  Y-system and discontinuity relations}},
  \href{https://doi.org/10.1007/JHEP08(2011)095}{\emph{JHEP} {\bfseries 08}
  (2011) 095} [\href{https://arxiv.org/abs/1104.4054}{{\ttfamily 1104.4054}}].

\bibitem{Gromov:2011cx}
N.~Gromov, V.~Kazakov, S.~Leurent and D.~Volin, \emph{{Solving the AdS/CFT
  Y-system}}, \href{https://doi.org/10.1007/JHEP07(2012)023}{\emph{JHEP}
  {\bfseries 07} (2012) 023} [\href{https://arxiv.org/abs/1110.0562}{{\ttfamily
  1110.0562}}].

\bibitem{Gromov:2017blm}
N.~Gromov, \emph{{Introduction to the Spectrum of $N=4$ SYM and the Quantum
  Spectral Curve}},  \href{https://arxiv.org/abs/1708.03648}{{\ttfamily
  1708.03648}}.

\bibitem{Levkovich-Maslyuk:2019awk}
F.~Levkovich-Maslyuk, \emph{{A review of the AdS/CFT Quantum Spectral Curve}},
  \href{https://doi.org/10.1088/1751-8121/ab7137}{\emph{J. Phys. A} {\bfseries
  53} (2020) 283004} [\href{https://arxiv.org/abs/1911.13065}{{\ttfamily
  1911.13065}}].

\bibitem{Kazakov:2015efa}
V.~Kazakov, S.~Leurent and D.~Volin, \emph{{T-system on T-hook: Grassmannian
  Solution and Twisted Quantum Spectral Curve}},
  \href{https://doi.org/10.1007/JHEP12(2016)044}{\emph{JHEP} {\bfseries 12}
  (2016) 044} [\href{https://arxiv.org/abs/1510.02100}{{\ttfamily
  1510.02100}}].

\bibitem{Marboe:2019wyc}
C.~Marboe and E.~Wid\'en, \emph{{The fate of the Konishi multiplet in the
  $\beta$-deformed Quantum Spectral Curve}},
  \href{https://doi.org/10.1007/JHEP01(2020)026}{\emph{JHEP} {\bfseries 01}
  (2020) 026} [\href{https://arxiv.org/abs/1902.01248}{{\ttfamily
  1902.01248}}].

\bibitem{Volin:2010xz}
D.~Volin, \emph{{String hypothesis for gl(n$|$m) spin chains: a particle/hole
  democracy}}, \href{https://doi.org/10.1007/s11005-012-0570-9}{\emph{Lett.
  Math. Phys.} {\bfseries 102} (2012) 1}
  [\href{https://arxiv.org/abs/1012.3454}{{\ttfamily 1012.3454}}].

\bibitem{Frolov:2010wt}
S.~Frolov, \emph{{Konishi operator at intermediate coupling}},
  \href{https://doi.org/10.1088/1751-8113/44/6/065401}{\emph{J. Phys. A}
  {\bfseries 44} (2011) 065401}
  [\href{https://arxiv.org/abs/1006.5032}{{\ttfamily 1006.5032}}].

\bibitem{Roiban:2011fe}
R.~Roiban and A.A.~Tseytlin, \emph{{Semiclassical string computation of
  strong-coupling corrections to dimensions of operators in Konishi
  multiplet}},
  \href{https://doi.org/10.1016/j.nuclphysb.2011.02.016}{\emph{Nucl. Phys. B}
  {\bfseries 848} (2011) 251}
  [\href{https://arxiv.org/abs/1102.1209}{{\ttfamily 1102.1209}}].

\bibitem{Frolov:2009in}
S.~Frolov and R.~Suzuki, \emph{{Temperature quantization from the TBA
  equations}},
  \href{https://doi.org/10.1016/j.physletb.2009.06.069}{\emph{Phys. Lett. B}
  {\bfseries 679} (2009) 60} [\href{https://arxiv.org/abs/0906.0499}{{\ttfamily
  0906.0499}}].

\bibitem{Gromov:2009zb}
N.~Gromov, V.~Kazakov and P.~Vieira, \emph{{Exact Spectrum of Planar ${\cal
  N}=4$ Supersymmetric Yang-Mills Theory: Konishi Dimension at Any Coupling}},
  \href{https://doi.org/10.1103/PhysRevLett.104.211601}{\emph{Phys. Rev. Lett.}
  {\bfseries 104} (2010) 211601}
  [\href{https://arxiv.org/abs/0906.4240}{{\ttfamily 0906.4240}}].

\bibitem{Gromov:2015vua}
N.~Gromov, F.~Levkovich-Maslyuk and G.~Sizov, \emph{{Pomeron Eigenvalue at
  Three Loops in $\mathcal N=$ 4 Supersymmetric Yang-Mills Theory}},
  \href{https://doi.org/10.1103/PhysRevLett.115.251601}{\emph{Phys. Rev. Lett.}
  {\bfseries 115} (2015) 251601}
  [\href{https://arxiv.org/abs/1507.04010}{{\ttfamily 1507.04010}}].

\bibitem{Arutyunov:2009ax}
G.~Arutyunov, S.~Frolov and R.~Suzuki, \emph{{Exploring the mirror TBA}},
  \href{https://doi.org/10.1007/JHEP05(2010)031}{\emph{JHEP} {\bfseries 05}
  (2010) 031} [\href{https://arxiv.org/abs/0911.2224}{{\ttfamily 0911.2224}}].

\bibitem{Hegedus:2016eop}
A.~Heged\'{u}s and J.~Konczer, \emph{{Strong coupling results in the AdS$_{5}$
  /CFT$_{4}$ correspondence from the numerical solution of the quantum spectral
  curve}}, \href{https://doi.org/10.1007/JHEP08(2016)061}{\emph{JHEP}
  {\bfseries 08} (2016) 061}
  [\href{https://arxiv.org/abs/1604.02346}{{\ttfamily 1604.02346}}].

\bibitem{Marboe:2018ugv}
C.~Marboe and D.~Volin, \emph{{The full spectrum of AdS$_5$/CFT$_4$ II: Weak
  coupling expansion via the quantum spectral curve}},
  \href{https://doi.org/10.1088/1751-8121/abd59c}{\emph{J. Phys. A} {\bfseries
  54} (2021) 055201} [\href{https://arxiv.org/abs/1812.09238}{{\ttfamily
  1812.09238}}].

\bibitem{Alfimov:2014bwa}
M.~Alfimov, N.~Gromov and V.~Kazakov, \emph{{QCD Pomeron from AdS/CFT Quantum
  Spectral Curve}}, \href{https://doi.org/10.1007/JHEP07(2015)164}{\emph{JHEP}
  {\bfseries 07} (2015) 164} [\href{https://arxiv.org/abs/1408.2530}{{\ttfamily
  1408.2530}}].

\bibitem{Gromov:2015dfa}
N.~Gromov and F.~Levkovich-Maslyuk, \emph{{Quantum Spectral Curve for a cusped
  Wilson line in $ \mathcal{N}=4 $ SYM}},
  \href{https://doi.org/10.1007/JHEP04(2016)134}{\emph{JHEP} {\bfseries 04}
  (2016) 134} [\href{https://arxiv.org/abs/1510.02098}{{\ttfamily
  1510.02098}}].

\end{thebibliography}\endgroup
\end{document}